\documentclass[numberedappendix,appendixfloats]{emulateapj}
\usepackage{graphicx}
\usepackage{amssymb}
\usepackage{amsmath}
\usepackage{ulem}
\usepackage[colorlinks=true,linkcolor=blue,anchorcolor=blue,citecolor=blue,urlcolor=blue]{hyperref}  

\shorttitle{Resolving the High Energy Universe with Strong Gravitational Lensing}
\shortauthors{Barnacka et al.}

\begin{document}
\title{Resolving the High Energy Universe with Strong Gravitational Lensing: \\ The Case of PKS 1830-211}
\author{Anna Barnacka$^{1,2}$}
\author{Margaret J. Geller$^1$}
\author{Ian P. Dell'Antonio$^3$}
\author{Wystan Benbow$^1$}
\affil{$^1$Harvard-Smithsonian Center for Astrophysics, 60 Garden St, MS-20, Cambridge, MA 02138, USA\\
$^2$Astronomical Observatory, Jagiellonian University, Cracow, Poland \\
$^3$Department of Physics, Brown University, Box 1843, Providence, RI 02912}

\email{abarnacka@cfa.harvard.edu}

\begin{abstract}

Gravitational lensing is a potentially powerful tool for elucidating  the origin of gamma-ray  emission from distant  sources.
Cosmic lenses magnify the emission from distance sources
and produce  time delays between mirage images.
Gravitationally-induced time delays depend on the position of the emitting regions in the source plane.
The {\it Fermi}/LAT telescope continuously monitors the entire sky and detects gamma-ray flares, 
including those from gravitationally-lensed blazars.
Therefore,  temporal resolution at gamma-ray energies  can  be used to measure these time delays,
which, in turn, can be used  to resolve the origin of the gamma-ray flares spatially. We
provide a guide to the application and Monte Carlo simulation of three techniques for analyzing these unresolved light curves: 
the Autocorrelation Function, the Double Power Spectrum, and the Maximum Peak Method.
We apply these methods to derive time delays from the gamma-ray light curve of the  gravitationally-lensed blazar PKS 1830-211.
The result of temporal analysis combined with the properties of the lens from radio observations  yield an improvement in  spatial resolution  at gamma-ray energies  by a factor of 10000.
We analyze four active periods. 
For two of these periods, the emission is consistent with origination from the core 
and for the other two, the data suggest that the emission region is displaced from the core by more that $\sim 1.5$ kpc. 
For the core emission, the gamma-ray time delays, $23\pm0.5\,$days and $19.7\pm1.2\,$days, are consistent with the radio time delay $26^{+4}_{-5}\,$days.

\end{abstract}


%%%%%%%%%%%%%%%%%%%%%%%%%%%%%%%%%%%%%%
\section{Introduction}
%%%%%%%%%%%%%%%%%%%%%%%%%%%%%%%%%%%%%%

Our ability to study  gamma-ray radiation from distant sources
is observationally limited by the poor ($\sim 0.1^\circ$) angular resolution of the detectors. 
We demonstrate a technique for using a cosmic lens to enhance the angular resolution at gamma-ray energies.

%%%%%%%%%%%%%%%%%%%%%%%%%%%%%%%%%%%%%%
\subsection{History and Challenges} 
%%%%%%%%%%%%%%%%%%%%%%%%%%%%%%%%%%%%%%

The high-energy sky is dominated by the most extreme and puzzling objects in the universe. 
These sources harbor powerful jets, the largest particle accelerators in the universe;
they produce radiation ranging from radio wavelengths up to very high-energy gamma rays.
Observations of resolved jets 
reveal very complex structures  including bright knots, blobs and filaments with sizes ranging from sub-pc up to dozens of kpc
\citep{1991AJ....101.1632B,2002ApJ...570..543S,2005ApJS..156...13M,2006ARA&A..44..463H,2007ApJ...662..900T,2008Natur.452..966M}.  
Observations with improved angular resolution reveal that the variability of these sources is very complex;
the flaring emission  originates from regions close to the core, or from knots along the jet, 
as in  M87 \citep{2003ApJ...586L..41H,2006ApJ...640..211H}.

At gamma-ray energies, the Large Area Telescope onboard the Fermi mission ({\it Fermi}/LAT)  
continuously detects flares from hundreds of blazars. 
{\it Fermi}/LAT scans  the entire sky in a few hours with excellent time resolution. 
However, the angular resolution of {\it Fermi}/LAT 
is at best $\sim 0.1^\circ$, precluding localization of the flares along the jets,  even for nearby sources. 
To resolve gamma-ray emission from  M87, an improvement in angular resolution by a factor of 1000 is required. 
The angular resolution of pair-creation gamma-ray detectors is typically a strong function of gamma-ray energy and viewing angle, 
and is limited by  multiple scattering of electron-positron pairs and by bremsstrahlung \citep{2009ApJ...697.1071A}. 
For medium-energy gamma-rays, other physical effects, including the nuclear recoil, 
limit the ultimate angular resolution of any nuclear pair-production telescope. 
The angular resolution is unlikely to improve by more than a factor of a few in future instruments  \citep{2014arXiv1406.4830B,2014arXiv1407.0710W,2014APh....59...18H}. 

Jets can be well resolved in the radio band down to sub parsec scales. 
The variability time scales at radio wavelengths are large, weeks to months, 
compared to gamma-ray time scales of hours to days. 
Radio monitoring shows that roughly 2/3 of the gamma-ray flares coincide  with 
the appearance of a new superluminal knot and/or a flare in the millimeter-wave core located parsecs from the central engine 
\citep{2010ApJ...710L.126M,2011ApJ...726L..13A,2012arXiv1204.6707M,2012arXiv1201.5402M,2014MNRAS.441.1899F,2014MNRAS.445.1636R,2015arXiv150503871C}.

Further study shows correlations and similarities  between multiwavelength  and gamma-ray observations \citep{2012ApJ...754...23L,2012ApJ...749..191C,2014A&A...562A..79S}.
However, \citet{2014MNRAS.445..428M} modeled the light curves of blazars as red noise processes, and found  that only 1 of 41 sources with high-quality data in both the radio and gamma-ray bands shows correlations with a significance larger than $3\sigma$. They thus demonstrate the difficulties of measuring statistically robust multiwavelength correlations even when the data span many years. 

%%%%%%%%%%%%%%%%%%%%%%%%%%%%%%%%%%%%%%
\subsection{The Importance of the Spatial Origin of Flares}
%%%%%%%%%%%%%%%%%%%%%%%%%%%%%%%%%%%%%%

The origin of gamma-ray flares is a subject of intense debate  \citep{2014ApJ...789..161N,2010MNRAS.405L..94T}.
In  theoretical modeling, 
it is generally assumed that  gamma-ray flares originate from 
 regions close to the central engine, typically on  parsec scales 
 \citep{2014ApJ...796L...5N,2014A&A...567A.113B,2014ApJ...789..161N,2015MNRAS.448.3121H,2011ApJ...733...19T}.  

Models where the emission originates from spatially distinct
knots, as in M87, differ in the underlying fundamental physics 
\citep{1992A&A...256L..27D,1996ApJ...461..657B,1994ApJ...421..153S,2013ApJ...779...68S,2009ApJ...704...38S,2009MNRAS.395L..29G,2003APh....18..593M,2013ApJ...768...54B,2004A&A...419...89R,2005ApJ...626..120S,2006MNRAS.370..981S,2014ApJ...780L..27M}.
Thus, the location of gamma-ray flares is crucial  for understanding  particle acceleration  and magnetic fields at both small and  large distances from  massive black holes.

Multiple variable emitting regions place limitations on the use of the most variable quasars 
for measurement of the Hubble parameter based on time delays.
 \citet{2015ApJ...799...48B} point out that even a small spatial offset, for example $5\%$ of Einstein radius,  
 between the resolved position of the core and  site of variable emission may result in
 a bimodal distribution of values of Hubble parameters characterized by an RMS of  $\sim12\, [\mbox{km}\, \mbox{s}^{-1}\,\mbox{Mpc}^{-1}]$. 
Complex structure  can be an important source of systematics in measurement of the Hubble parameter from  gravitationally  induced time delays. 

%%%%%%%%%%%%%%%%%%%%%%%%%%%%%%%%%%%%%%
\subsection{Gravitational Lensing as a High Resolution Cosmic Telescope }
%%%%%%%%%%%%%%%%%%%%%%%%%%%%%%%%%%%%%%
 
Time delays and magnification ratios derived from well-sampled light curves
gamma-ray sources can  elucidate the location of the emitting region. 
 \citet{2014arXiv1403.5316B} built a toy model placing an M87 analog at $z\sim1$, 
 with  a galaxy acting as a lens at $z\sim0.6$. 
 In the M87 case, the projected distance between the core and the flaring knot HST1 is 60 pc \citep{1991AJ....101.1632B} 
 or in the toy model $\sim 3$\% of the Einstein radius. 
In the toy model, 
the differences in time delay ($\sim 3$ days) and magnification ratio ($\sim 0.2$)  resulting from displacement of the radiation source to the jet location can be measured with current instruments. 
Thus cosmic lenses are, in principle, a tool for revealing the origin of gamma-ray flares once the time delays are measured. 
 
Measurement of  the time delays from unresolved light curves requires long, evenly-sampled time series with low photon noise.  
Thanks to the {\it Fermi}/LAT observation strategy, 
 the light curves are perfectly suited for these time delay measurements. 
 Thus, the temporal resolution of {\it Fermi}/LAT translates into spatial resolution 
with the help of cosmic lenses. 

%%%%%%%%%%%%%%%%%%%%%%%%%%%%%%%%%%%%%%
\subsection{Outline}
%%%%%%%%%%%%%%%%%%%%%%%%%%%%%%%%%%%%%%

We use PKS~1830-211 
as a prototypical example of a gravitationally-lensed, gamma-ray emitting system (Section~\ref{sec:PKS1830_lens};Section~\ref{sec:PKS1830_gamma}).
Our ability to resolve the gamma-ray sky relies on time delay measurement (Section~\ref{sec:timedelay}).  
We evaluate our methods with Monte Carlo simulations (Section~\ref{sec:MCSettings}).
We characterize the signal in Section~\ref{sec:Signal},
and  evaluate the probability of detecting gravitationally-induced time delays and distinguishing them 
from spurious fluctuations in the power law noise using approaches described in Sections ~\ref{sec:Significance} and~\ref{sec:Detectability}. 

We use three methods of  time delay estimation:
the Autocorrelation Function (Section~\ref{sec:ACF}), 
the Double Power Spectrum (Section~\ref{sec:DPS} and Appendix~\ref{app:DPS}),
and the Maximum Peak Method (Section~\ref{sec:MPM} and Appendix~\ref{app:MPM}).

We analyze four series  of gamma-ray flares in Sections~\ref{sec:Flare1},~\ref{sec:Flare2},~\ref{sec:Flare3}, and~\ref{sec:Flare4}. 
We use the time delays and properties of the lens to elucidate the origin of these flares which we  discuss in Section~~\ref{sec:discussion},
and we conclude in Section~\ref{sec:conclusions}.

%%%%%%%%%%%%%%%%%%%%%%%%%%%%%%%%%%%%%%
\section{ PKS~1830-211}
%%%%%%%%%%%%%%%%%%%%%%%%%%%%%%%%%%%%%%

\begin{figure*}
\begin{center} 
%http://colormap.org/
\includegraphics[width=8.9cm,angle=0]{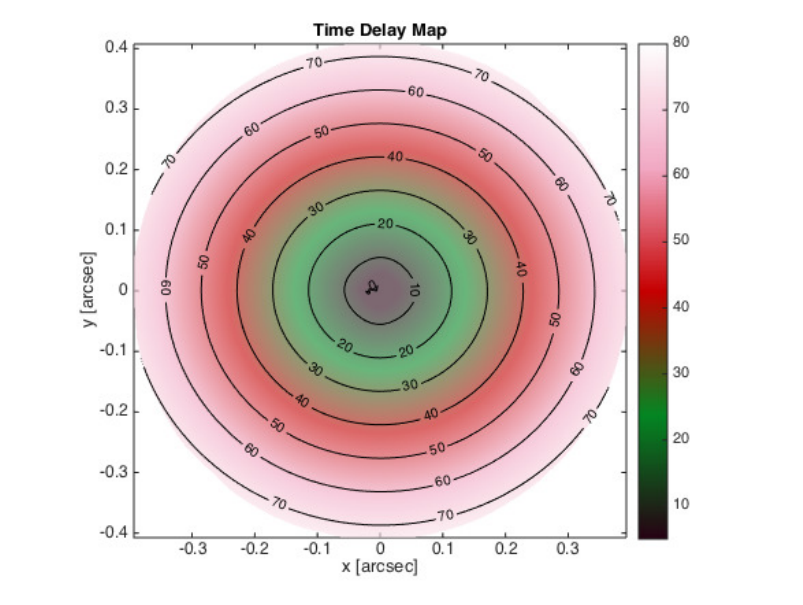}
\includegraphics[width=8.9cm,angle=0]{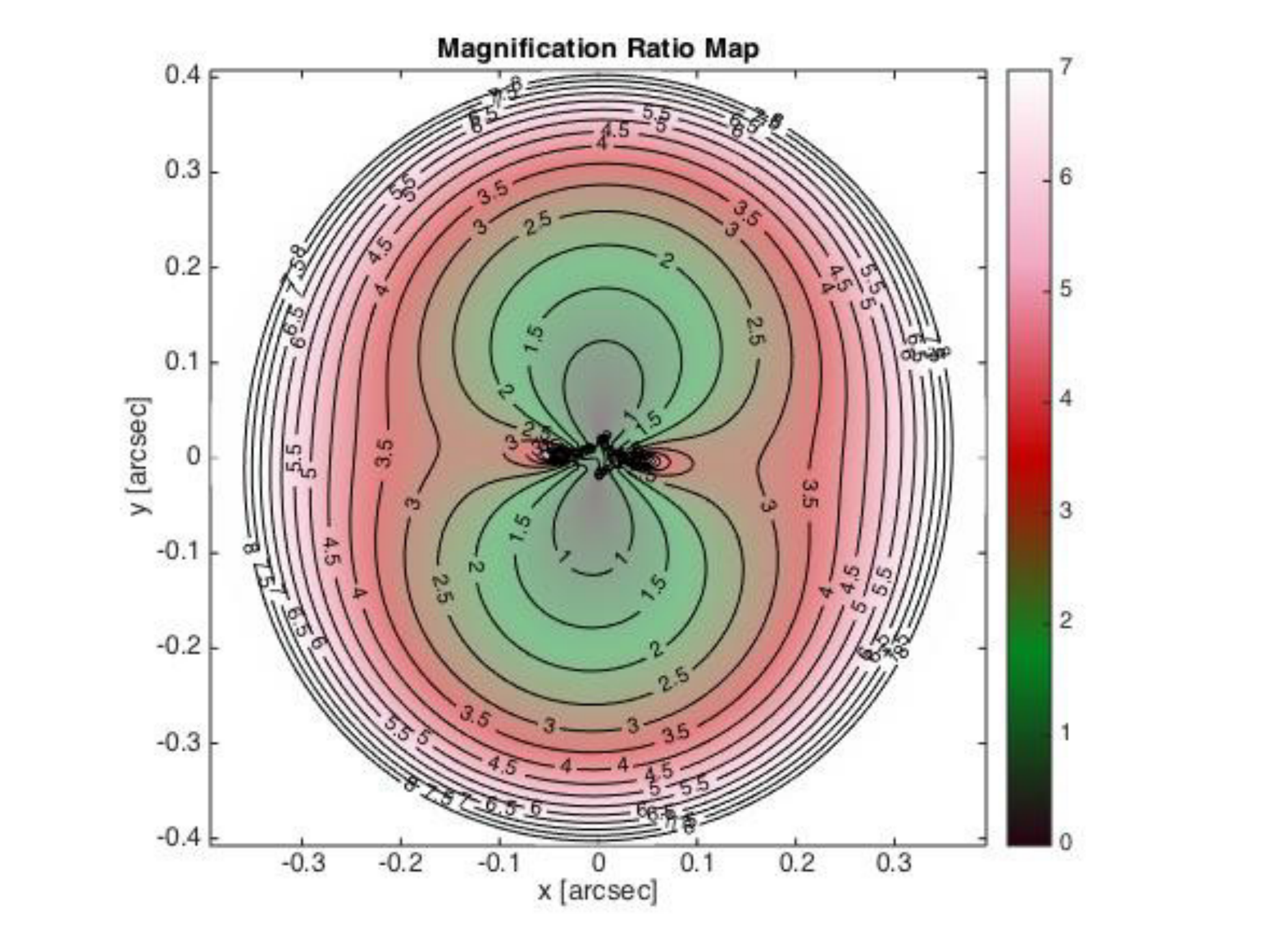}
\end{center}
\caption{\label{fig:dt_map}  Maps based on the lens model of the lensing galaxy for PKS~1830-211 (Section~\ref{sec:PKS1830_lens}). 
					   Left: Time delays between mirage images for different positions of the emitting region within  the Einstein radius.
					   The color scale bar represents time delay in days.
					   Right: Magnification ratio map for different positions of the emitting regions in the source plane.  }
\end{figure*}

PKS~1830-211 is a bright blazar 
lensed by a galaxy located close to the line-of-sight. 
The emission from PKS~1830-211 is detected  
from radio up to $\gamma$-ray wavelengths. 
In section~\ref{sec:PKS1830_lens},  we describe the gravitationally-lensed system 
and, in section~\ref{sec:PKS1830_gamma}, we detail the gamma-ray emission as observed by the Fermi satellite over a 6.5-year period.

%%%%%%%%%%%%%%%%%%%%%%%%%%%%%%%%%%%%%%
\subsection{PKS~1830-211 as Gravitationally-Lensed System }
\label{sec:PKS1830_lens}
%%%%%%%%%%%%%%%%%%%%%%%%%%%%%%%%%%%%%%

\citet{1988MNRAS.231..229P} first recognized PKS~1830-211 as a gravitationally-lensed flat spectrum radio quasar.
The lens is a face-on spiral galaxy  at redshift $z = 0.886$  \citep{1996Natur.379..139W,2001ASPC..237..155W, 2002ApJ...575..103W}. 
The quasi-stellar object, a blazar with a powerful jet, has
redshift $z=2.507$ \citep{1999ApJ...514L..57L}. 
 
The lens  distorts the extended radio emission from the jet 
and produces   semi-ring-like structure, along with two bright images separated
by roughly one arcsecond \citep{1991Natur.352..132J}. 
The Australia Telescope Compact Array observation 
at 8.6 GHz revealed these compact components, 
interpreted as  emission from the jet core.
  
The radio monitoring program lasted  for 18 months and 
resulted in the measurement of  a time delay  of $26^{+4}_{-5}\,$days,
and a magnification ratio of $1.52\pm0.05$ \citep{1998ApJ...508L..51L}. 
An independent measurement of the time delay using molecular absorption lines 
yields a delay of $24^{+5}_{-4}\,$days, consistent with the result of radio monitoring \citep{2001ASPC..237..155W}.

The images of PKS~1830-211 have prominent radio core-knot structures 
where the axis is roughly perpendicular to the line of separation of the two mirage images of the core.
These radio observations define the projection of the jet.
The mirage images of the core are separated by about 0.98 arcsec 
and aligned northeast (NE) and southwest (SW) in the plane of the sky 
\citep{2005MNRAS.362.1157N,1997VA.....41..281G,1996ApJ...470L..23J,1999A&A...346..392G}.
\citet{2003MNRAS.340.1309J} reported time dependent positional variations in the radio centroids at 43 GHz.
\citet{2013A&A...558A.123M}  investigated  the chromatic variability in resolved lensed images of PKS~1830-211 with ALMA to probe the jet base of this object. 
Recently, \citet{2015Sci...348..311M}  analyzed ALMA data and detected a polarization signal (Faraday rotation) related to the strong magnetic field at the jet base of PKS~1830-211. 

\citet{2002ApJ...575..103W} modeled the lens parameters. 
\citet{Sridhar2013} refinements yields  a best fit for
a singular isothermal ellipsoid (SIE)  with ellipticity $e=0.091$ and a lens oriented at 
$86.1^\circ$  \citep[for detailed definition of lens model see][]{2001astro.ph..2341K}.
We use these SIE lens parameters, the  redshifts of the source and lens, 
 and the Einstein radius of 0.491~arcseconds, 
 to build time delay and magnification ratio  maps. 
 We base our maps on the  {\it gravlens} code \citep{2001astro.ph..2341K,2001astro.ph..2340K}.
 
Figure~\ref{fig:dt_map} (left) shows the time delays between mirage images of 
the emitting region in the source plane.  
For PKS~1830-211, the maximum time delay between the mirage images, 
 which occurs when the source is close to the Einstein radius, is $\sim 70$ days.
In the outer region of the source plane, the magnification ratio between the mirage images  $\gtrsim 10$. 
One of the images can  be so faint that large time delays are undetectable.  
The right panel of  Figure~\ref{fig:dt_map} shows a
map of magnification ratios between mirage images of 
the emitting region located as a function of the  position in the source plane.

\begin{figure}
%\vskip 1cm
\begin{center}
\includegraphics[width=8.4cm,angle=0]{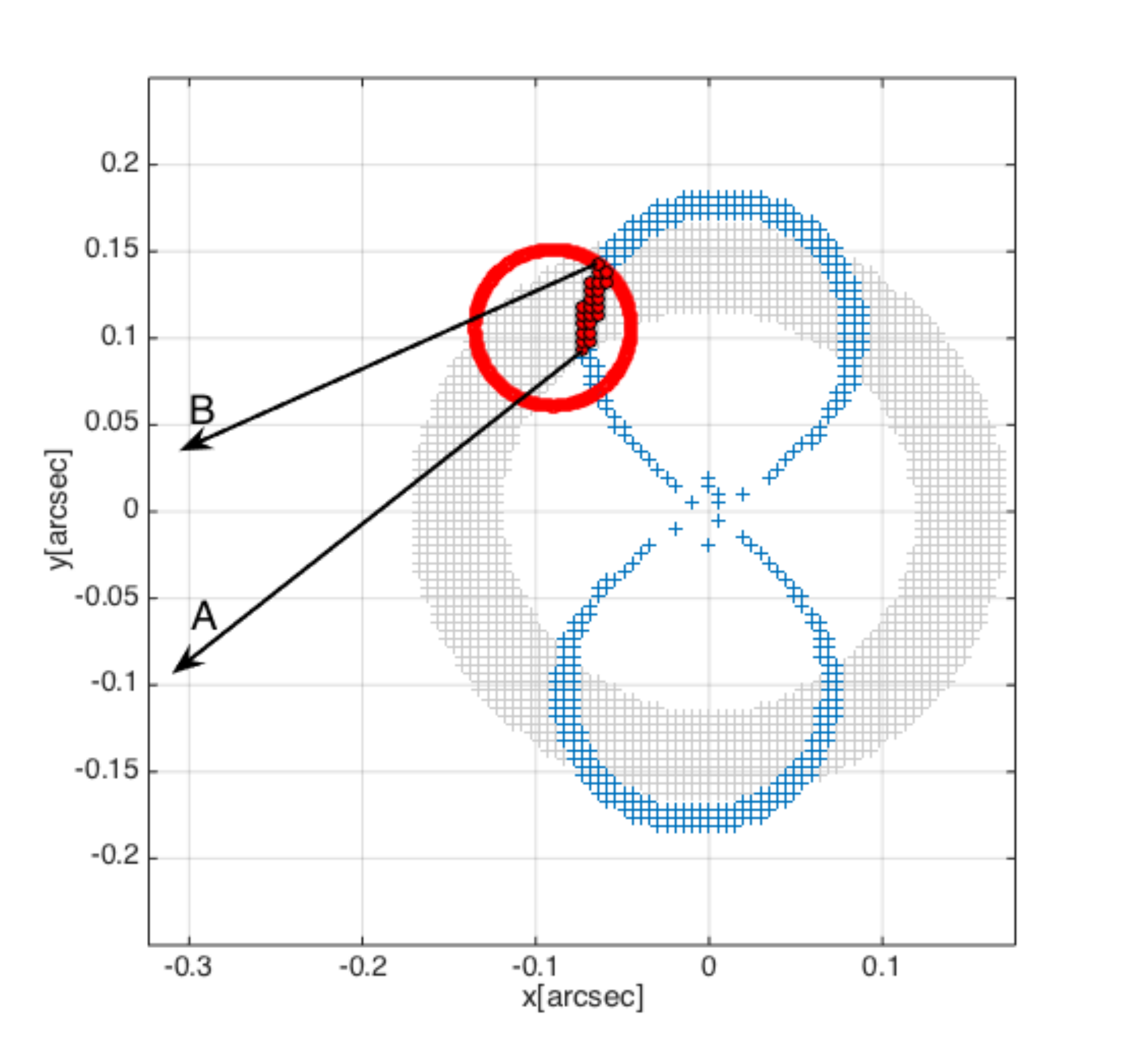}
\end{center}
\caption{\label{fig:core}  The range of possible core locations and the jet projections  in the source plane. 
The gray area shows the allowed range (1$\sigma$ boundary) of the core positions with time delays from 21 to 30 days \citep{1998ApJ...508L..51L}. 
The corresponding magnification ratio between the resolved images is  $1.52\pm0.05$. 
The blue area represents the positions of the core constrained by the magnification ratio measurement. 
The red circle delimits the allowed core positions derived by \citep{Sridhar2013}.
Arrows~A and~B indicate the limiting  jet projections constrained by resolved radio images.
}
\end{figure}

To determine the position of  emitting regions along the jet, 
we need to identify the position of the core and the alignment of the jet. 
\citet{Sridhar2013} determine the position of the core based on  
the  time delay \citep{2011A&A...528L...3B}  and magnification ratio reported by \citet{2013A&A...558A.123M} along with 
the Hubble constant \citep{2013ApJS..208...19H}. %of $70.2\pm2.2\,$km/s/Mpc . 
The red circle in Figure~\ref{fig:core} delimits the region where the core is located.

\begin{figure*}
%\vskip 1cm
\begin{center}
%Flare1_ACF.eps  UL not included in the CL
%/dane/Lensing/data/PKS1830/whole/mag_vs_distance_SIE.gp
\includegraphics[width=3.9cm,angle=-90]{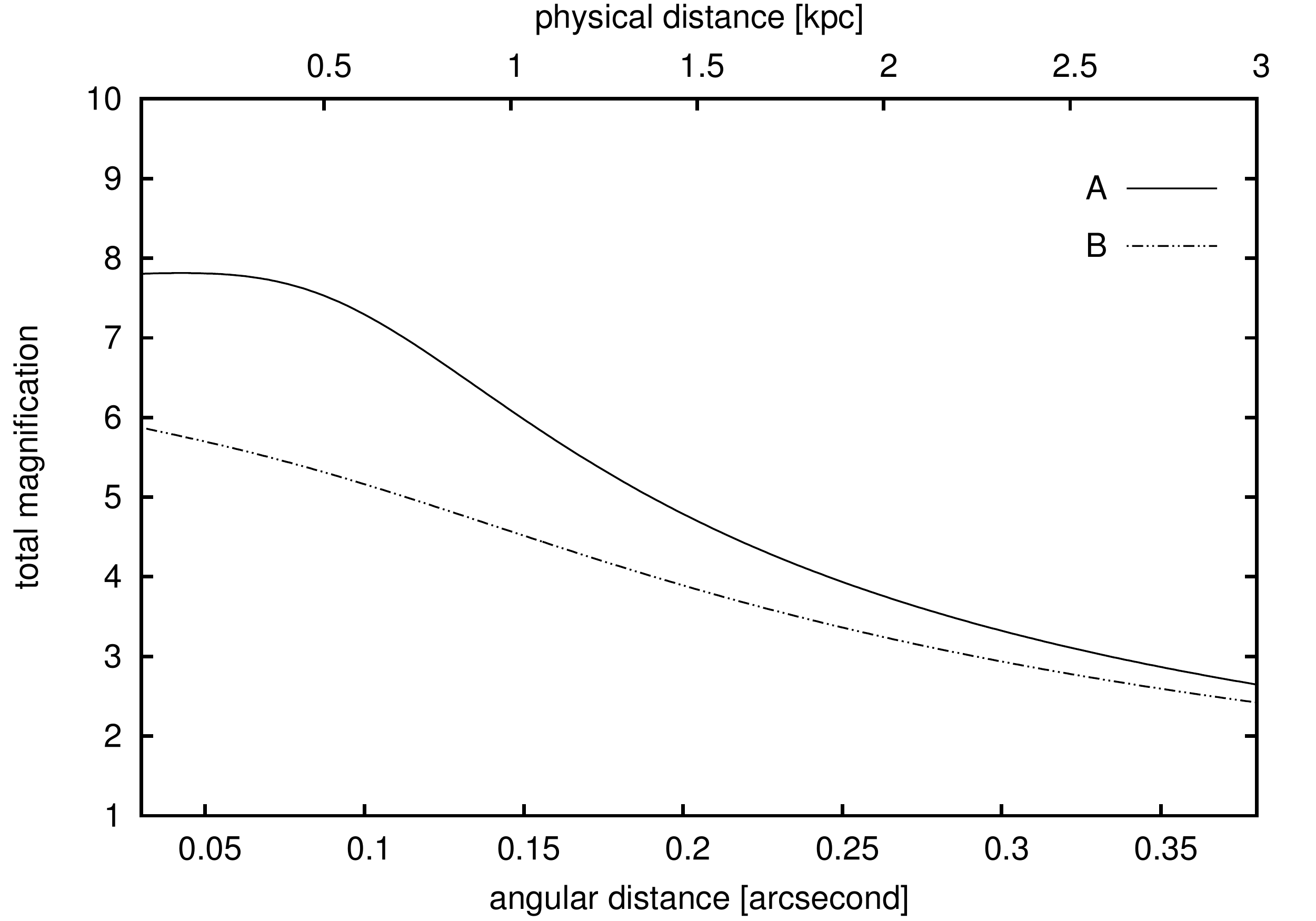}
%/dane/Lensing/data/PKS1830/whole/mr_vs_distance_SIE.gp?
\includegraphics[width=3.9cm,angle=-90]{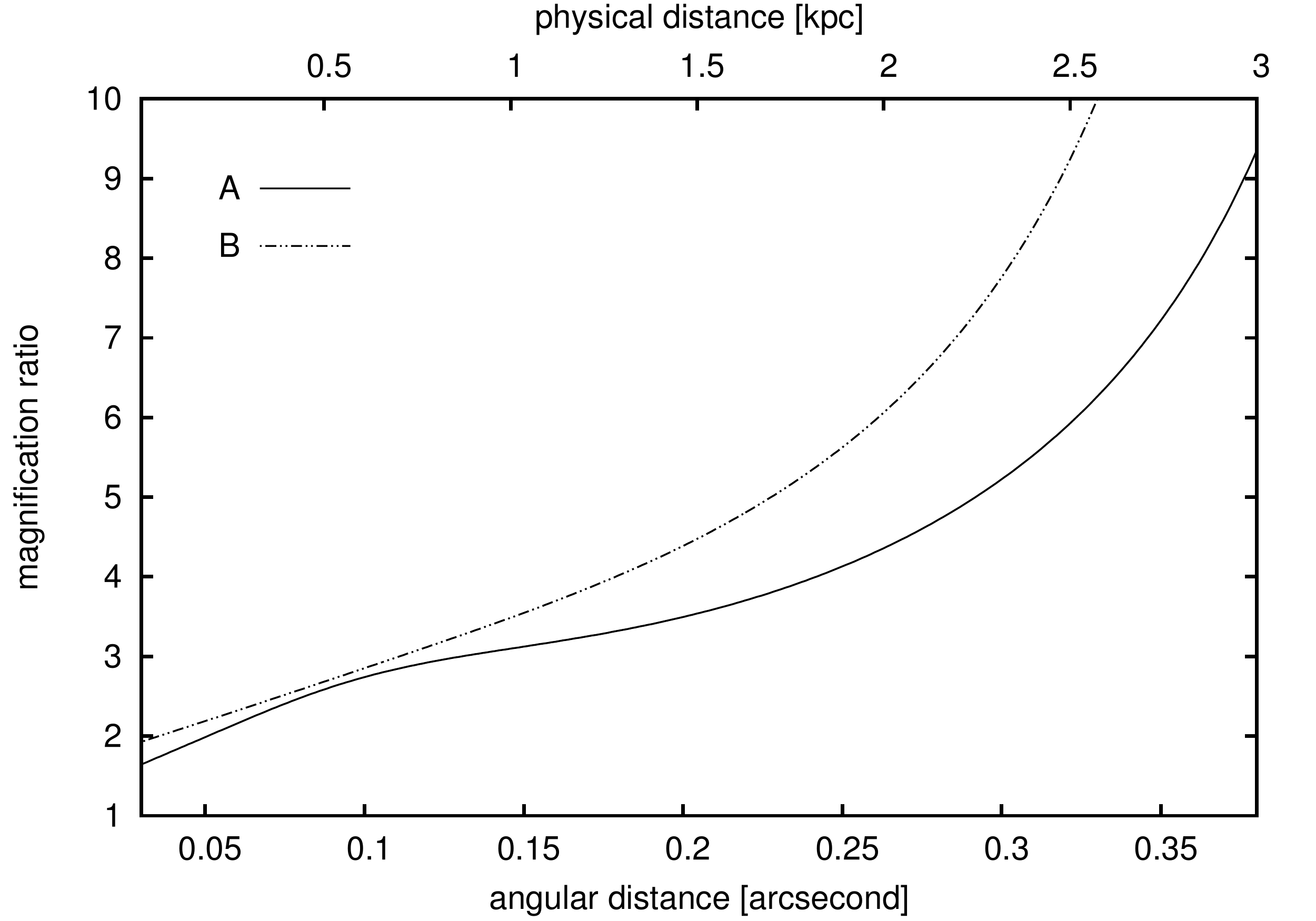}
%/dane/Lensing/data/PKS1830/whole/dt_vs_distance_SIE.gp
\includegraphics[width=3.9cm,angle=-90]{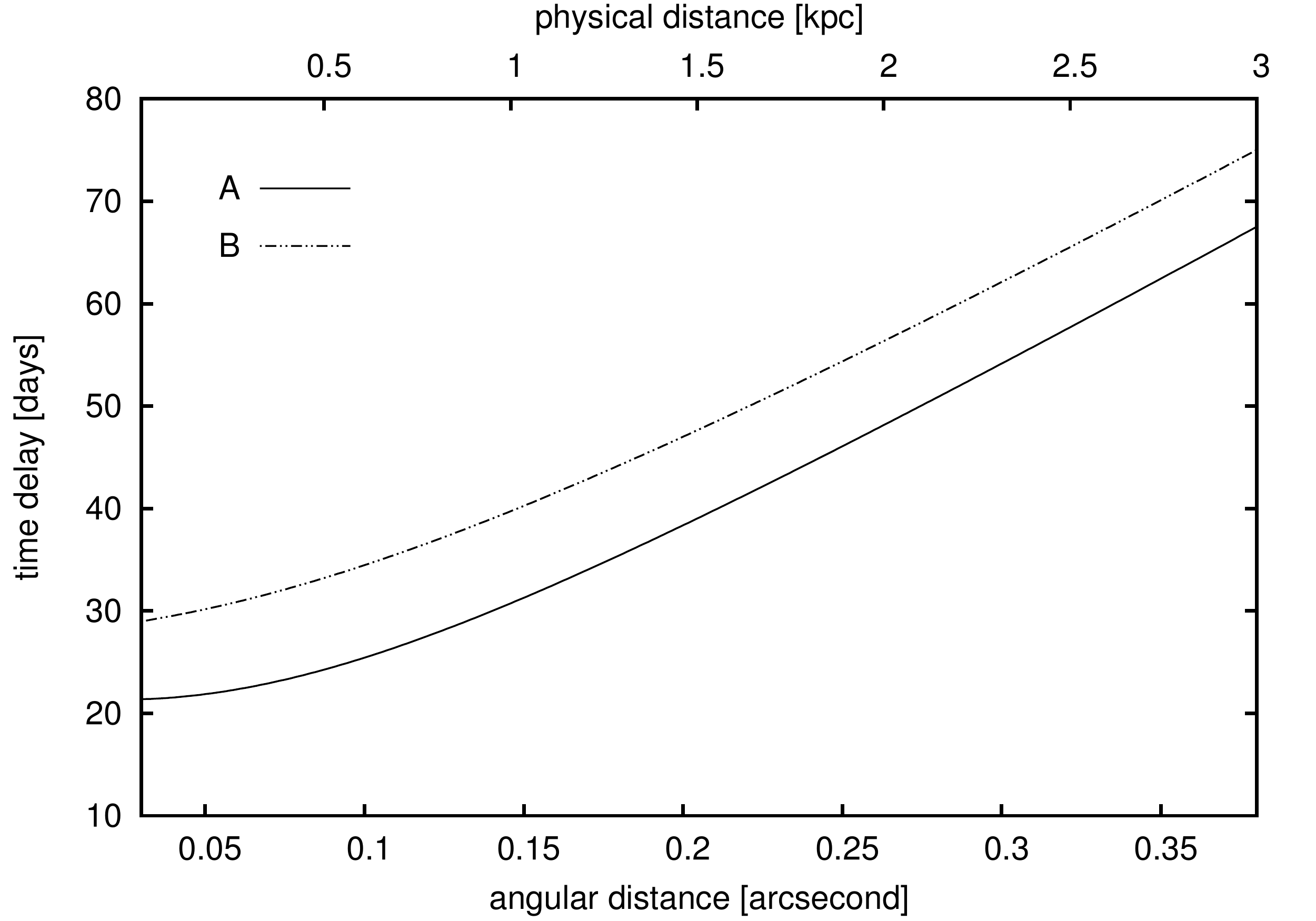}
\end{center}
\caption{\label{fig:jet_lens} Time delays and magnification ratios as a function of the distance between  the emitting region and the core.
					   {\bf Left}: Total magnification defined as the sum of the image magnifications. 
					   {\bf Middle}: Magnification ratios along the limiting jet projections (indicated by arrows in Figure~\ref{fig:core}) 
					   {\bf Right}: Time delays for emitting region located along the limiting jet projections.	
	       }
\end{figure*}

Figure~\ref{fig:jet_lens} shows 
the predicted total magnification,  the time delay, and  magnification ratios 
as a function of position along the jet. 
Once the time delays and magnification ratios are constrained by the light curve,
we  use Figure~\ref{fig:jet_lens} to identify the positions of the emitting regions.
 
%%%%%%%%%%%%%%%%%%%%%%%%%%%%%%%%%%%%%%
\subsection{PKS~1830-211 as a Gamma-ray Emitter}
\label{sec:PKS1830_gamma}
%%%%%%%%%%%%%%%%%%%%%%%%%%%%%%%%%%%%%%

%-------------------------------- Count Map --------------------------------%
%/dane/Lensing/data/PKS1830/whole/PKS1830_cmap.eps
\begin{figure}
%\vskip 1cm
\begin{center}
\includegraphics[width=9.2cm,angle=0]{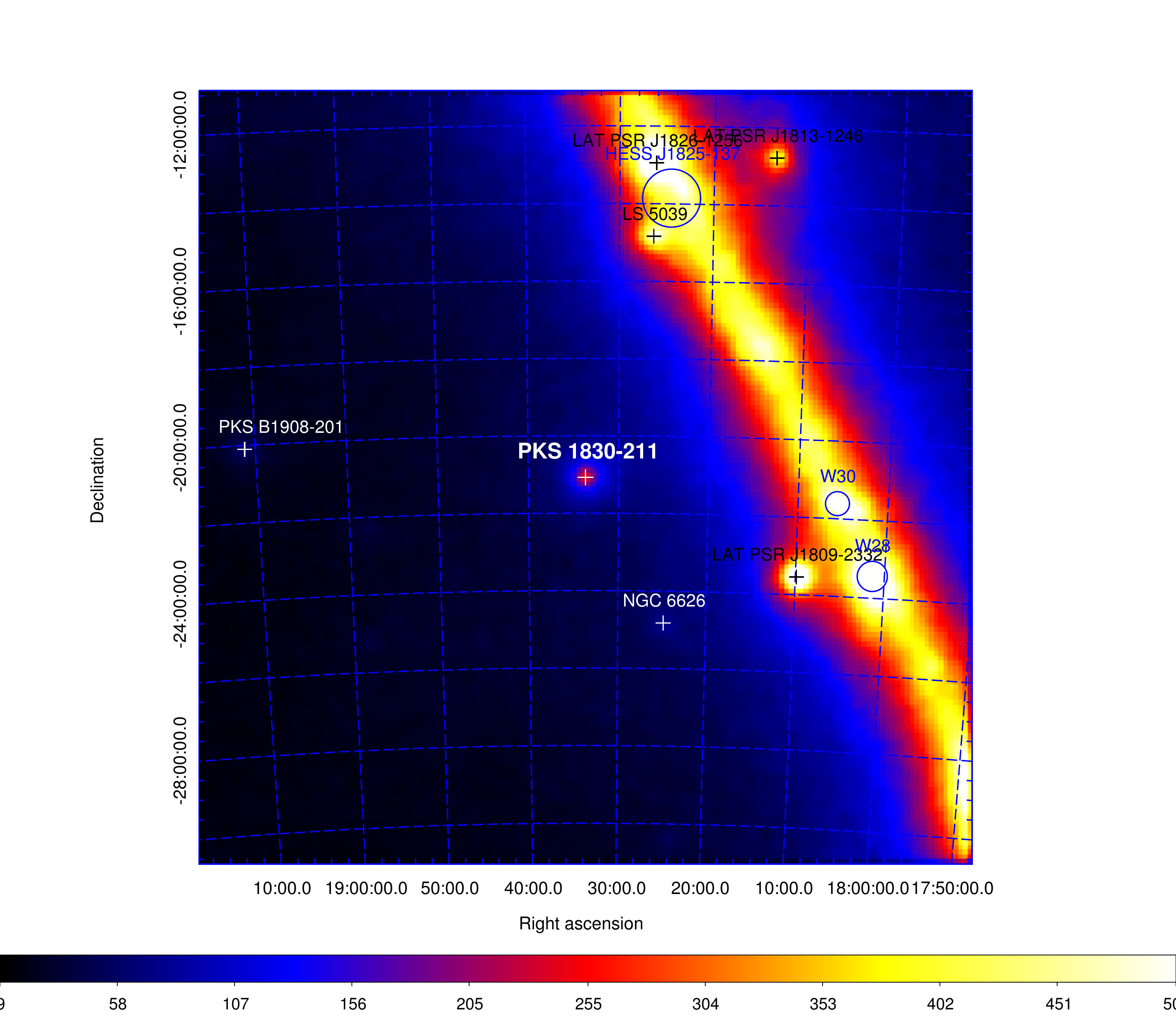}
\end{center}
\caption{\label{fig:cmap_whole} 
                           {\it Fermi}/LAT count map around PKS~1830-211.
                           The map contains photons in the energy range from 200 MeV to 300 GeV. 
                           The counts map is smoothed by a Gaussian kernel of $\sigma = 0.2\,$deg,
                           with the pixel size of $\sigma = 0.025\,$deg.
                           }
\end{figure}

%%%%%%%%%%%%%%%%%%%%%%%%%%%%%%%%%%%%%%
\subsubsection{{\it Fermi}/LAT Data Analysis}
%%%%%%%%%%%%%%%%%%%%%%%%%%%%%%%%%%%%%%

PKS~1830-211 was
detected at gamma rays \citep{1997ApJ...481...95M,1998A&AS..128..423C,2012ApJS..199...31N,2015ApJ...799..143A}. 
The {\it Fermi}/LAT telescope is sensitive to photons in the energy range from 20~MeV to $>300$~GeV \citep{2009ApJ...697.1071A}. 
Our data analysis from 54682~MJD to 57044~MJD,
and in the energy range from $0.2-300$~GeV,
detects the source at a level of  $94\, \sigma$.  
We analyze the {\it Fermi}/LAT {\tt P7REP}  events and spacecraft data 
using standard likelihood tools distributed with 
the {\tt Science Tools v9r32p5} package available at the {\it Fermi} Science Support Center webpage.

For source detection and for detection of the time delay, 
we used the  {\tt P7\_CLEAN} event selection 
and the associated {\tt P7\_CLEAN\_V6} instrument response function (IRFs).
The events in the  {\tt P7\_CLEAN} class have a high probability of being photons. 
We exclude events with zenith angles $>$~100$^\circ$  from the analysis to 
limit contamination by Earth albedo gamma rays.
In addition, we remove events with a rocking angle of $>$~52$^\circ$  
to eliminate time intervals when the Earth entered the LAT Field of View (FoV).

PKS~1830-211 is about $\sim 5^\circ$ from the  galactic plane (Galactic coordinates, $l=12.^{\circ}2,\,b=-5.^{\circ}7$).
To avoid large background contamination, we analyze only 
events with reconstructed energies above 200~MeV and
selected within a square region of 10$^\circ$ radius 
centered on the coordinates of PKS~1830-211 ($R.A.$=$278.41333,\, Dec$=$-21.07492$).
Figure~\ref{fig:cmap_whole} shows %the {\it Fermi}/LAT 
the count map around PKS~1830-211. 
The source is clearly well-separated from the Galactic plane and there are no significant nearby sources.

To build the gamma-ray light curves, we use
a binned-maximum likelihood method\footnote{http://fermi.gsfc.nasa.gov/ssc/data/analysis/scitools/\\ binned\_likelihood\_tutorial.html} \citep{1996ApJ...461..396M}.
This method accounts for sources in the  region of interest (ROI), including pulsar PSR J1809-2332. 
We model the background using a galactic diffuse emission model ({\tt gll\_iem\_v05}), 
and an isotropic component ({\tt iso\_clean\_v05}) available at the {\it Fermi} Science Support Center webpage. 
The fluxes are derived from the post-launch instrument response functions {\tt P7REP\_CLEAN\_V15}. 

The {\tt XML} source model input to  the binned maximum analysis 
contains all the sources  included in the Second {\it Fermi}/LAT catalogue \citep{2012ApJS..199...31N} 
within an annulus of 20$^\circ$ around the ROI.
We first analyze the data  based on the {\tt XML} source model with free parameters for the sources within 7$^\circ$; the parameters at larger radius
are fixed to their 2FGL values. 
We use this {\tt XML} source model to produce the light curves. 

%%%%%%%%%%%%%%%%%%%%%%%%%%%%%%%%%%%%%%
\subsubsection{Gamma-Ray Light Curves}
%%%%%%%%%%%%%%%%%%%%%%%%%%%%%%%%%%%%%%

\begin{figure*}
%\vskip 1cm
\begin{center}
\includegraphics[width=18.4cm,angle=0]{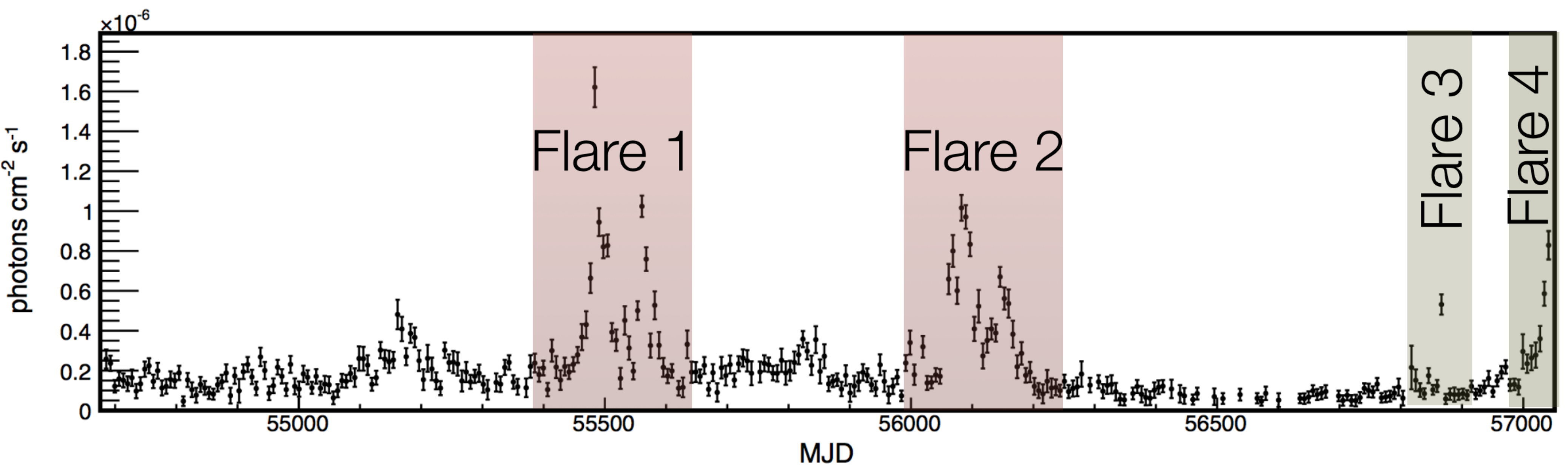}
\end{center}
\caption{\label{fig:lc_whole} 
                           {\it Fermi}/LAT light curve of PKS 1830-211 from August 2008 through February 2015.
                           The fluxes are based on  seven-day binning.
                           The energy range is 200 MeV to 300 GeV. }
\end{figure*}

 Figure~\ref{fig:lc_whole} shows 
the light curve for August 2008 through  February 2015 with 7 day binning.
The energy spectrum  is well described by a power law with 
$\Gamma = 2.54 \pm 0.01$ 
and an integral flux of 
$F(0.2-300\,\mbox{GeV})=(1.94\pm0.02)\times10^{-7}\,\mbox{ph}\,\mbox{cm}^{-2}\mbox{s}^{-1}$.
The highest energy event  was 50~GeV, 
detected in the time window 55389~MJD - 55395~MJD. 
 These energies, in principle, are accessible by the H.E.S.S. II telescope.
Thus detection of PKS~1830-211 may be possible with H.E.S.S.~II.

\begin{figure*}
%\vskip 1cm
\begin{center}
%-------------------------------- Flare 1--------------------------------%
\includegraphics[width=8.9cm,angle=0]{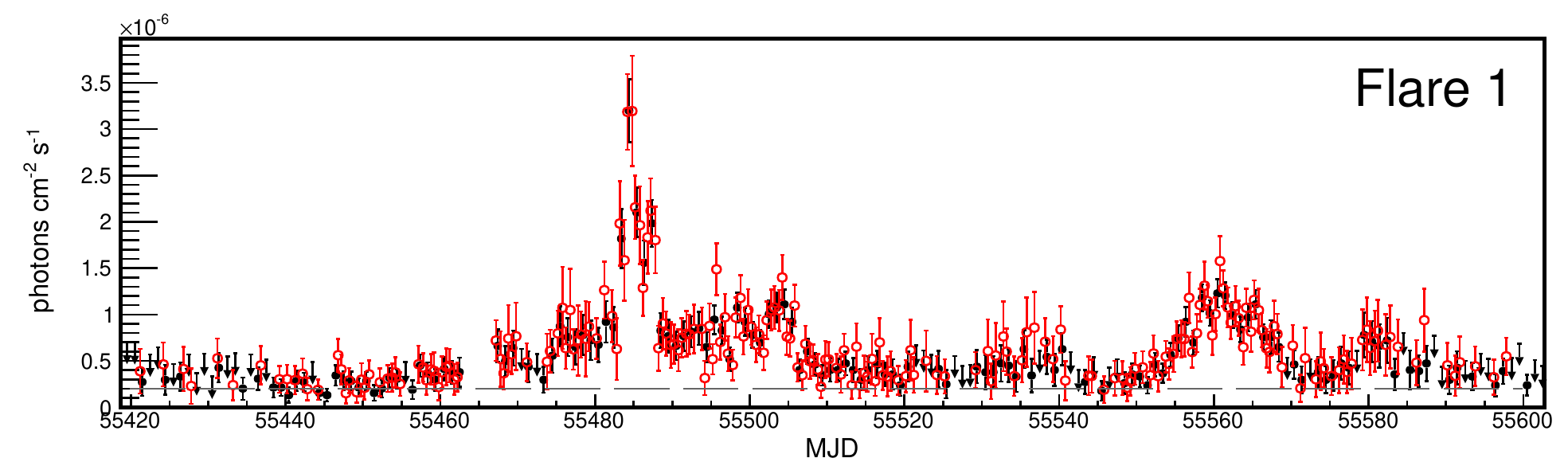} 
%\includegraphics[width=18.5cm,angle=0]{plots/d1_PKS1830_flare1.pdf} \\
%-------------------------------- Flare 2--------------------------------%
%/dane/Lensing/data/PKS1830/flare2/d1_d05_PKS1830_flare2.pdf
\includegraphics[width=8.9cm,angle=0]{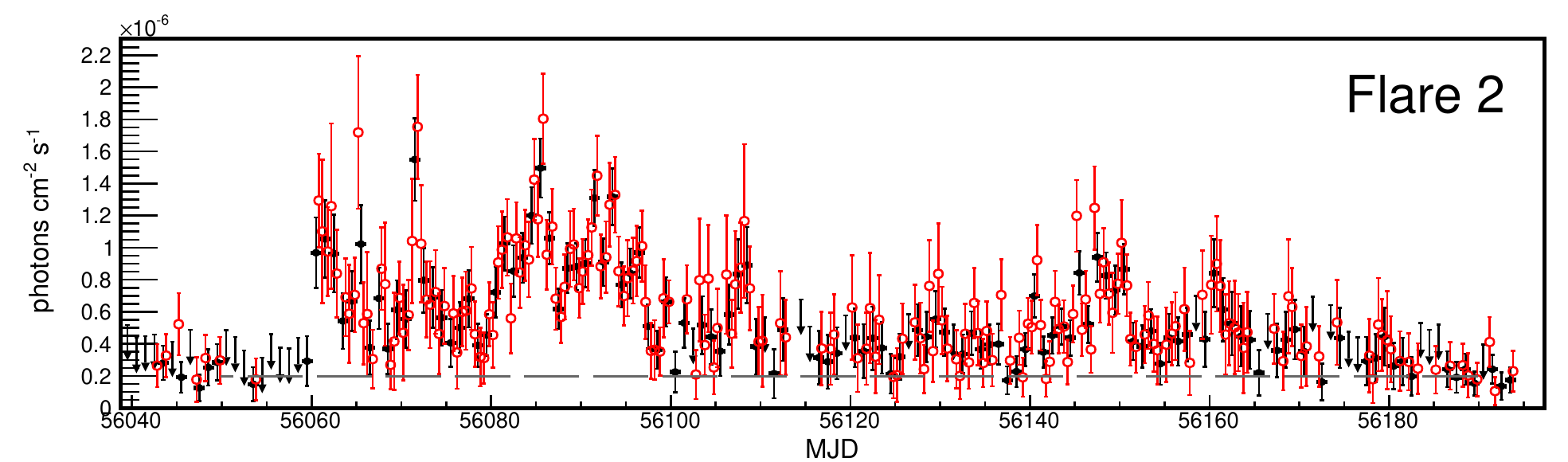} \\
\includegraphics[width=8.9cm,angle=0]{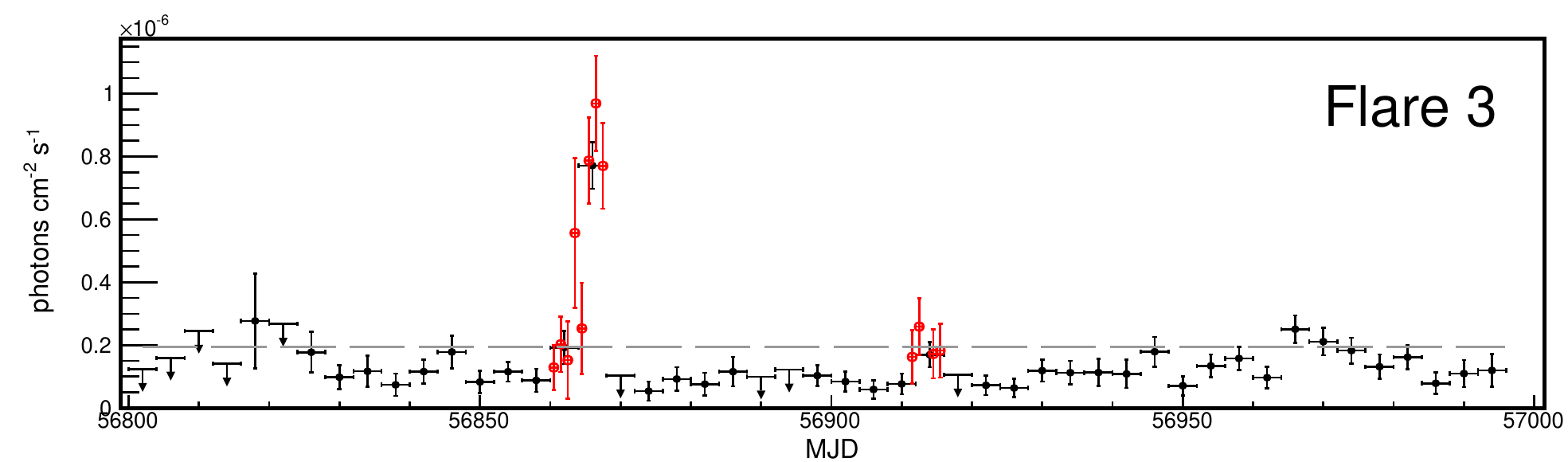} 
\includegraphics[width=8.9cm,angle=0]{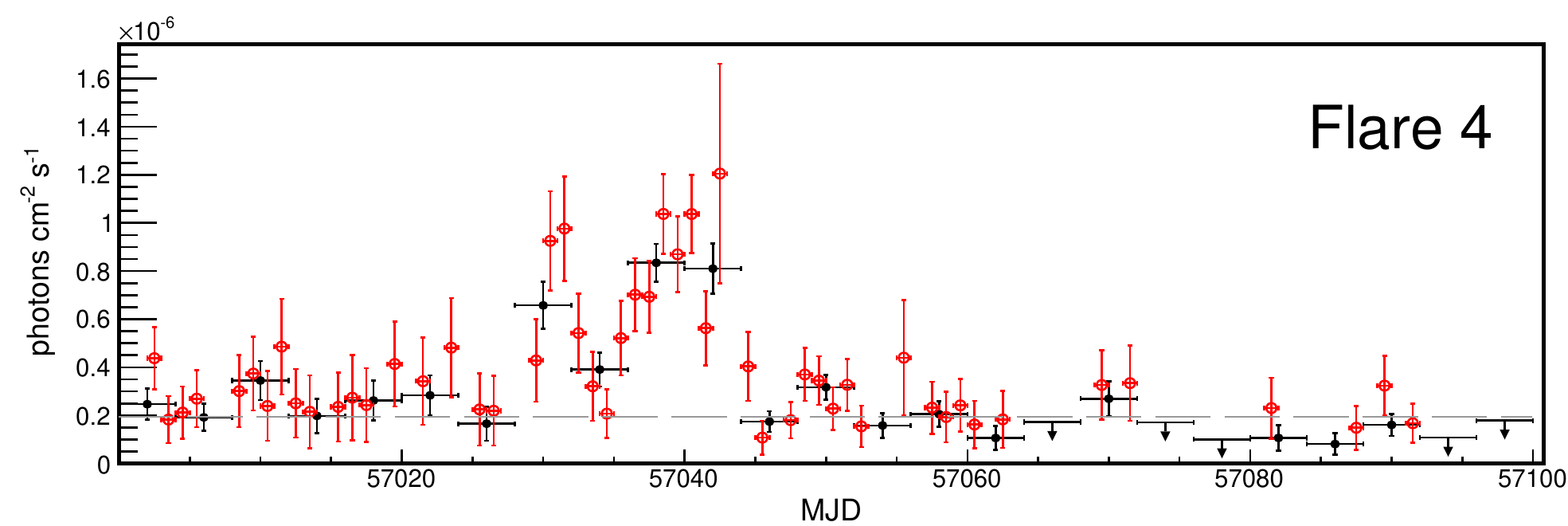} \\
\end{center}
\caption{\label{fig:lc_flares} 
                           {\it Fermi}/LAT light curves of flaring PKS 1830-211. We show
                           Flare~1 and Flare~2 with  one-day binning  (black-filled circles),
                           and with 12-hour binning (red open circles). We show
                           Flare~3 and Flare~4  with  four-day binning  (black-filled circles),
                           and  one-day binning (red open circles). We show
                           red points only for bins with at least $2\,\sigma$ detection. 
                           The green dashed line represents the average flux ($1.94\pm0.02 \times 10^{-7}\, \mbox{photons}\, \mbox{cm}^{-2}\,\mbox{s}^{-1}$) 
                           measured from the 7 years light curve of PKS~1830-211 in the energy range  from 200 MeV to 300 GeV. }
\end{figure*}

Figure~\ref{fig:lc_whole} shows several active periods. 
We define active  periods as times when the gamma-ray emission exceeds the average flux by least  $2\,\sigma$.
This approach yields four active periods.   
The  first series of very bright flares 
occurs in the period  55420~MJD to 55620~MJD. 
The second series of flares occurs in the period 56050~MJD to 56200~MJD.
Next, a bright single flare occurs around July 28, 2014. 
Recently, on January 8, 2015, another flare occurred.
Figure~\ref{fig:lc_flares}  shows the light curves of these bright flares.

%%%%%%%%%%%%%%%%%%%%%%%%%%%%%%%%%%%%%%
\section{Time Delay Measurement}
\label{sec:timedelay}
%%%%%%%%%%%%%%%%%%%%%%%%%%%%%%%%%%%%%%

Gravitationally-induced time delays are fundamental measurements in cosmology. 
In principle, they provide a measurement of  the Hubble constant 
independent of the distance ladder 
\citep{1964MNRAS.128..307R,1997ApJ...475L..85S,2002MNRAS.337L...6T,2002ApJ...578...25K,2003ApJ...599...70K,2007ApJ...660....1O,2013ApJ...766...70S,2014MNRAS.437..600S}. 

Monitoring of gravitationally lensed sources at both radio
and optical wavelength where the mirage images are resolved provide  a basis for 
a number of measured time delays 
\citep{2002ApJ...581..823F,2011A&A...536A..44E,2013A&A...557A..44R,2013A&A...556A..22T,2013A&A...553A.121E}. 
Unevenly spaced data resulting from, for example, weather and/or observing time allocation,
are a challenge for light-curve analysis. 
A number of techniques have been specially developed to utilize 
these multiple light curves of mirage images  with unevenly sampled data 
\citep{1988ApJ...333..646E,1992ApJ...385..404P,1992ApJ...398..169R,2001A&A...380..805B,1998A&A...337..681P,2005ApJ...626..649P,1982ApJ...263..835S,1987AJ.....93..968R,1996MNRAS.282..530G,2014MNRAS.441..127G,2011arXiv1105.5991H}.

%Gamma-ray observations have very low spatial resolution but long, nearly uniform time coverage. In particular,
 {\it Fermi}/LAT   provides a very long, evenly sampled, light curve with low photon noise. 
%
%At gamma-ray energies, the mirage images cannot be resolved.
The observed light curve of lensed blazars is  a sum of the mirage images. 
%The mirage images have  similar time evolution, but they appear  shifted in time and with different magnification. 
%For any particular position of the emitting region the time delay and corresponding magnification ratio are fixed. 
The challenge is to extract the time delay
and magnification ratio from the time series informed by the model results based on shorter wavelength data (Figure~\ref{fig:jet_lens}).

In the following sections, we investigate three different methods of determining time delays from unresolved light curves:
the standard Autocorrelation Function (Section~\ref{sec:ACF}), 
the Double Power Spectrum method (Section~\ref{sec:DPS}),
and the Maximum Peak Method (Section~\ref{sec:MPM}). 
Using Monte Carlo simulations, 
we evaluate the significance levels for these methods,
and their sensitivity in detecting the gravitationally-induced time delays. 
The Appendices show the detailed steps for the Double Power Spectrum (Appendix~\ref{app:SPS1}) and for the
Maximum Peak Method (Appendix~\ref{app:MPM}). 
We use PKS 1830-211 as a prototype for broader application of these techniques.

%%%%%%%%%%%%%%%%%%%%%%%%%%%%%%%%%%%%%%
\subsection{Settings for the Monte Carlo Simulations}
\label{sec:MCSettings}
%%%%%%%%%%%%%%%%%%%%%%%%%%%%%%%%%%%%%%

Monte Carlo simulations are a traditional and powerful tool for calibrating the analysis of time series. 
They are important in the case of sparsely sampled data 
and they are necessary for evaluating the significance of an apparent time delay detection
\citep{2005A&A...431..391V}. 

In this section, we describe the settings for our Monte Carlo simulations. 
We include the  characteristics of the time series,  
the procedures for evaluating the significance of time delay detections, 
and the sensitivity of the methods for detecting gravitationally-induced time delays in unresolved light curves. The full analysis takes advantage of the physical relationship between the time delay and the magnification ratio. 

%%%%%%%%%%%%%%%%%%%%%%%%%%%%%%%%%%%%%%
\subsubsection{Characteristics of the Signal}
\label{sec:Signal}
%%%%%%%%%%%%%%%%%%%%%%%%%%%%%%%%%%%%%%

The observed temporal behavior in blazars is represented 
by  power law noise  \citep{2014ApJ...791...21F,2012ApJ...754..114H,2013ApJ...773..177N,2014ApJ...786..143S}
where the power spectral density (PSD) is inversely proportional to the frequency, $f$,  of the signal to the power $\alpha$:
\begin{equation}
\label{eq:PSD}
S(f) \propto 1/f^{\alpha} \,.
\end{equation}
%
%where $f$ is the signal frequency.

This random variability is often referred to as {\it noise} intrinsic  to the source (not measurement error),  
which is a result of stochastic processes  \citep{2003MNRAS.345.1271V}.
Astronomers refer to these stochastic fluctuations as {\it signal};
in other fields, the most common terminology is {\it noise} \citep{1978ComAp...7..103P}. 
For ease of presentation, we adopt this more general terminology and explore properties of various types of noise. 

Typically, quasars  have $\alpha \sim 1\, - \,2$ (Equation~\ref{eq:PSD}).
The average slopes for gamma radiation from the brightest 22 flat spectrum radio quasars 
and from the 6 brightest BL Lacs are 1.5 and 1.7, respectively \citep{2010ApJ...722..520A}.
During the gamma-ray quiescent state, where blazars remain  most of the time, the fluctuations in the flux are small, 
and the temporal behavior is characterized by power law noise with index $\sim1$.
During flaring, the amplitude of the fluctuation of the flux can increase by  a few to dozens.
In general, the signal is still represented by power law noise, but with a greater  index $\alpha$.

In our simulations, we produce artificial light curves with time series represented by  red and  pink noise.
Red noise, also known as a Brown noise, has $\alpha = 2$, consistent with the
observed behavior of many gamma-ray active periods.   
Figures~\ref{fig:lc_flares} shows flaring periods of PKS~1830-211,
and  Figure~\ref{fig:lc_MC} shows  an example of an artificial light curve based on red noise,
with and without an artificially induced gravitational time delay. 
The time structure of the observed and simulated light curves is remarkably similar by eye. 

Pink noise has a power spectrum inversely proportional to the frequency of the signal ($\alpha = 1$); this type of noise 
describes the temporal behavior in the gamma-ray quiescent state.
For demonstration purposes, we  also construct artificial light curves of white noise,  $\alpha \sim 0$. 
We use these types of noise to demonstrate the sensitivity of time delay detection to 
the nature of the underlying signal along with the method of analysis.

We conducted our simulations and analysis using the {\tt Matlab} environment. 
We generated  samples of power law noise using  the \citet{Little2007} code. 

%In general, the temporal behavior of blazars is simply represented by power law noise. 
The lensed light curve is still power law noise, but it contains information about the time delay.
The lens itself is not a gamma-ray emitter at a detectable level.
Therefore, we can construct the observed gamma-ray light curve as a sum of the lensed components of the blazar:
\begin{equation}
\label{eq:lc}
S(t) = s(t)+s(t+a)/b\,,
\end{equation}
where $S(t)$ is the  unresolved light curve of the lensed blazar, 
composed of the sum of the mirage images. The 
temporal behavior of individual images is determined  by the source,
but the images are shifted in time 
by the gravitationally-induced time delay, $a$, 
and with the magnification ratio between mirage images, $b$.

%Data from  {\it Fermi}/LAT  allows construction of an 6.5 year-long light curve (see Figure~\ref{fig:lc_whole}). 
We focus on the nature of the gamma-ray emission during flaring activity.
The durations of these active periods
range from  a few to hundreds of days (see Figure~\ref{fig:lc_flares}).  

The lens model predicts time delays  up to $\sim70\,$days. 
To have a chance of investigating the entire permitted  range of time delays, 
the sample has to be at least twice as long as the maximum time delay. 
In our simulations, we produced time series of 155 days, exactly the duration of the active period of Flare 1.

{\it Fermi}/LAT  continuously monitors the entire sky, 
but,  sometimes, the photon flux of the source can be too low to  be 
detected significantly in a  time bin chosen {\it a priori}. There is no gap in the time series, but the flux from the source cannot be detected at $\gtrsim$2$\sigma$.  
In these cases,we compute an upper limit on the flux. 
The value of upper limits does depend, 
for example, on the exposure within the time bin.
Thus, if we use upper limits as a measure of a flux, 
satellite-related periodicities can appear in the signal. 
Another approach is to set the flux to zero in these low photon flux bins.
We check the impact of both approaches (Section~\ref{sec:Flare1}).

One can also interpolate the flux
or set the flux to an average value. 
However, this approach could  introduce a flux, in a particular, 
that exceeds the observed upper limit,  thus invalidating  the time series analysis. 
We thus do not investigate this approach. 

Monte Carlo simulations  provide a strong test of the method for treating data points with upper limits. We can readily
introduce the estimated  upper limits into simulated time series and treat the detections and upper limits in an internally consistent way. 
This approach  gives us a measure of the impact of bins with upper limits only
on the significance of the detection. %, and can allow selecting the right approach to dealing with the data.
We discuss this issue further in the results section, 
where we describe the temporal analysis of the PKS~1830-211 light curves. 

%%%%%%%%%%%%%%%%%%%%%%%%%%%%%%%%%%%%%%
\subsubsection{Statistical Significance}
\label{sec:Significance}
%%%%%%%%%%%%%%%%%%%%%%%%%%%%%%%%%%%%%%
%
A crucial element of the analysis is evaluation of the statistical significance of the detection of a time delay. 
An observed modulation of the signal could be a real time delay
or it could arise purely by chance. 
Monte Carlo simulations provide a way to compute 
the probability of detections as opposed to false positives \citep[see e.g. ][]{2005A&A...431..391V}.

We produce $N=10^6$ artificial light curves of power law noise characterized by $\alpha=\{ 0,1,2\}$.
We calculate the chance that a particular time delay signal will appear randomly in the simulated light curve which contains no intrinsic time delays.
For each simulated light curve, the output consists of time delays and the relative power of the signal as a function of the time delay. 
%NEED SOME EQUATIONS... 
%
We construct the cumulative probability distribution of false detection (CPD: p(POWER)) for each time delay as a function of the power. The cumulative probability of a false detection is less than $p$.
We investigate $p=0.317$ ($1\,\sigma$),  $p=0.0455$ ($2\,\sigma$), $p=0.0027$ ($3\,\sigma$), and $p=0.00006$ ($4\,\sigma$), respectively, for each analysis method and for each of the three representative noise spectra. We show the results in Sections~\ref{sec:ACF} and~\ref{sec:DPS}. 
The CPDs are sensitive both to  the underlying noise spectrum and to the analysis method.  

%%%%%%%%%%%%%%%%%%%%%%%%%%%%%%%%%%%%%%
\subsubsection{Detectability of the Time Delay}
\label{sec:Detectability}
%%%%%%%%%%%%%%%%%%%%%%%%%%%%%%%%%%%%%%
%

We also use the Monte Carlo simulations to assess the chance of detecting a real signal at a given significance level. 
Here the simulated light curves contain artificial time delays with the appropriate magnification ratio. For each combination
of noise spectrum, time delay and magnification ratio, the detectability depends on the analysis method. In some cases where, 
for example, the magnification range is large and the time series is too short, some methods can not detect the time delay at all.

The lens model  predicts a range of time delays
and corresponding magnification ratios. 
For each time delay in the range from 1 to 80 days along with the corresponding magnification ratios, 
we produce $10^5$ artificial light curves of power law noise with artificially induced time delays. 
We distribute the time delays uniformly on the interval 1 to 80 days with a 1 day interval (the same as the light curve binning). 
These light curves simulate the observations of PKS 1830-211.

%The observed light curve is a sum of components.
%The first component (the first mirage image), $s(t)$, is just power law noise.
%The second component is the same noise, 
%but shifted by the time delay, $a$,
%and with magnification ratio,  $b$. 
%Thus, the second component is $s(t+a)/b$. 
%The final observed light curve is the sum of these components. 
We  simulate analogous artificial light curves with red noise ($\alpha = 2$)
to evaluate the probability of time delay detection in the observed flaring light curve. 
To assess time delay detectability in quiescent light curves we analyze simulated light curves based on pink noise ($\alpha = 1$). 
For demonstration purposes, we also check the detectability of a time delay for white noise light curves ($\alpha = 0$).

Analysis of the artificial light curves gives 
the probability of detecting the time delay 
at $1\,\sigma$ ,$2\,\sigma$, $3\,\sigma$, and $4\,\sigma$ levels. 
From the simulations with no inherent time delays, 
we know the power where the detection rate of false positives is at a given significance level. 
At each power (corresponding to a particular significance level), 
we then check how many simulated light curves have power above this limit. 
We do this analysis for each time delay and significance level for false positive detection.  

We discuss the results of this analysis in Sections~\ref{sec:ACF} and~\ref{sec:DPS}. 
As in the case of the analysis of false positives, 
the results for detection of a true time delay depend on the input spectrum, 
the analysis method, and the combination of time delay and magnification ratio.
 
%%%%%%%%%%%%%%%%%%%%%%%%%%%%%%%%%%%%%%
\subsection{Methods}
\label{sec:Methods}
%%%%%%%%%%%%%%%%%%%%%%%%%%%%%%%%%%%%%%

%We investigate different approaches to detect the time delays in unresolved light curves.
%We start with standard approach of searching for time delays, which is the Autocorrelation Function. 

%%%%%%%%%%%%%%%%%%%%%%%%%%%%%%%%%%%%%%
\subsubsection{Autocorrelation Function}
\label{sec:ACF}
%%%%%%%%%%%%%%%%%%%%%%%%%%%%%%%%%%%%%%
%-------------------------------- ACF --------------------------------%
%/dane/Lensing/data/PKS1830/whole/pinknoise_sig_ACF.eps
\begin{figure*}
%\vskip 1cm
\begin{center}
\includegraphics[width=3.9cm,angle=-90]{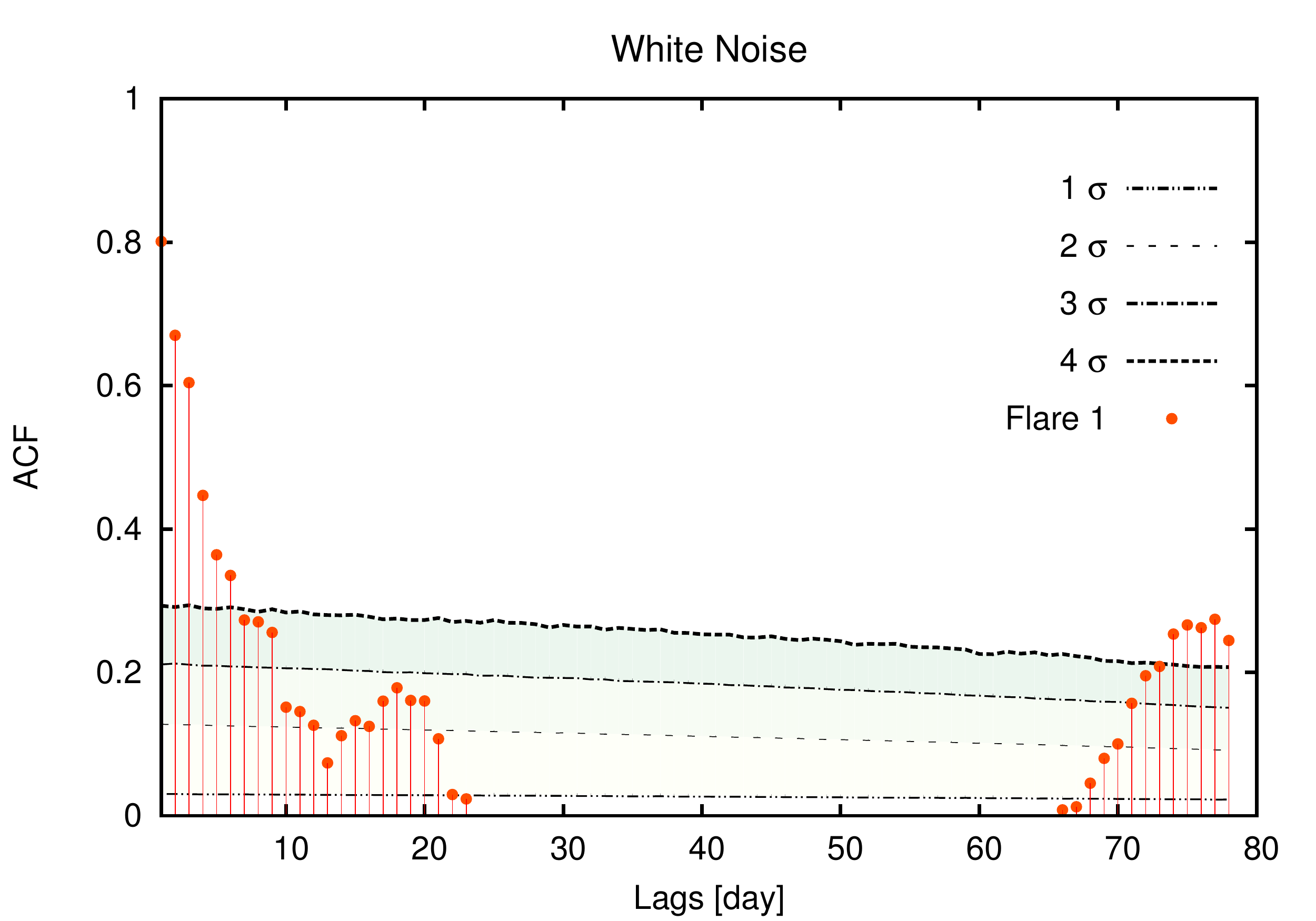}
\includegraphics[width=3.9cm,angle=-90]{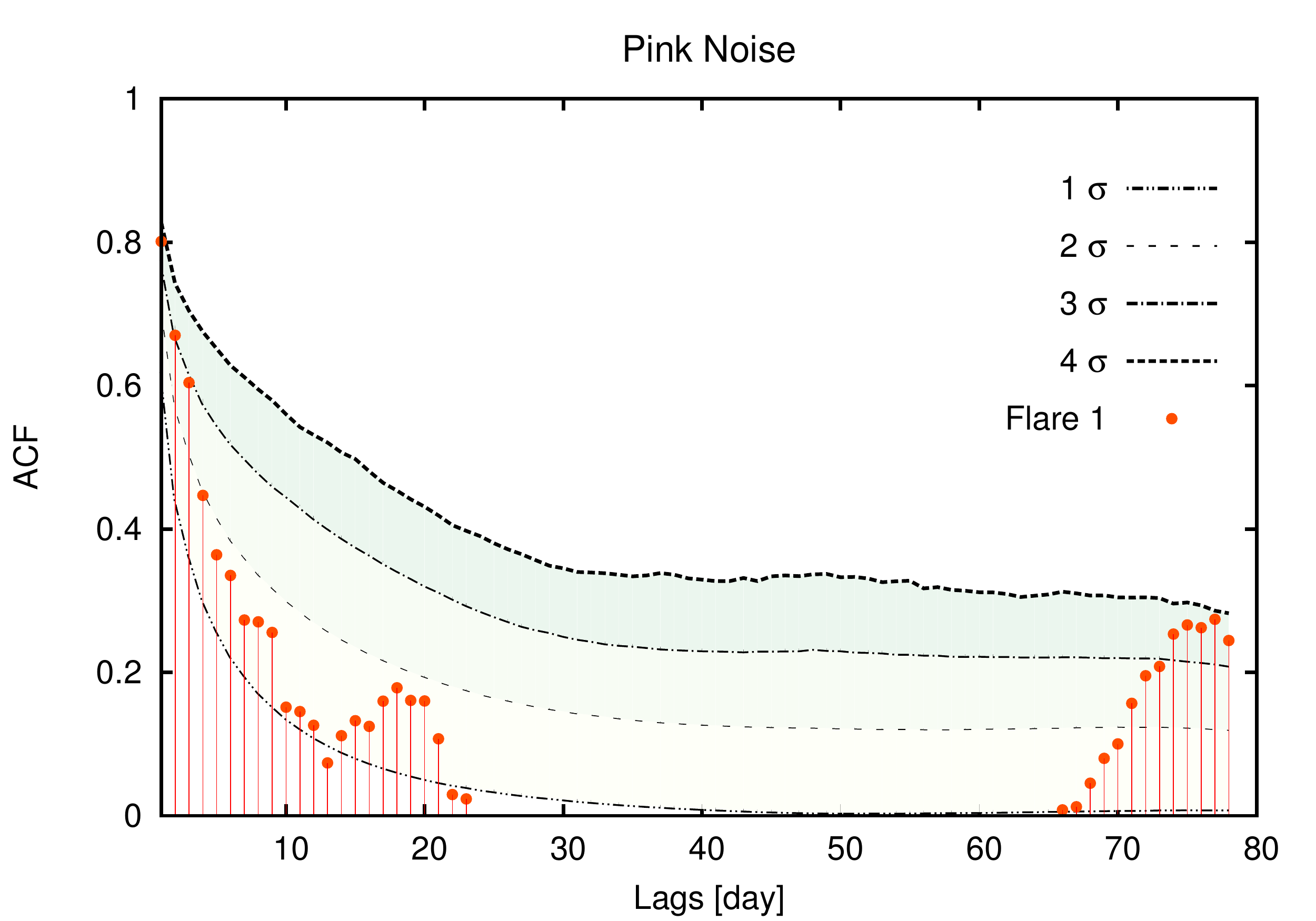}
\includegraphics[width=3.9cm,angle=-90]{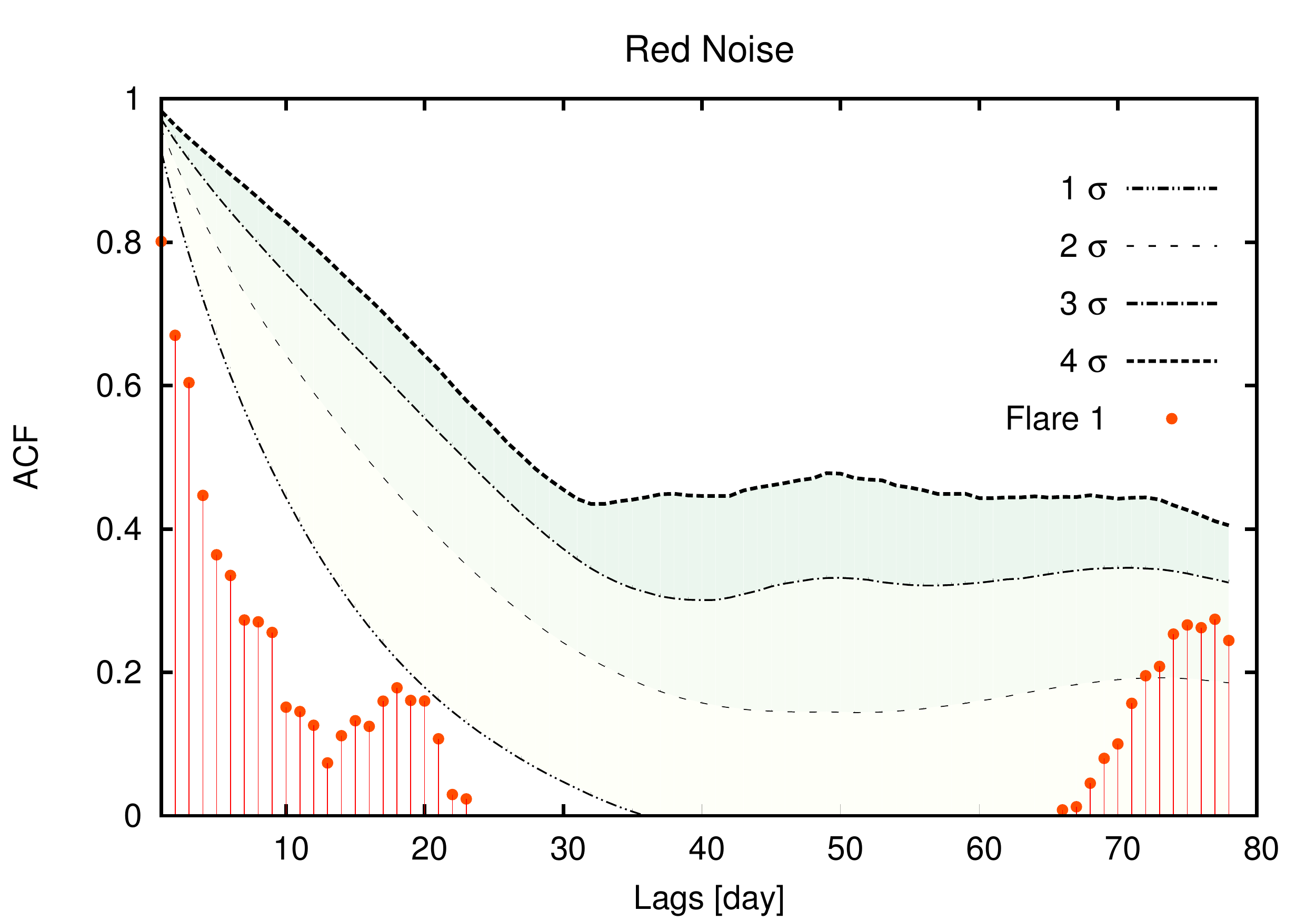}
\end{center}
\caption{\label{fig:ACF_CL} 
		Confidence level for the autocorrelation method derived from 
		Monte Carlo simulations of $10^5$ light curves for
		white (left), pink (middle), and red (right) noise.
		Autocorrelation coefficients as a function of the time delay.
		The lines indicate $1\,\sigma$, $2\,\sigma$, $3\,\sigma$, and $4\,\sigma$ confidence levels. 
		The red points show the ACF for Flare 1.  }
\end{figure*}

\begin{figure*}
%\vskip 1cm
\begin{center}
\includegraphics[width=3.9cm,angle=-90]{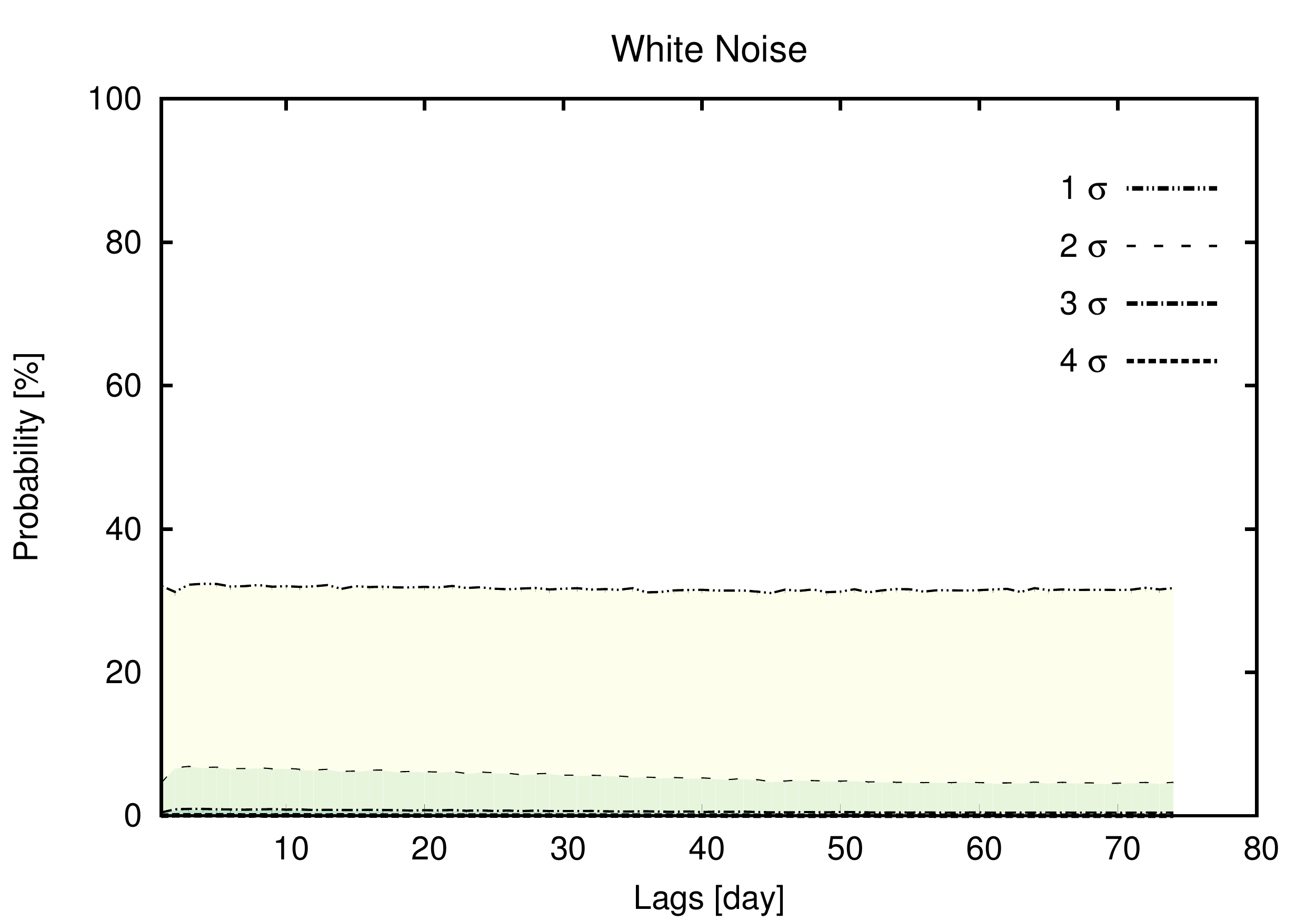}
\includegraphics[width=3.9cm,angle=-90]{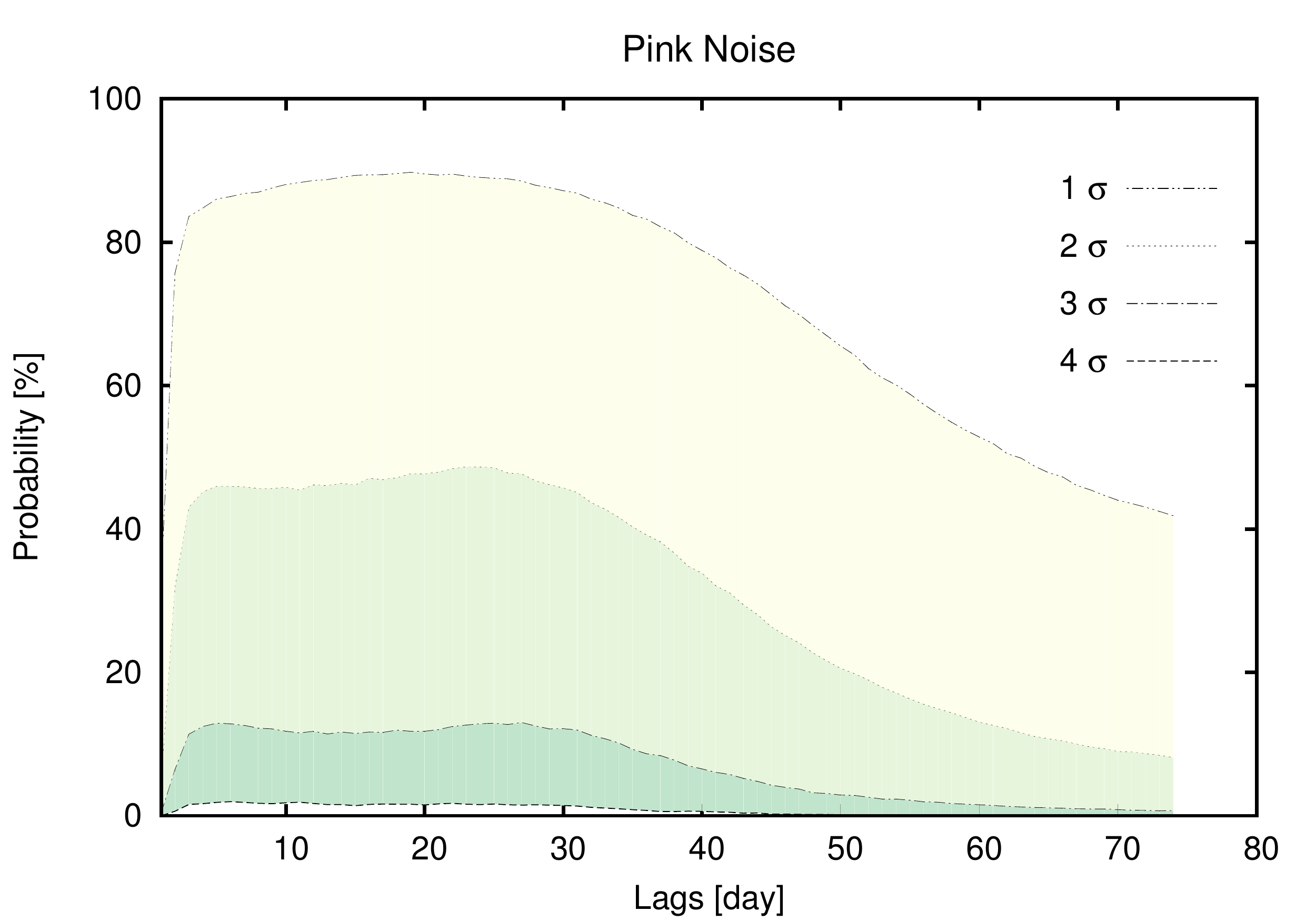}
\includegraphics[width=3.9cm,angle=-90]{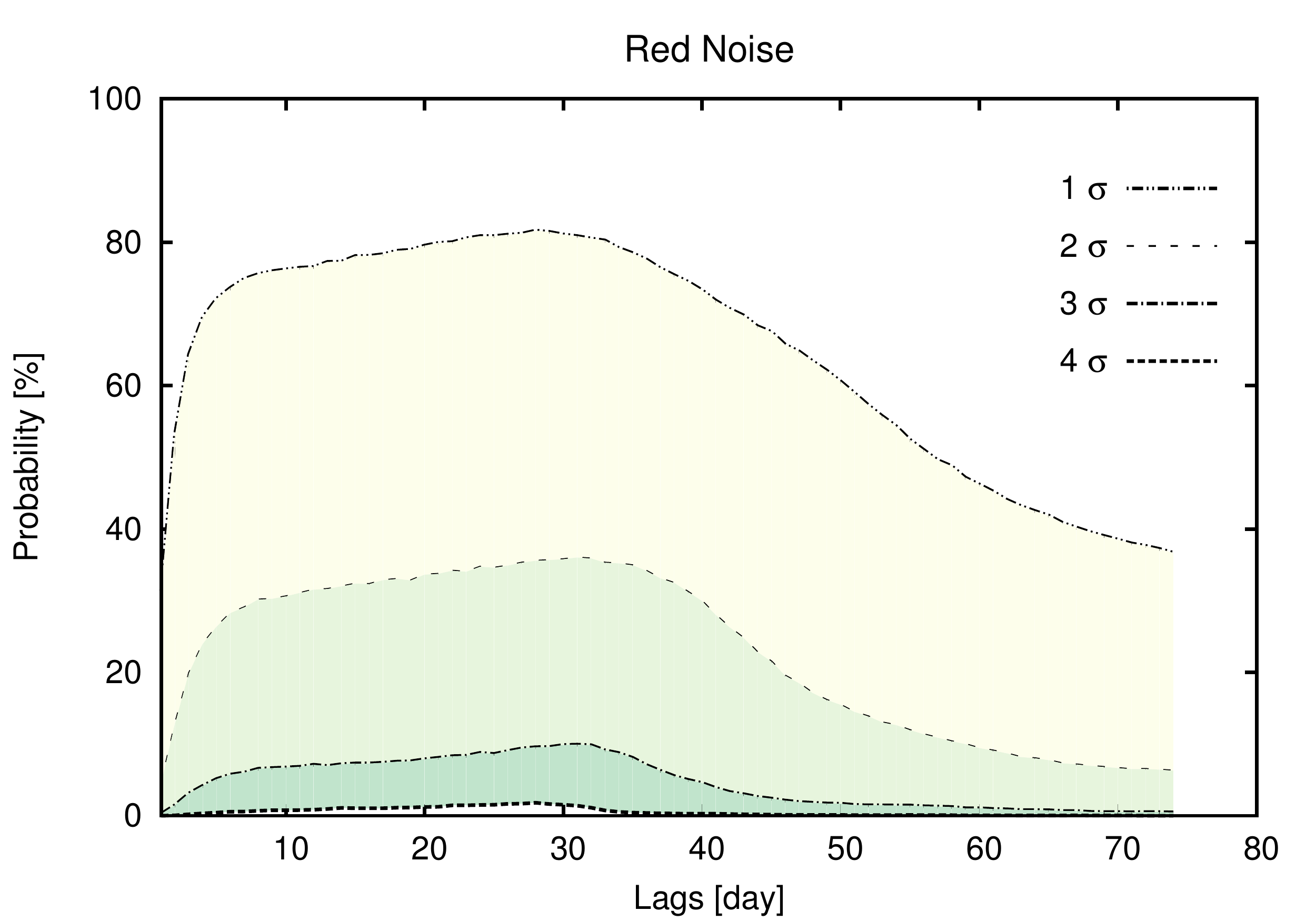}
\end{center}
\caption{\label{fig:ACF_probability} Probability of detecting a time delay at a given  significance level with the Autocorrelation method. 
							 }
\end{figure*}
The Autocorrelation Function (ACF)
is a  standard tool for estimating time lags in light curves. 
The ACF of a  signal describes the correlation of
the values of the samples at one time on the values at another.

For larger $\alpha$ (Eq.~\ref{eq:PSD}), the time series contains more  apparent structure; 
there are more and larger amplitude variations on shorter time scales. 
Thus a time delay detection is, in principle, easier for a source characterized by larger $\alpha$.  
However, the steep spectrum of power law noise can also more easily  produce spurious peaks in the ACF, 
thus raising the confidence levels artificially.
 
Figures~\ref{fig:ACF_CL} and \ref{fig:ACF_probability}
show ACF applied to  Monte Carlo simulations 
of $10^5$ time series for white, pink, and red noise. 
White noise does not cause spurious time delays with  large power: 
the confidence boundaries are low ($\sim 0.3$ for $1\,\sigma$), 
and flat over the full range of time delays we explore.
However, the probability of detecting  time delays in the signal characterized by white noise 
using the ACF  is very low. 
The probability of detecting a signal at the $1\,\sigma$ level does not exceed $35$\%, 
or 5\% at the $2\,\sigma$ confidence level.

Red and pink noise have a larger potential for producing spurious peaks in the ACF, 
(figure~\ref{fig:ACF_CL}); in these cases, 
the confidence  boundaries  are elevated.
For simulated time series, the ACF is most sensitive for time delays in the range $\sim5$~days to $\sim40$~days. 
The detection of  long time delays is  limited by the finite length of the sample; 155~days. 
Furthermore, mirage images separated by longer time delays also have larger magnification ratio
resulting in an attenuation effect that hinders the detection.

%%%%%%%%%%%%%%%%%%%%%%%%%%%%%%%%%%%%%%
\subsubsection{Double Power Spectrum}
%%%%%%%%%%%%%%%%%%%%%%%%%%%%%%%%%%%%%%
\label{sec:DPS}

\begin{figure*}
%\vskip 1cm
\begin{center}
\includegraphics[width=3.9cm,angle=-90]{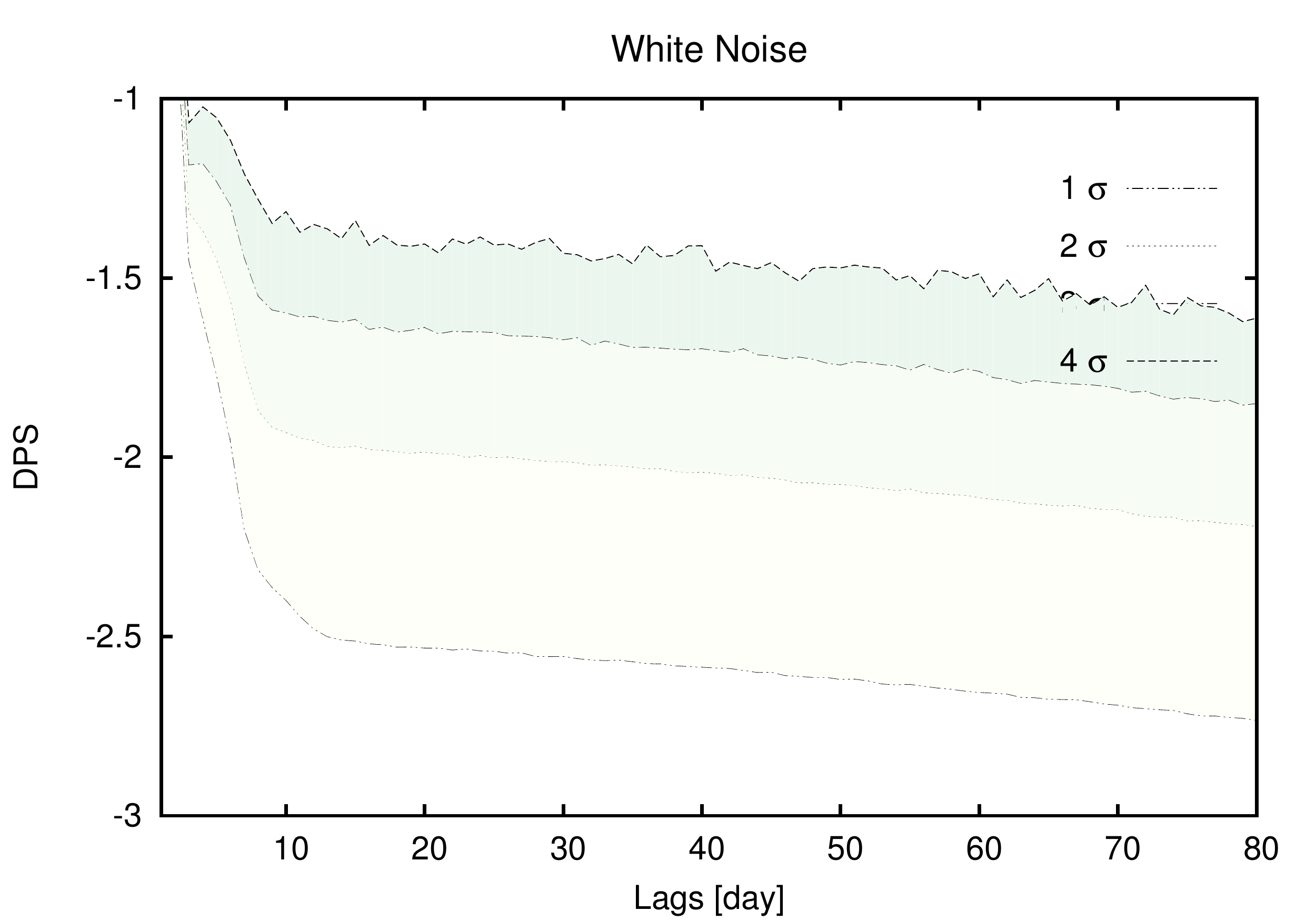}
\includegraphics[width=3.9cm,angle=-90]{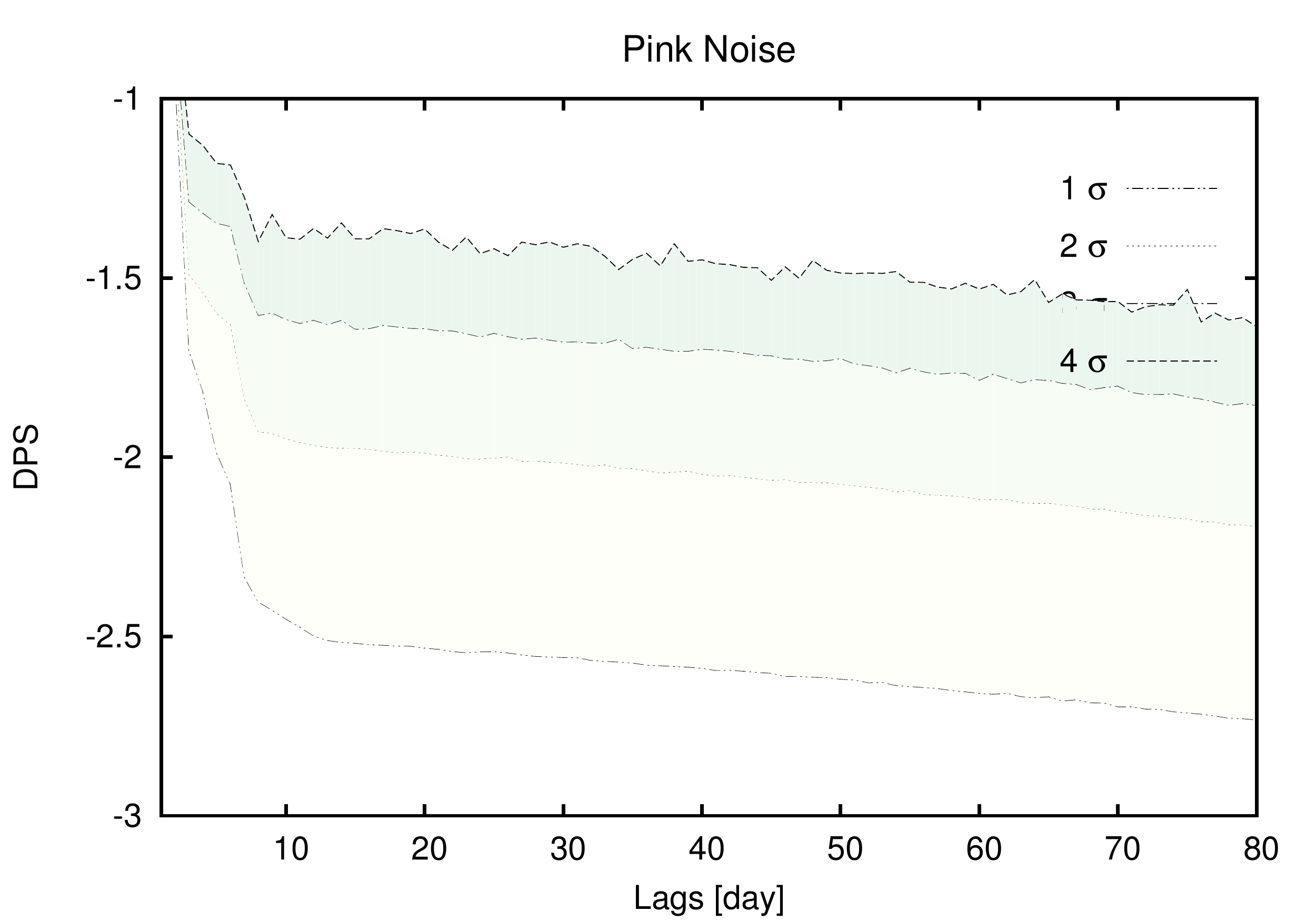}
\includegraphics[width=3.9cm,angle=-90]{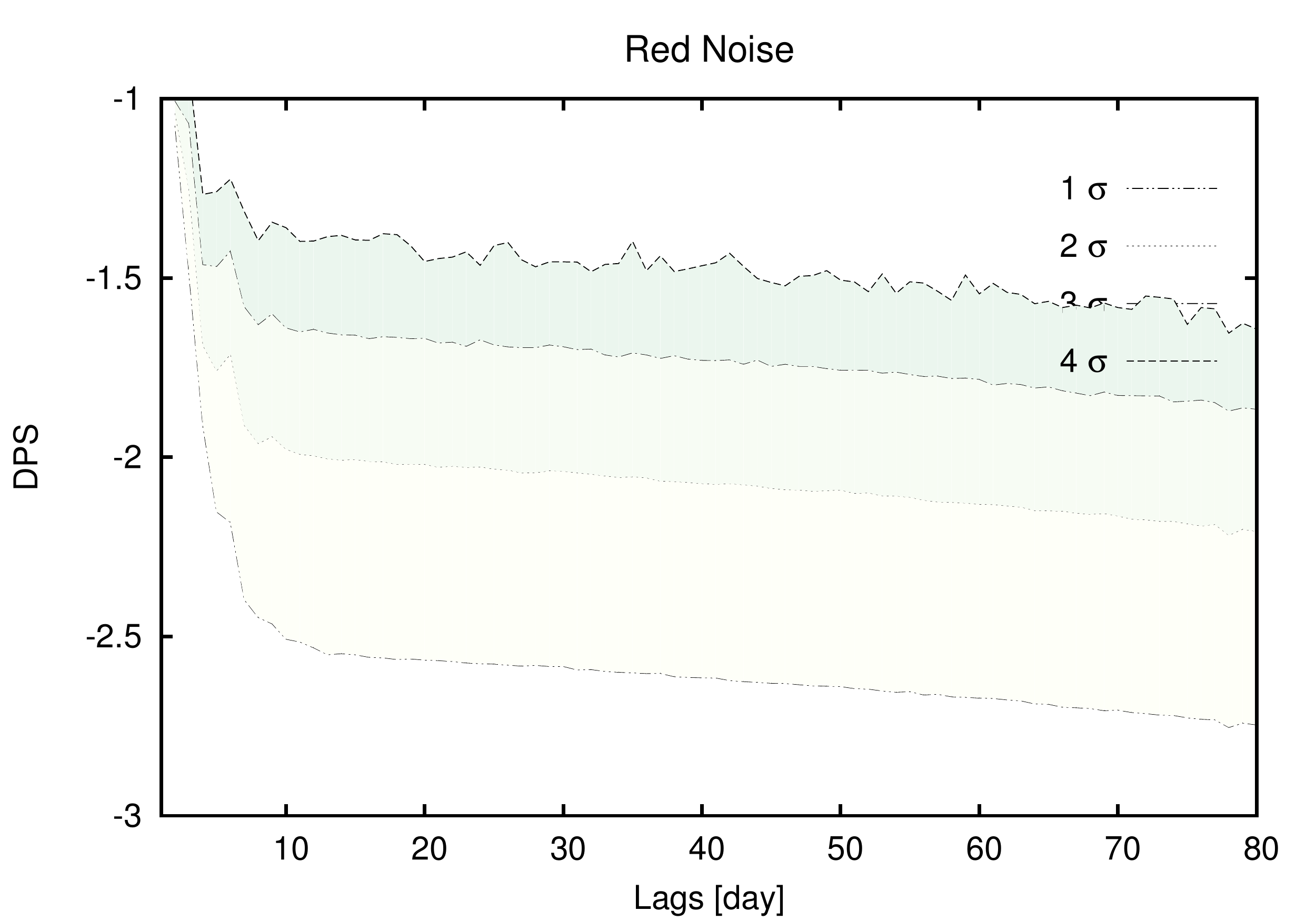}
\end{center}
\caption{\label{fig:DPS_CL} Confidence levels for the Double Power Spectrum method. }
\end{figure*}

\begin{figure*}
%\vskip 1cm
\begin{center}
\includegraphics[width=3.9cm,angle=-90]{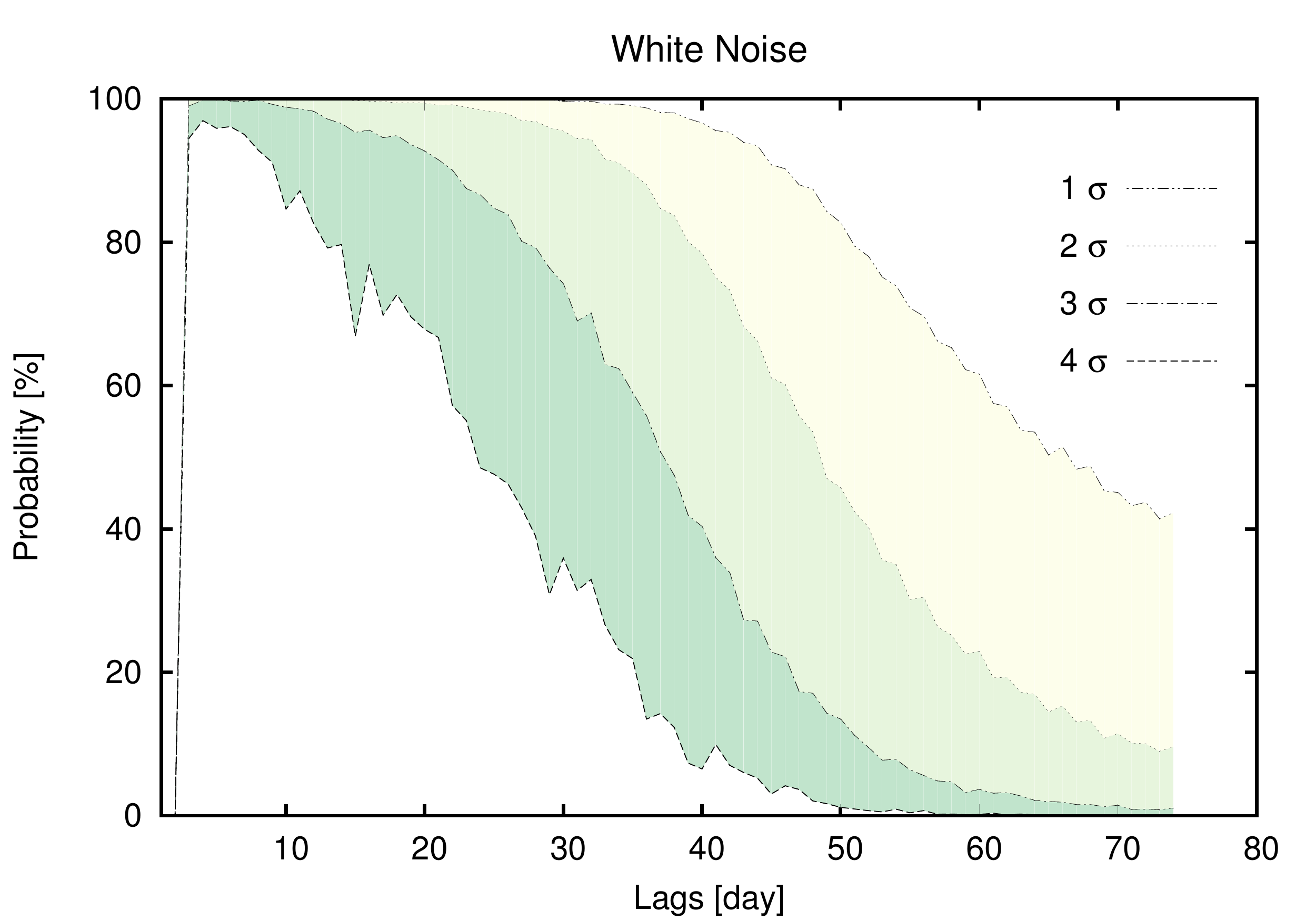}
\includegraphics[width=3.9cm,angle=-90]{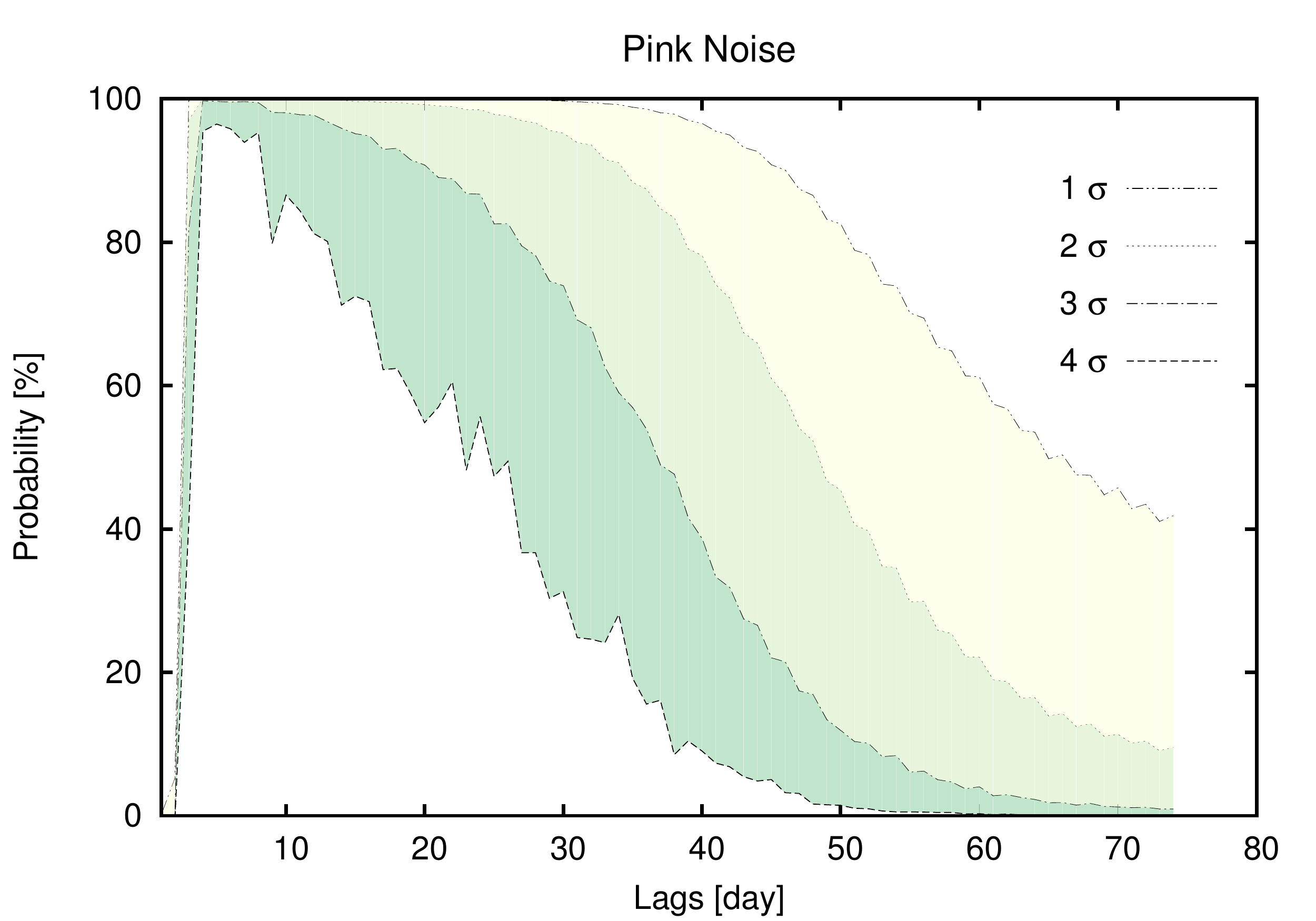}
\includegraphics[width=3.9cm,angle=-90]{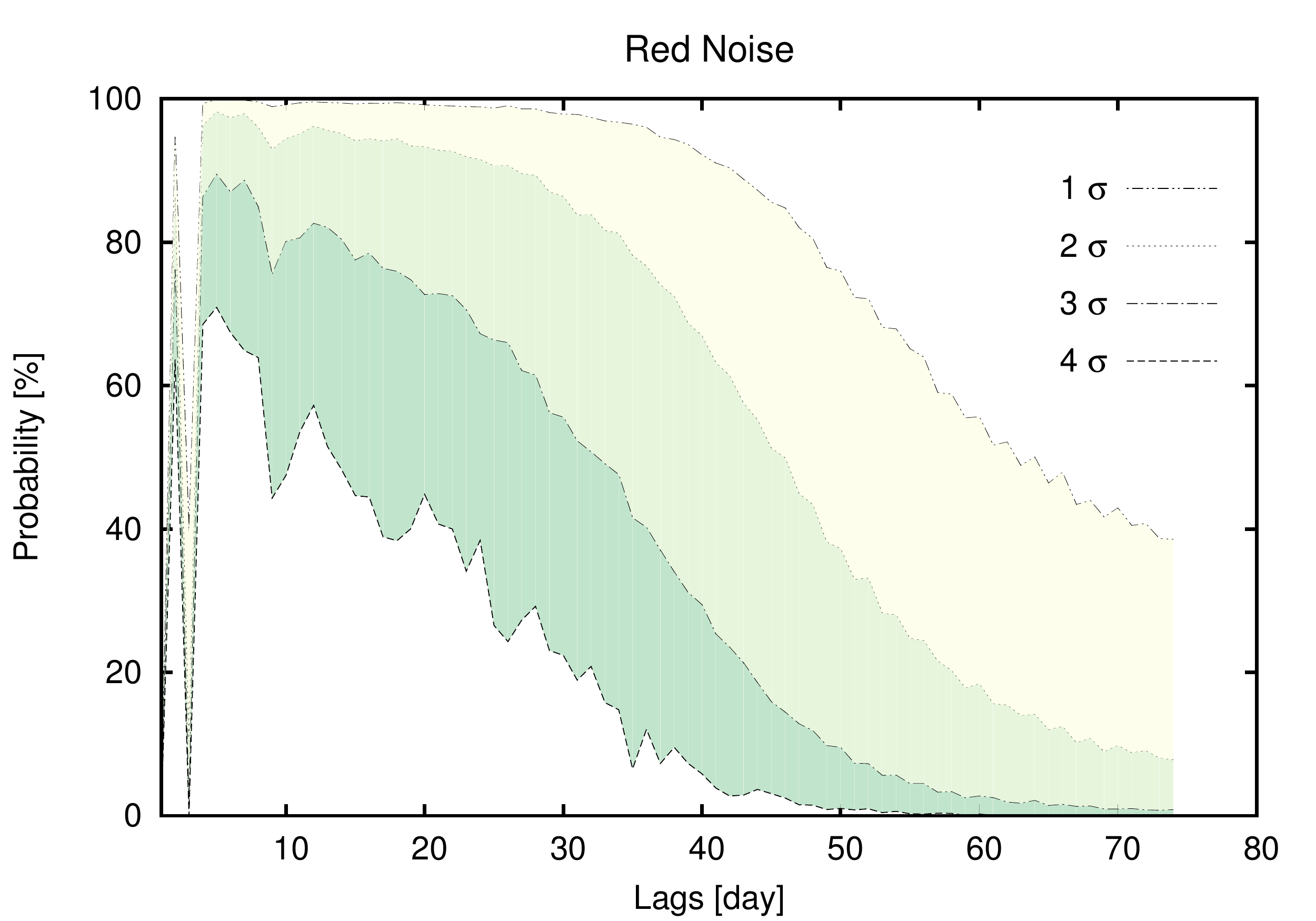}
\end{center}
\caption{\label{fig:DPS_probability} Probability of detecting a time delay at a given significance level based on the DPS method. }
\end{figure*}

The Double Power Spectrum (DPS) was the basis for detection of 
the first gravitationally-induced time delay at gamma-rays \citep{2011A&A...528L...3B}. 
%Here we review the method and we show the individual step in Appendix~\ref{app:DPS}. 
The steps in this signal processing are based on widely used methods \citep{1975dsp..book.....O,1971A&A....13..169B},
and have been described in \citet{2011A&A...528L...3B,2013arXiv1307.4050B}.

The signal in the time domain is equation~(\ref{eq:lc}).
The  Fourier transform of the first component, $s(t)$, is $\tilde{s}(f)$,
and the second component transforms to the frequency domain as $\tilde{s}(f) e^{-2\pi if a}$.
Therefore, the observed signal $S(t)$ transforms into 
\begin{equation}
\label{eq:fft}
\tilde{S}(f) = \tilde{s}(f) (1 + b^{-1} e^{-2\pi i f a}) \,,
\end{equation}
in Fourier space.

The first power spectrum of the source is the square modulus of $\tilde{S}(f)$:
\begin{equation}
\label{eq:fps}
|\tilde{S}(f)|^{2} = |\tilde{s}(f)|^{2}(1 + b^{-2} + 2b^{-1} cos(2\pi f a)) \,.
\end{equation}

The first power spectrum is the product of the "true" power spectrum of the source, $|\tilde{s}(f)|^2$,
%The first power spectrum is the product of the "true" power spectrum of the source, $|\tilde{s}(f)|^{2}$,
 times a periodic component with a period (in the frequency domain) equal to the inverse
of the relative time delay $a$.
Thus, to find  time delay $a$, we need to find the period of the pattern in the first power spectrum. 
The Fourier transform of the first power spectrum, which is in the frequency domain,
brings us back to the time domain, where the amplitude of 
the signal corresponds to the power of the time delay signal present in the original time series.
The method is similar, in spirit, to the Cepstrum method 
widely used in speech processing and seismology \citep{Bogert1963}. 

We use a Monte Carlo procedure to evaluate the performance of the DPS 
for power law noise. 
Figure~\ref{fig:DPS_CL} shows  the confidence levels of $1\,\sigma$, $2\,\sigma$, $3\,\sigma$, and $4\,\sigma$,
for white, pink and red noise, respectively. 

The Monte Carlo simulation demonstrates the effectiveness of the signal processing.
The DPS method emphatically enhances the probability of significant detection of time delays. 
The probability of detecting a time delay   $5 - 40\,$days, at  $3\,\sigma$ level 
is in the range from 90\% to 40\%  
in contrast with the ACF method where the probability is 10\% at best.
The signal processing in DPS eliminates the large dependence of 
significance levels and the probability of detection of time delay, 
on the index $\alpha$ of power law noise. 

The efficiency of detecting longer time delays is limited only by 
the sample length and the magnification ratio between the mirage images. 
We follow the same procedure independent of the time delay.
In principle, one can 
optimize the signal processing procedure for different ranges of time delays 
to account for the effects of signal attenuation and discontinuity at the begin and end of time series.

%%%%%%%%%%%%%%%%%%%%%%%%%%%%%%%%%%%%%%
\subsubsection{Maximum Peak Method}
%%%%%%%%%%%%%%%%%%%%%%%%%%%%%%%%%%%%%%
\label{sec:MPM}

%The temporal behavior of the active period of gamma-ray radiation from jets
%is characterized by distinctive flares where the emission can rise by a factor of a few to dozens relative to  the average gamma-ray flux from the source. 
%The duration of a single  flare is usually short, from a few hours to a few days.
%Thus, the duration of the flares is much shorter than the typical time delay. 
The short duration of the flares relative to the expected time delays are an important element in analysis of unresolved $\gamma$-ray light curves. 
Because gamma-ray flares can be identified as distinct events in the time series, 
and we know the range of expected time delays and corresponding magnification ratios,
we can search for echo flares in successive bins directly.  

The gamma-ray active state can last for a dozen to even hundreds of days; 
these periods consist of a  series of individually identifiable flares. 
Methods like the ACF or the DPS are well suited for analyzing these longer periods of activity
when  light curve can be extracted with shorter binning ranging from 12 hours to 1 day. 
When the active period consists of a single short flare, bins with a longer integration time are necessary.
The gamma-ray flux before and after these single flares corresponds to the quiescent state.
Thus to detect the signal  at a significant level we must increase the size of the bin (exposure). 

The MPM method complements the DPS method  for these isolated flares.
 We  identify the first brightest flare, 
and calculate the flux ratio between the bin with the largest flux (the flare) and flux in successive bins.
These flux ratios constrain the magnification ratios which are not constrained by the DPS method. 

The MPM method enables us to extract additional physical constraints from the time series.
We compare calculated ratios, as a function of a time delay between a particular bin and the position of the brightest flare, 
to the magnification ratios as a function of the time delay predicted from the model. 
We identify the time delays where the ratio of fluxes is consistent with the predicted magnification ratio. These bins might or might not exist. If there are bins consistent with the expected delays, the data support the picture based on the model. It is important to note here that the model takes constraints from data at other wavelengths into account.
This method of analysis of the light curve especially allows us to exclude ranges of time delays where there is no consistent magnification ratio observed ($\sim 80\%$ of the range). 

We demonstrate the method with Monte Carlo simulations in Appendix~\ref{app:MPM}.
We investigate this approach for cases with a single flare and with a series of flares.

%%%%%%%%%%%%%%%%%%%%%%%%%%%%%%%%%%%%%%

%%%%%%%%%%%%%%%%%%%%%%%%%%%%%%%%%%%%%%
\section{Results}
%%%%%%%%%%%%%%%%%%%%%%%%%%%%%%%%%%%%%%

Here, we use PKS 1830-211 as an example of elucidating  
the spatial origin of the gamma ray flares. 
In particular, we demonstrate that the flares probably do not all originate from the same location in the source. 

In the light curve (Figure~\ref{fig:lc_whole}) 
there are two long active periods (red area; Flares 1 and 2) of more than 100 days 
and two isolated individual flares (green area; Flares 3 and 4).
We analyze each of these four flaring periods separately.
%
%------------------------------------------Figures----------------------------------------------%
\begin{figure}
%\vskip 1cm
\begin{center}
%Flare1_ACF.eps  UL not included in the CL
%\includegraphics[width=3.9cm,angle=-90]{plots/Flare1_ACF.eps}
\includegraphics[width=5.cm,angle=-90]{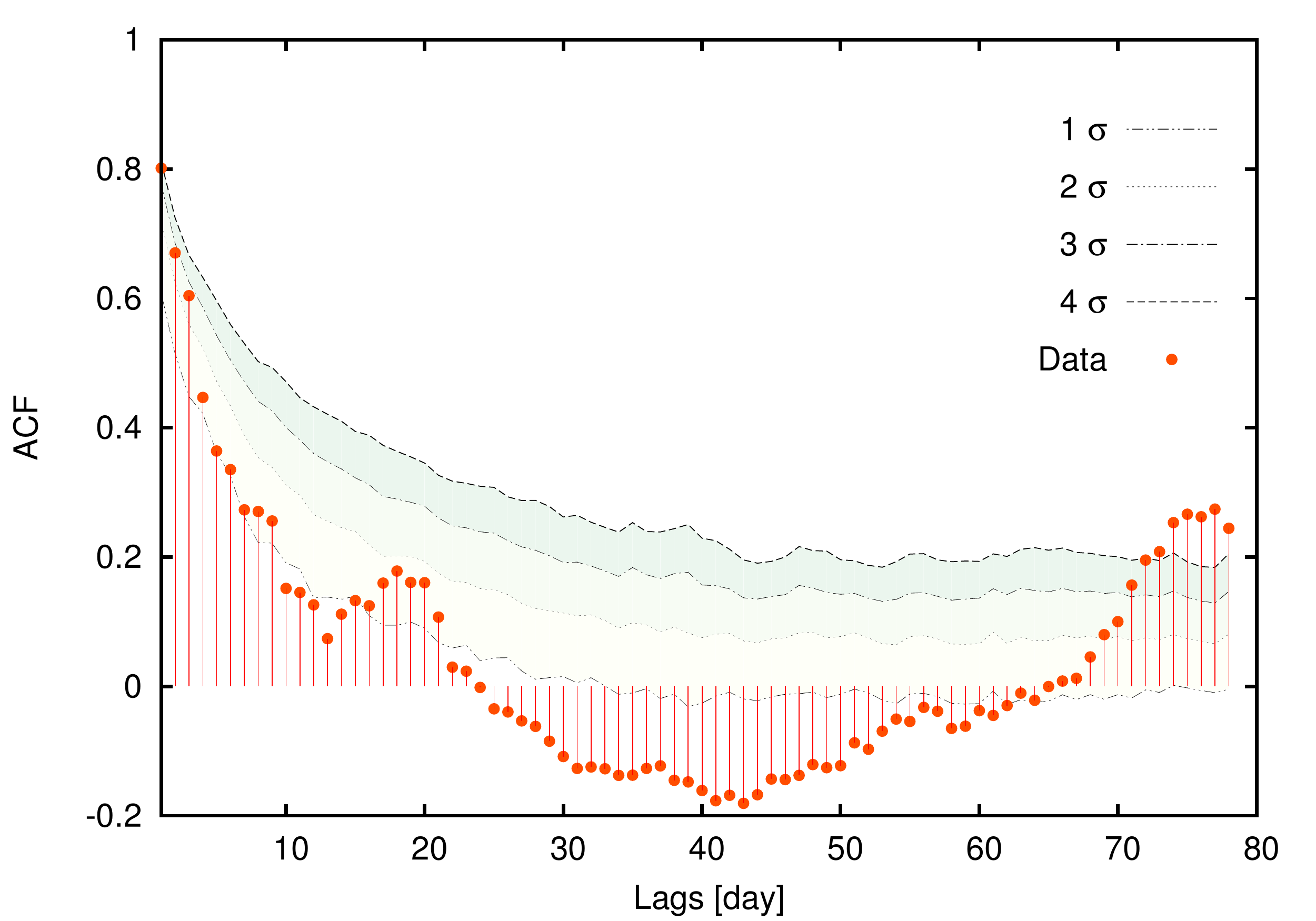}
\includegraphics[width=5.cm,angle=-90]{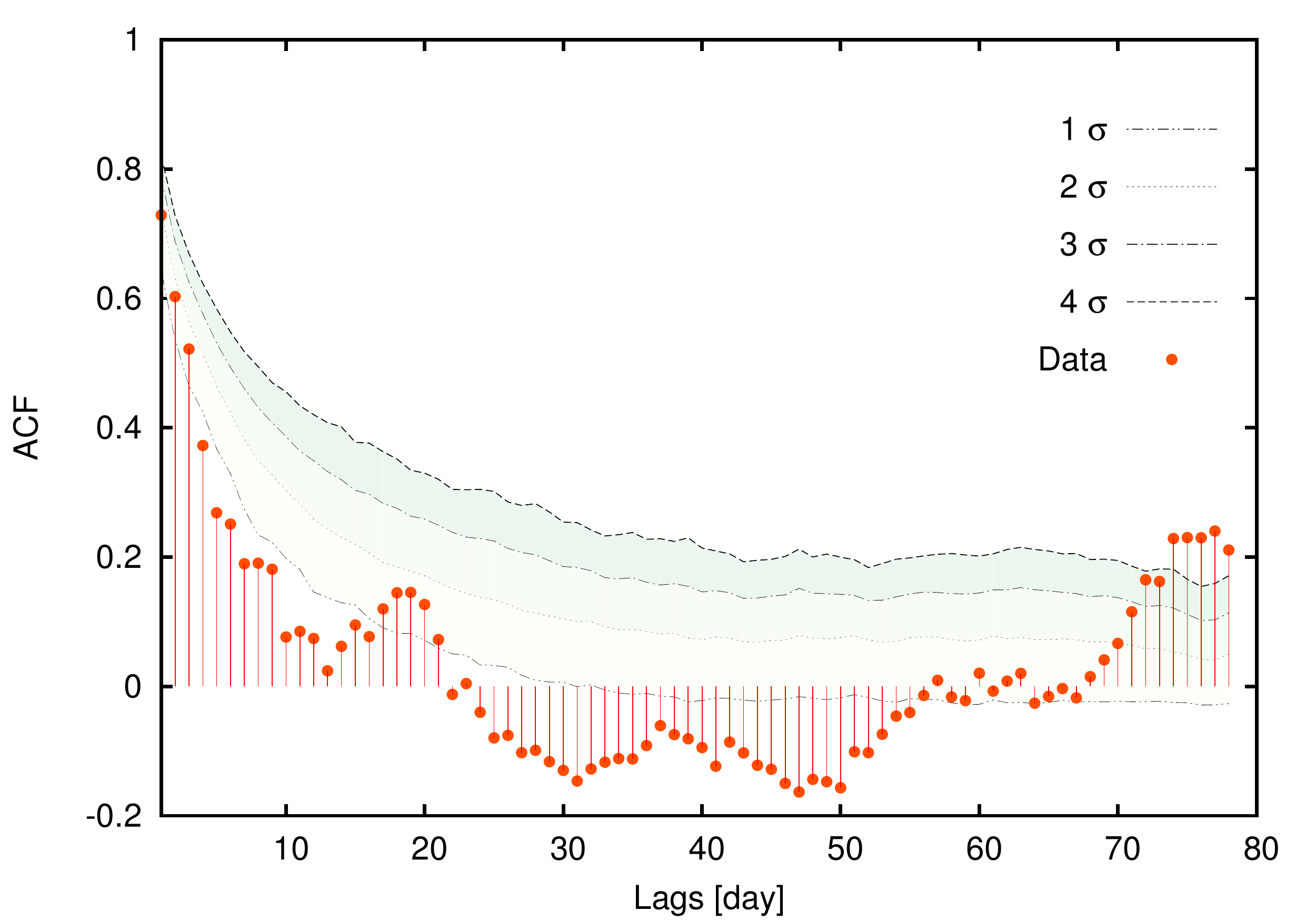}
\end{center}
\caption{\label{fig:ACF_CL_Flare1}  ACF for {\bf Flare 1} along with confidence levels. \\
	    %   A: Autocorrelation function calculated for the flare 1 based on upper limits as measures of a flux. 
	    %   The confidence levels are based on power law noise. \\
	       {\bf Top}: Autocorrelation function for Flare 1 based on upper limits as measures of  the flux. 
	       The confidence levels are based on MC simulations of  power law noise, 
	       with upper limits as measured for Flare 1. \\
	       {\bf Bottom}: Autocorrelation function for the light curve of Flare 1 with 
	       the flux set to 0 in time bins with upper limits. 
	       The confidence levels are derived by generating time series of power law noise,
	       with values set to zero in bins that have measured upper limits.}
\end{figure}

\begin{figure}
%\vskip 1cm
\begin{center}
%Flare1_ACF.eps  UL not included in the CL
%\includegraphics[width=3.9cm,angle=-90]{plots/Flare1_DPS.eps}
\includegraphics[width=5.6cm,angle=-90]{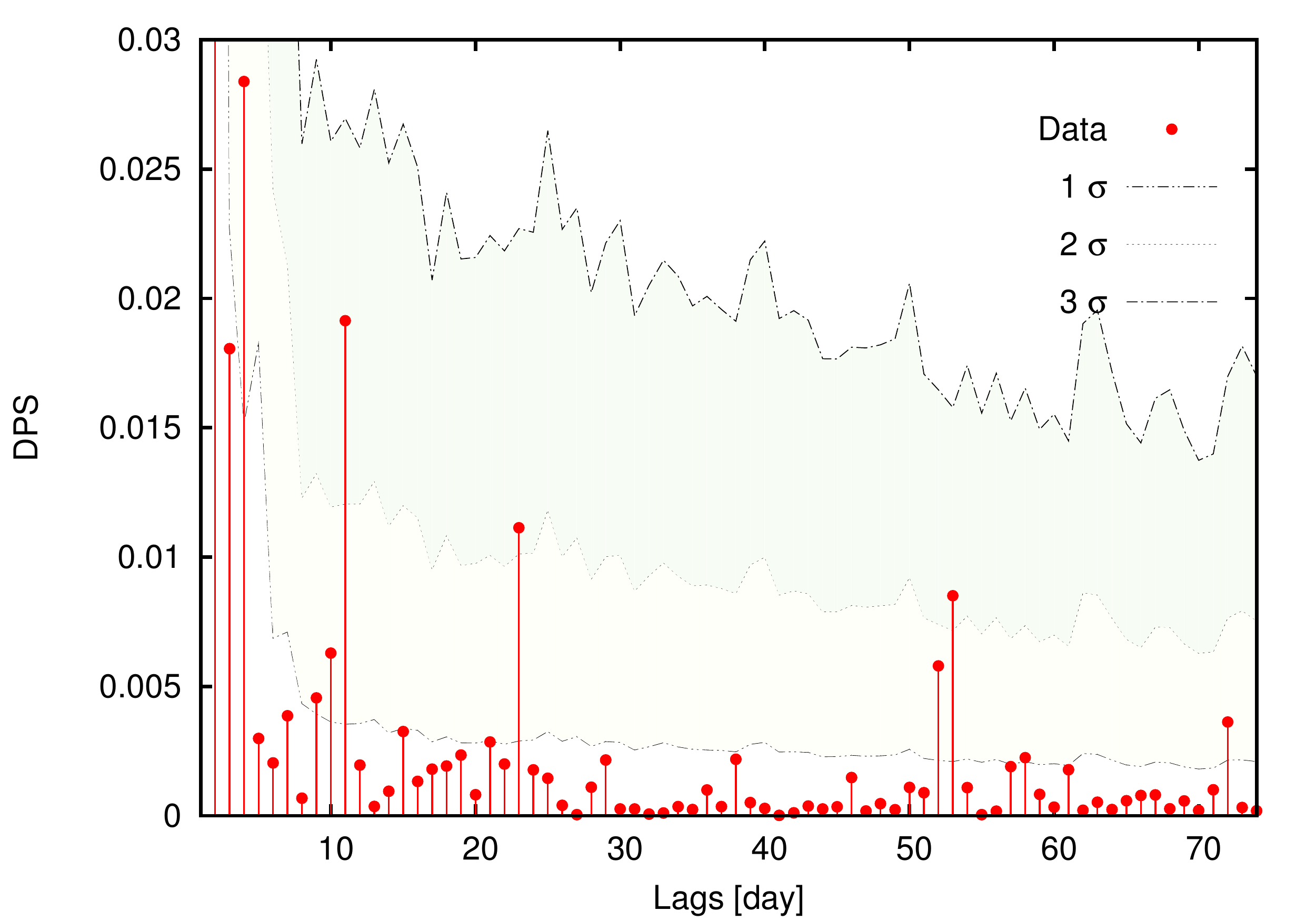}
\includegraphics[width=5.6cm,angle=-90]{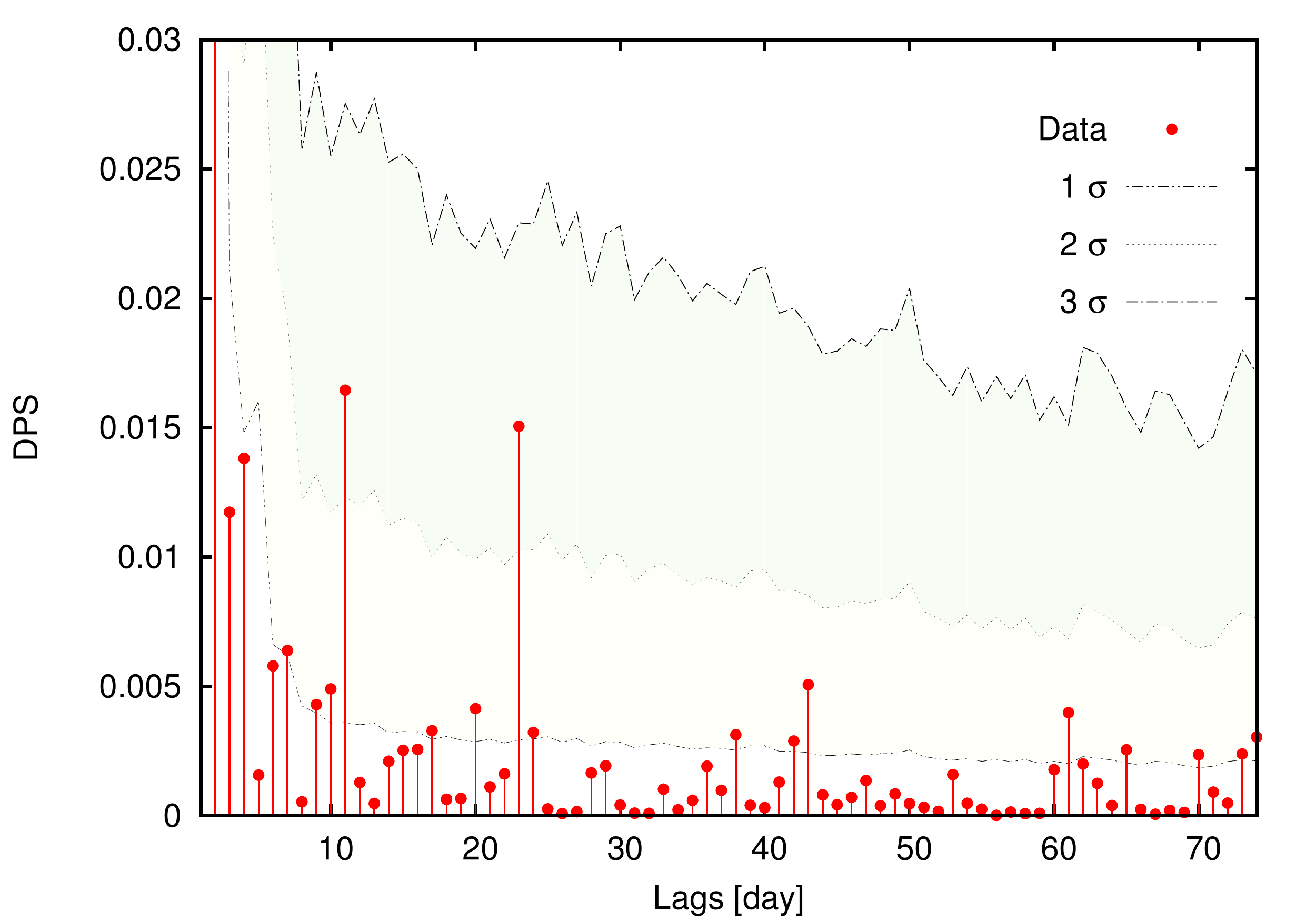}
\end{center}
\caption{\label{fig:DPS_CL_Flare1}  DPS for {\bf Flare 1} in arbitrary units with confidence levels. 
					    {\bf Top}: DPS based on upper limits as a measure of the flux.  
					    Confidence level are based on MC simulations of power law noise, 
					    including the upper limits as measured for Flare 1.\\
					    {\bf Bottom}: DPS with the flux set to zero for time bins with upper limits. Confidence levels are based on MC simulations also with zeros corresponding to upper limits.	  }
\end{figure}

\begin{figure}
%\vskip 1cm
\begin{center}
%Flare1_ACF.eps  UL not included in the CL
\includegraphics[width=5.6cm,angle=-90]{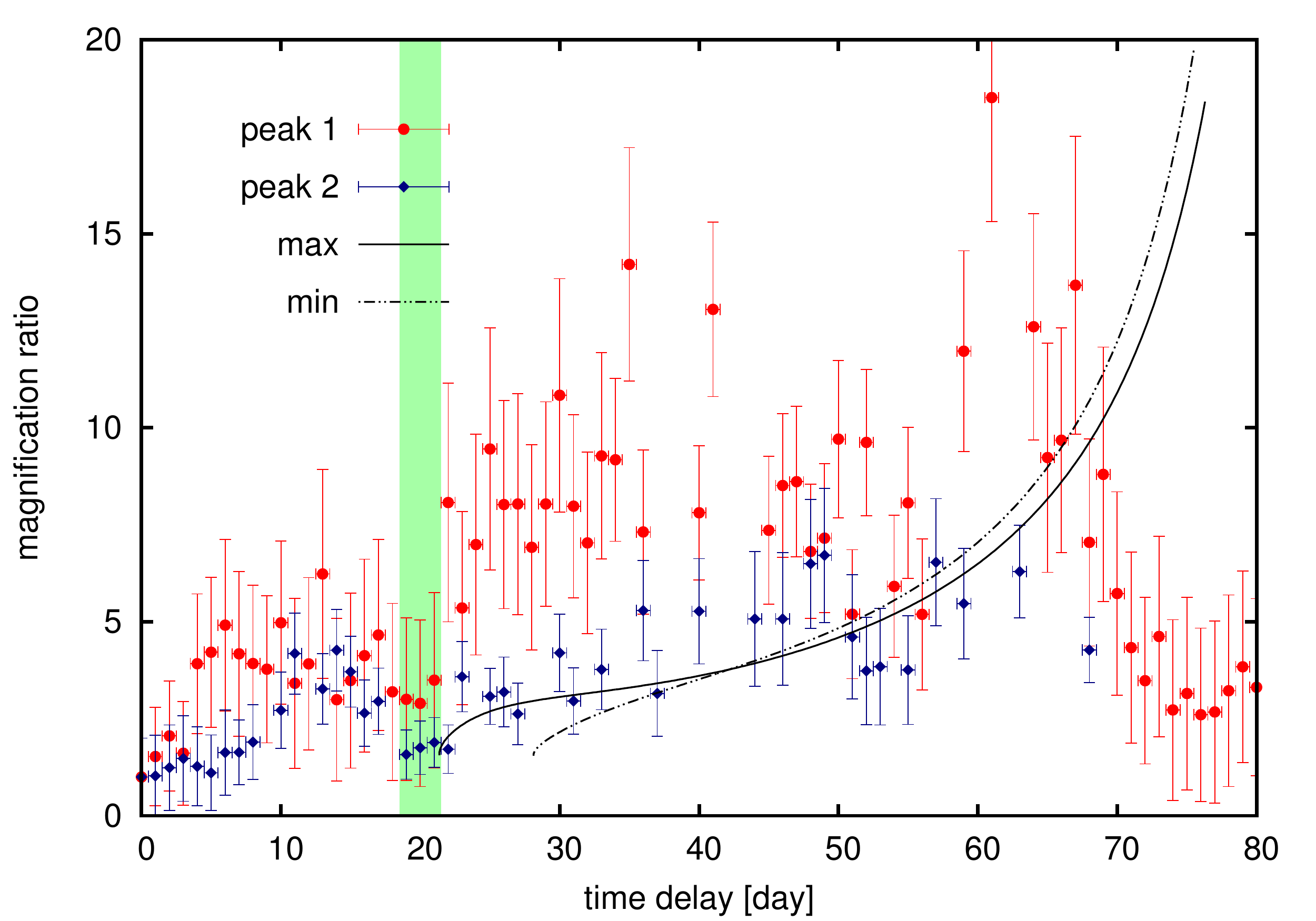}
\end{center}
\caption{\label{fig:MPM_Flare1} The MPM method applied to {\bf Flare~1}.
						  Red points correspond to  successive outbursts detected on  55484 MJD. 
						  Blue points represent the ratio between the outburst at  55560 MJD and  successive bins. 
						  To estimate the errors in the magnification ratio we use the flux error at the maxima. 
						  The area between the solid and dashed lines represents the allowed range of magnification 
						  ratios in the parameter space defined by the possible projections of the jet in the lens plane (Figure~\ref{fig:core}). 
						  The green area represents the range of time delays where the observed magnification ratio is consistent   
						   with the model predictions. }
\end{figure}
%%%%%%%%%%%%%%%%%%%%%%%%%%%%%%%%%%%%%%
\subsection{Gamma-Ray Flare 1}
\label{sec:Flare1}
%%%%%%%%%%%%%%%%%%%%%%%%%%%%%%%%%%%%%%

Figure~\ref{fig:lc_flares} shows Flare 1 with 1-day and 12-hour binning. 
The length of the light curve is  155~days.   
The temporal behavior is characterized by a set of very bright flares. 
Between the flares the flux  is close to the long-period average (covering the entire light curve); 
there are  upper limit detections for  $\sim60$~days in total. 
The fit to the power spectral density results in $\alpha = 1.45$, between pink and red noise.
Figure~\ref{fig:ACF_CL_Flare1} shows the ACF of  Flare 1.
%We have generated $10^6$ artificial light curves to evaluate the confidence levels. 

We  investigate two approaches  to  upper limits;
setting an upper limit as the flux, 
setting the flux to 0 in time bins with upper limits. 
The ACF does not show a significant difference in time delays 
and  confidence level estimates for  these two ways of treating the upper limits. 

The intrinsic variability of the source is consistent with the 1$\,\sigma$ confidence level. 
The ACF shows a broad feature at a time delay of $17.9\pm7.1$~days at $\sim2\,\sigma$ level. 
This result  agrees with time delay estimated with the ACF performed by \citet{2015ApJ...799..143A}, $19\pm1\,$days.

The other broad feature appears  at $76\pm20$~days and exceeds the 4$\,\sigma$ level. 
However, given the model of the lens, this value reaches the maximum allowed time delay.
At time delay of $\sim70\,$days, the magnification ratio between the mirage images is larger than 10, 
and, as we have demonstrated  with Monte Carlo simulations (Figure~\ref{fig:ACF_probability}), 
the probability of detecting  such gravitationally-induced  time delays 
at the 4$\,\sigma$ level using the  ACF is close to zero.
Thus, this feature is probably  not produced by a gravitationally-induced time delay,
but rather reflects the time difference between subsets of flares around 55485~MJD and 55560~MJD.

Figure~\ref{fig:DPS_CL_Flare1} shows the analysis of the same time period
but with the  DPS. 
The DPS method is much more sensitive to signal detection resulting in sharp peaks around the time delays. 
Introducing the values of upper limits as a measure of the flux results in a peak at a time delay of $\sim52\pm1.5$~day. 
This time variation corresponds to the precession period of the {\it Fermi} spacecraft of 53.4~days\footnote{http://fermi.gsfc.nasa.gov/ssc/data/analysis/LAT\_caveats\_temporal.html}.  
This  result demonstrates the sensitivity of the DPS in detecting even a faint  signal in the time series.

The DPS method, using a 1-day binned light curve,  detects two time delays at $11\pm0.5$~days and $23\pm0.5$~days above the $2\sigma$ level. 
The significance of the detection is consistent with expectations for these time delays (see Figure~\ref{fig:DPS_probability}).
For comparison, the DPS method calculated for a 12-hour binned light curve yields consistent results,
with a time delay of $22.5\pm0.5$~days. The time delay at $11\pm0.5$~days  is also present, but at lower significance. 
The lower significance is direct consequence of smaller bin.

To further investigate whether the  time delays that appear in the DPS method are induced by the gravitational lensing of a flaring emission region, 
we use the MPM which combines the observations with the predictions of the lens model. 
Figure~\ref{fig:MPM_Flare1} shows magnification ratios between the two successive periods 
following the two largest outburst in the Flare 1. 
To conclude that the detected time delay is indeed induced by the gravitational potential of the lens, 
we require that both subsets of flares have  magnification ratios consistent with the time delay. 
The time delay of $11\pm0.5$~days is inconsistent with the model. This time delay may be a harmonic of the 23-day delay or it may be a previously undetected instrumental effect. 
Figure~\ref{fig:SP_SPSstep3} shows an analysis of a randomly selected simulated light curve where a harmonic appears at this delay.

The  time delay of $\sim23$~days is consistent with the magnification ratio for both subsets of flares. 
Thus, this time delay of $23\pm0.5$~days is probably gravitationally induced, 
and constrains the spatial origin of the Flare 1. 

%------------------------------------------Figures----------------------------------------------%
\begin{figure*}[ht!]
%\vskip 1cm
\begin{center}
\includegraphics[width=4.1cm,angle=-90]{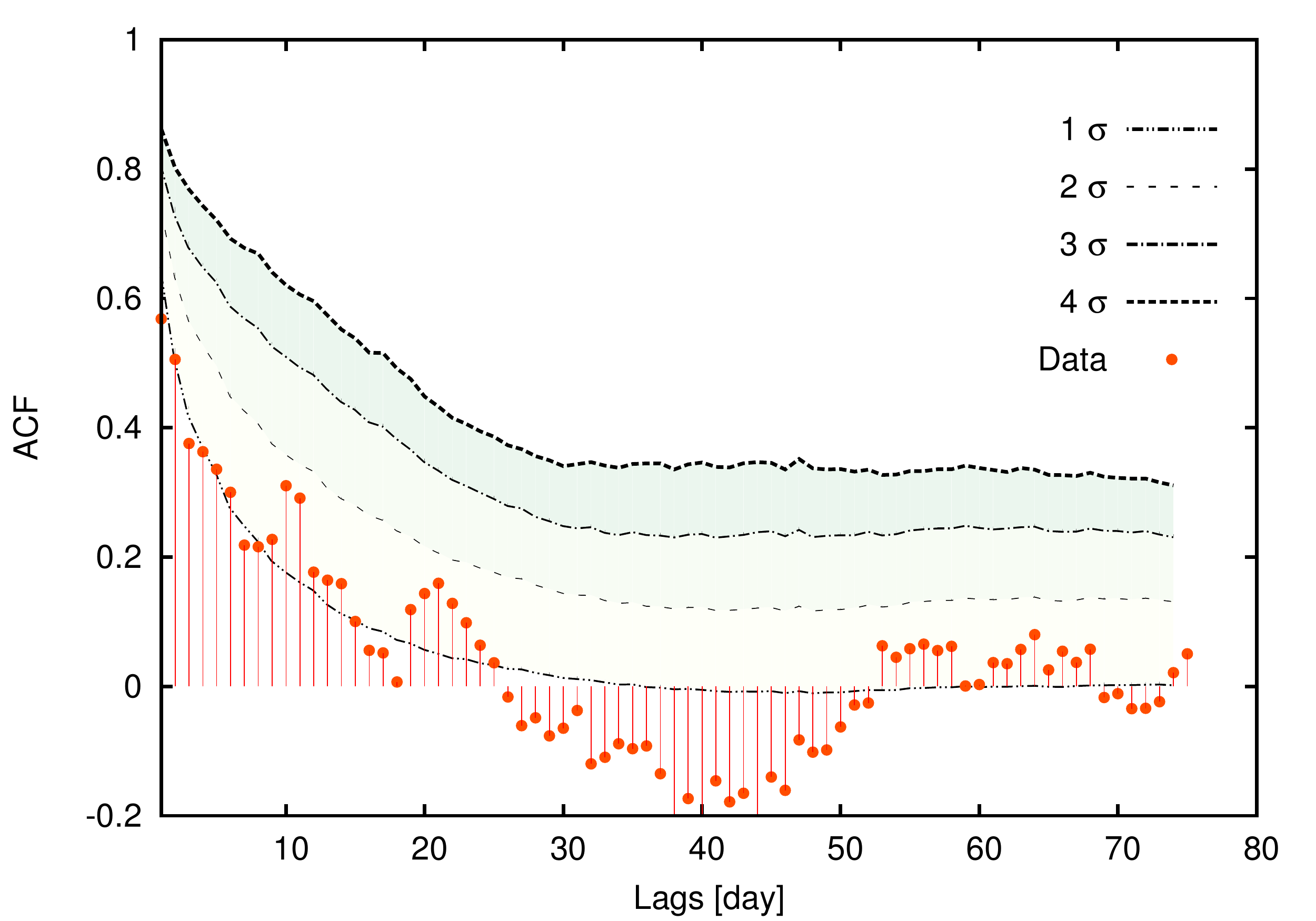}
\includegraphics[width=4.1cm,angle=-90]{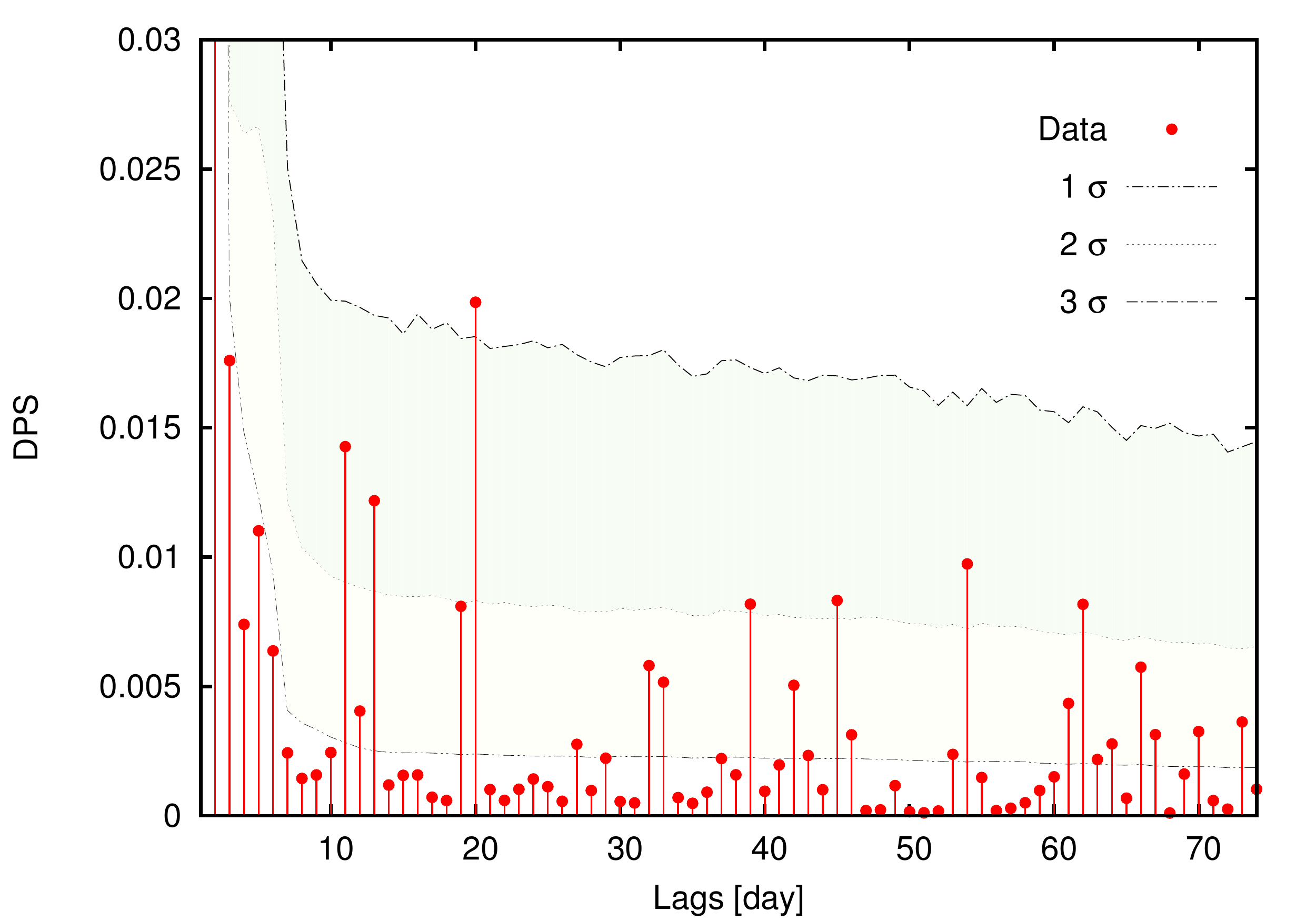}
\includegraphics[width=4.1cm,angle=-90]{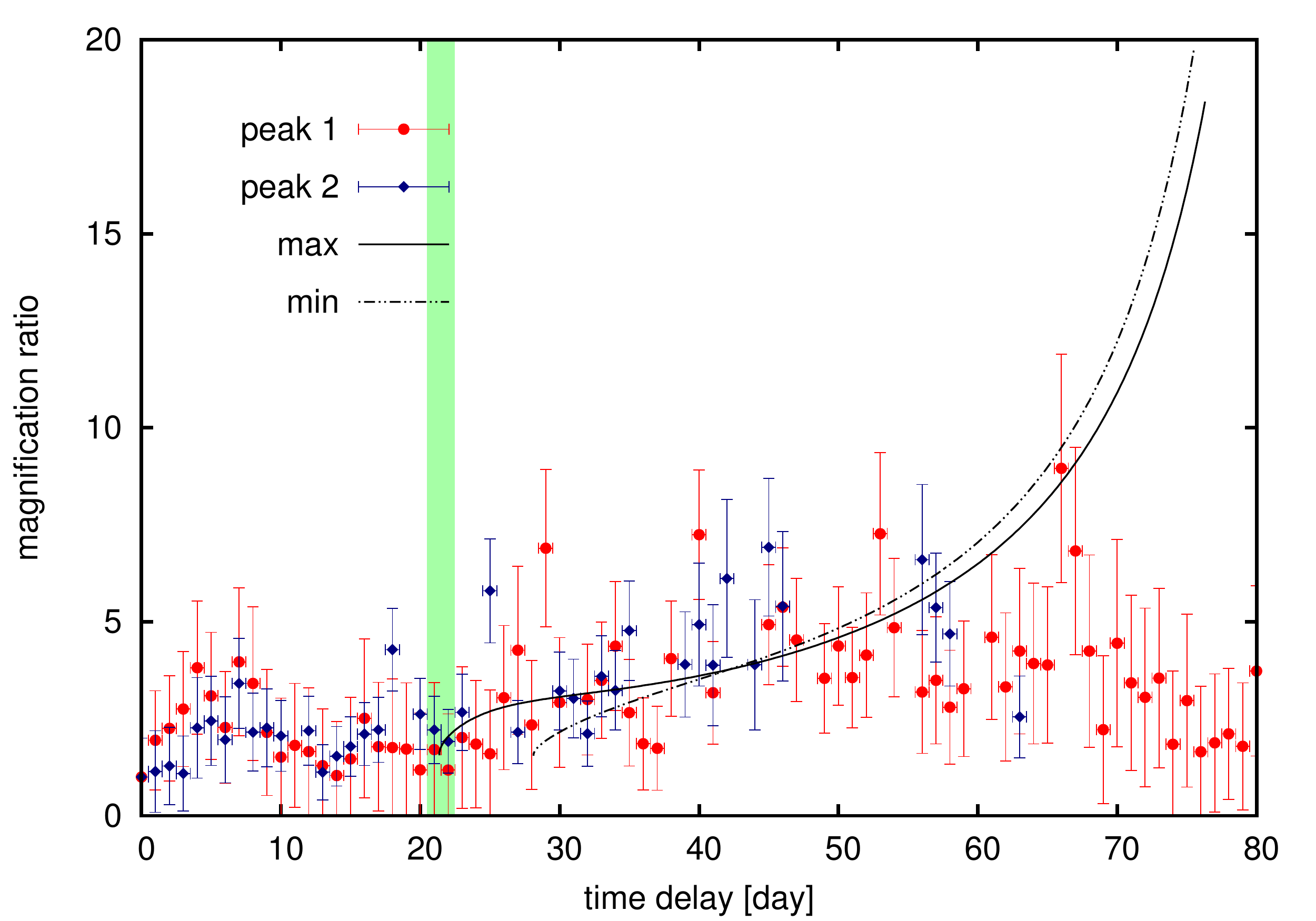}
\end{center}
\caption{\label{fig:ACF_CL_F2} Time delay estimation for {\bf Flare 2}. 
			{\bf Left:}  Autocorrelation Function with  confidence levels. 
			{\bf Middle:} Double Power Spectrum method with confidence levels.
			{\bf Right:} Maximum Peak Method, peak 1 corresponds to the flare at  56072~MJD,
			and peak 2 is a ratio calculated relative to the flare at 56146~MJD.
			Solid and dotted lines indicate predicted magnification ratios along 
			the jet indicated as the arrows A and B in Figure~\ref{fig:core}.  }
\end{figure*}

%%%%%%%%%%%%%%%%%%%%%%%%%%%%%%%%%%%%%%
\subsection{Gamma-Ray Flare 2}
\label{sec:Flare2}
%%%%%%%%%%%%%%%%%%%%%%%%%%%%%%%%%%%%%%

Figure~\ref{fig:lc_flares} shows the light curve for MJD $56043 - 56194$.
The power spectral density is represented by a power  law with an index $\alpha = 1.3$. 
We use this index in Monte Carlo simulations to evaluate the confidence levels 
for signal detection. 

The ACF (Figure~\ref{fig:ACF_CL}) shows two features at a significance level close to 2; 
the first occurs at a time delay of $10.1\pm2.5\,$days, 
and the second at $21.1\pm2.7\,$days. 

The DPS method, using 1-day binned light curve,  shows detection of the same features (see Figure~\ref{fig:DPS_CL}).
The first feature appears as a double peak at 11 and 13 days at a significance level greater that $2\,\sigma$. As in the case of Flare 1, the 11-13 day delay is inconsistent with the lens model and may be a harmonic or instrumental effect.
The other peak at $19.7\pm1.2\,$days is detected at a significance level greater then  $3\,\sigma$.
For comparison, the DPS method  for a 12-hour binned light curve yields consistent results,
with a time delay of $19\pm1.0$~days.

For Flare 2,  MPM, (see   Figure~\ref{fig:ACF_CL_F2}), 
shows a magnification ratio consistent with the model predictions 
for time delays in the range from  20 to 23 days. 
Thus, the time delay of $19.7\pm1.2\,$days is consistent  with the time delay expected for the position of the core,
and is probably a result of gravitational lensing of the flaring gamma-ray region.

%------------------------------------------Figures----------------------------------------------%
\begin{figure}[ht!]
%\vskip 1cm
\begin{center}
%Flare1_ACF.eps  UL not included in the CL
\includegraphics[width=5.5cm,angle=-90]{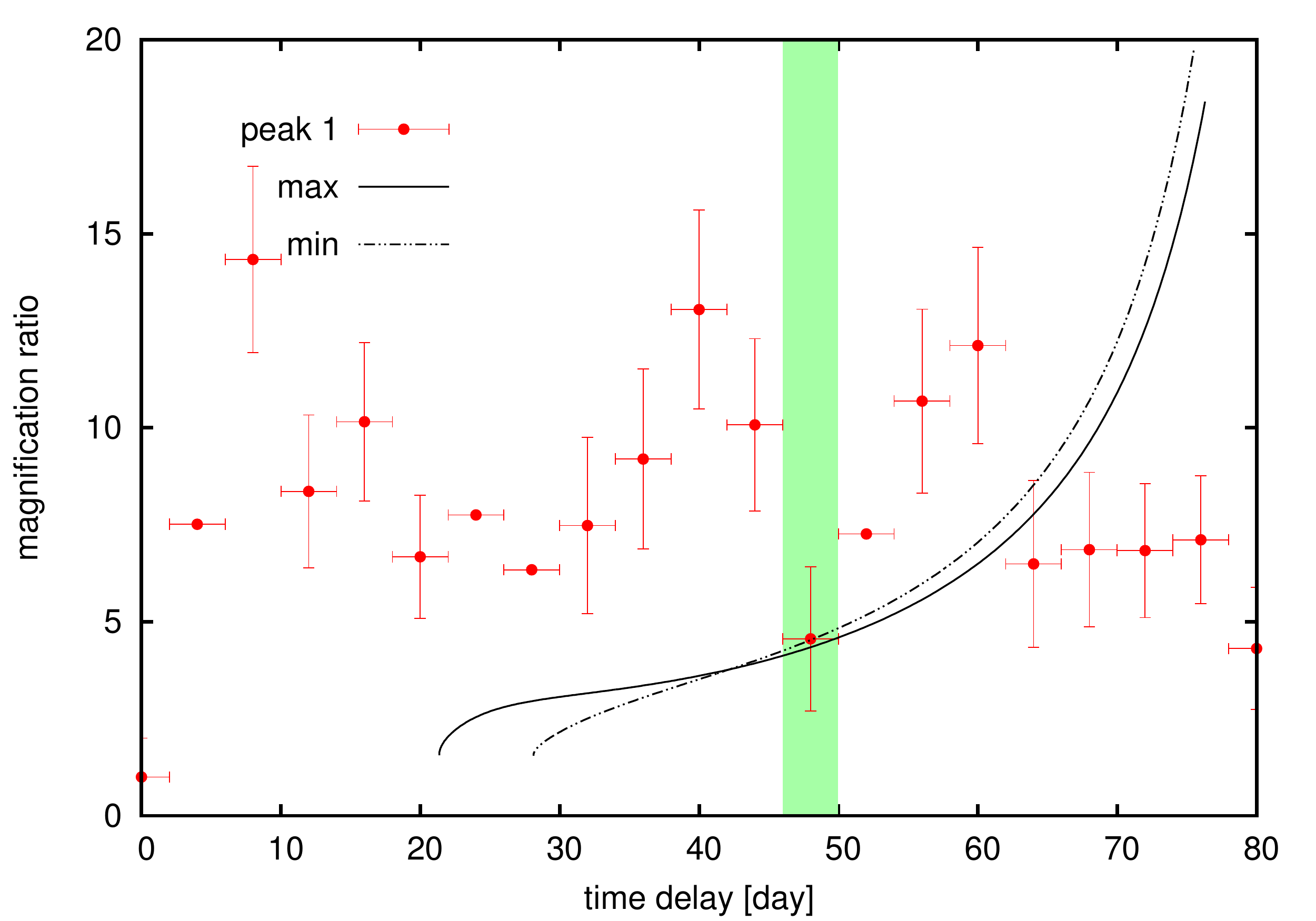}
\end{center}
\caption{\label{fig:MPM_Flare3} MPM applied to {\bf Flare 3} on July 28. }
\end{figure}

%%%%%%%%%%%%%%%%%%%%%%%%%%%%%%%%%%%%%%
\subsection{Gamma-Ray Flare 3}
\label{sec:Flare3}
%%%%%%%%%%%%%%%%%%%%%%%%%%%%%%%%%%%%%%

Figure~\ref{fig:lc_flares} shows the gamma-ray light curve of flare, which occurred on July 28. 
During the flare, the emission increased by a factor of 5 relative to the average flux. 
The flux for a period of at least 80~days before and after  Flare 3 is at or below the average flux. 

Figure~\ref{fig:MPM_Flare3} shows the results of the MPM.
The time delay range consistent with the expected magnification ratio appears at  46 - 50 days, %, in spite of the large error bars. This delay
and corresponds to an  increase in the flux recorded at 56912~MJD. 
To further investigate whether this period of activity is indeed an echo of 
the flare which occurred at 56865 MJD, 
we construct a light curve around that period with a time binning of 1 day. 

The delayed counterparts should have similar time evolution. 
The red points in Figure~\ref{fig:MPM_Flare3} show that 
these two episodes do not have identical time evolution. 
The bin around 56865 MJD consists of flux close to the average for the source; 
thus the bin may contain significant contribution from the photons originating from the quiescent state. 

Flare 3 must have a time delay equal or larger than 48 days.
The secure detection of such a long time delay is beyond the sensitivity of the ACF and  the DPS for such a short light curve with time bins of 4 days. 

If Flare 3 originated from the core, we  expect a time delay in the range 20 to 30 days.
Even with the short light curve, we  expect to detect the echo flare; in the initial flare the flux increases by a factor of $\sim 5$ 
and the echo flare should appear with a flux at least twice the average.
Absence of a detection in this range between 20 and 30 days
makes it clear that Flare 3 does not originate from the core region.
For a time delay $\gtrsim50$~days, 
Flare 3 must originate  at a projected distance from the core $\gtrsim 1.5$~kpc (see Figure~\ref{fig:jet_lens}).

%------------------------------------------Figures----------------------------------------------%
\begin{figure*}[ht!]
%\vskip 1cm
\begin{center}
\includegraphics[width=4.1cm,angle=-90]{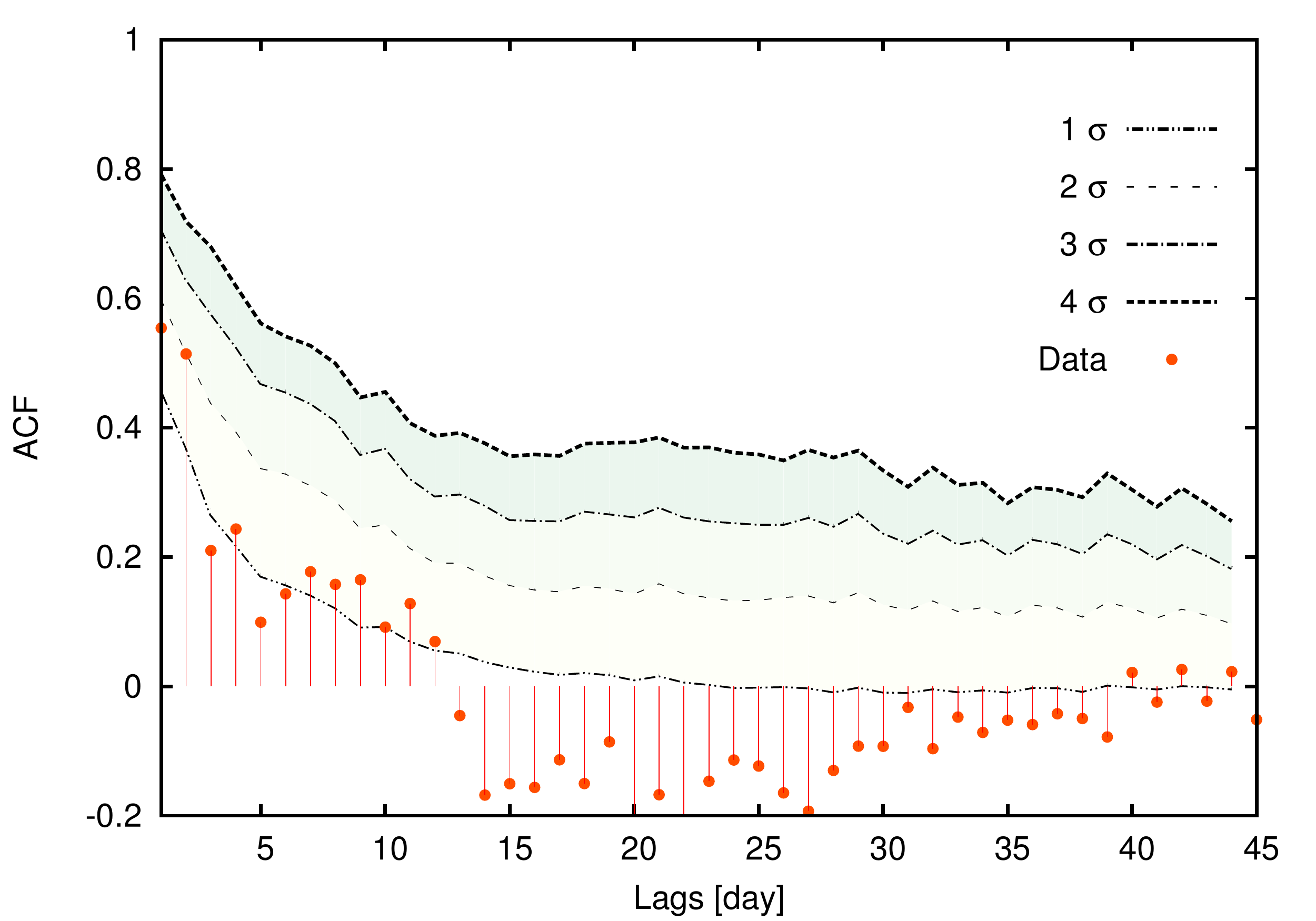}
\includegraphics[width=4.1cm,angle=-90]{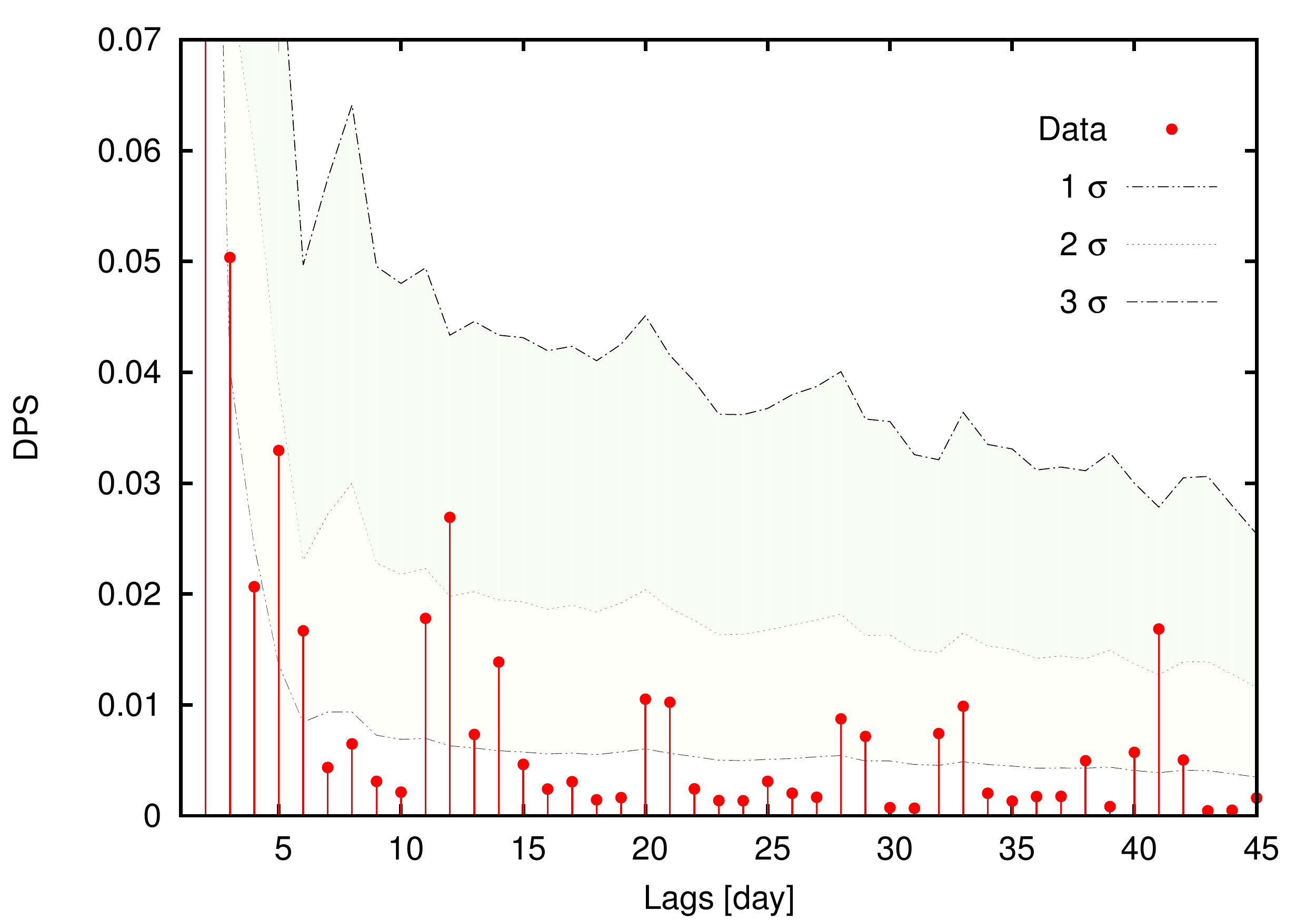}
\includegraphics[width=4.1cm,angle=-90]{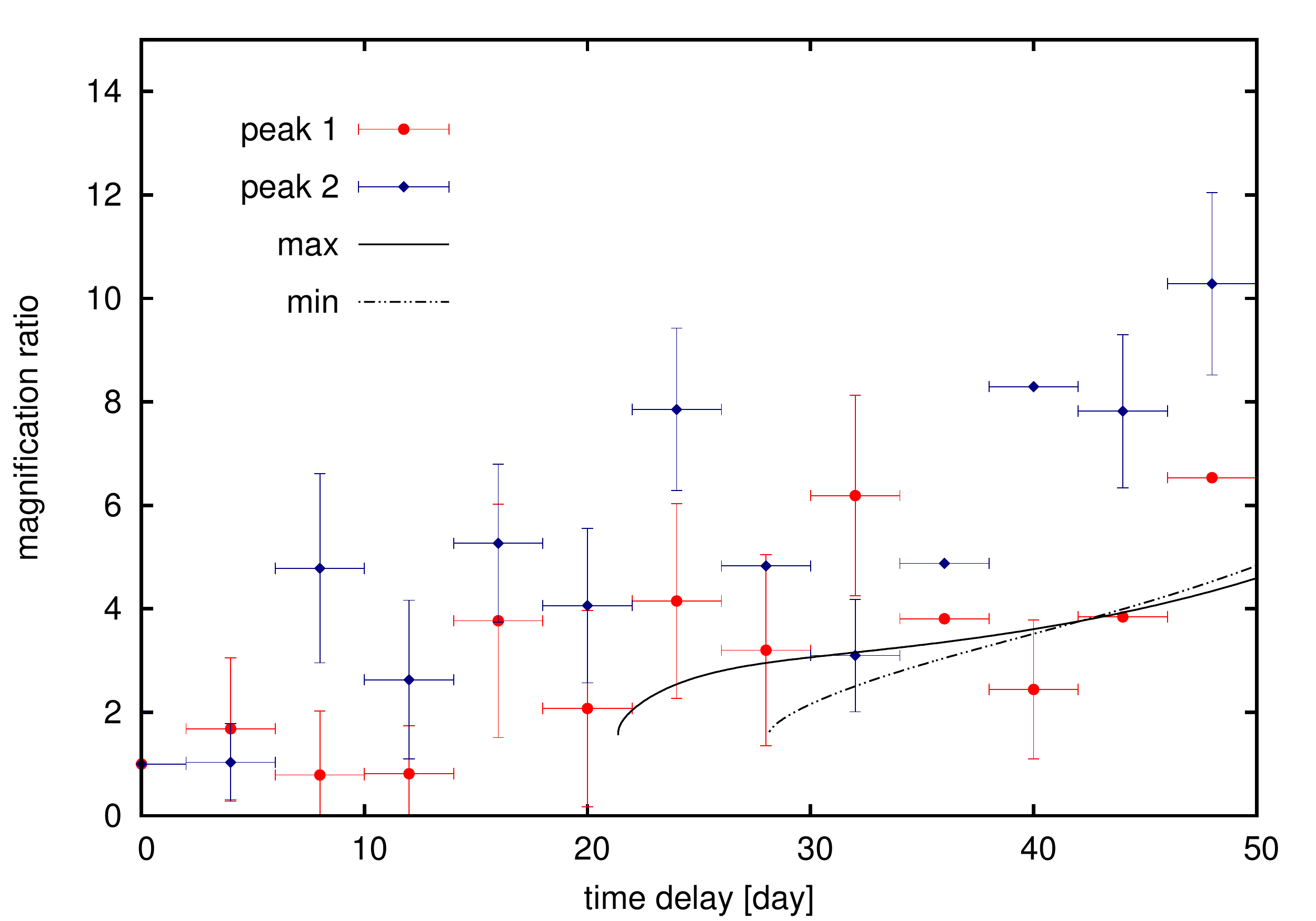}
\end{center}
\caption{\label{fig:TD_F4} Time delay estimation for {\bf Flare 4}. 
			{\bf Left:}  ACF with  confidence levels. 
			{\bf Middle:} DPS method with confidence levels.
			{\bf Right:} MPM, peak 1 corresponds to the bin centered at  57030~MJD,
			and the peak 2 is a ratio calculated relative to the time bin centered at 57038~MJD.
			Solid and dotted lines indicate the predicted magnification ratios along 
			the jet shown as the arrows A and B in Figure~\ref{fig:core}. }
\end{figure*}

%%%%%%%%%%%%%%%%%%%%%%%%%%%%%%%%%%%%%%
\subsection{Gamma-Ray Flare 4}
\label{sec:Flare4}
%%%%%%%%%%%%%%%%%%%%%%%%%%%%%%%%%%%%%%

The most recent gamma-ray activity of PKS~1830-211  (Figure~\ref{fig:lc_flares})
consists of two flares. 
For the temporal analysis, we  use a light curve with one-day binning consisting of 90 days.
The ACF, the DPS,  
and the MPM do not show  time delays consistent with  origination in the core (Figure~\ref{fig:TD_F4}). 
The DPS method  indicates a time delay at $11.8\pm0.8$~days with a significance of $\sim 2\sigma$; 
however this time delay is inconsistent with model based on radio observations
and is thus probably a false positive. 
The time delay of $\sim\,11\,$days accidentally  corresponds to the time between the two flares. 
The first flare was brighter than the average flux for about 4 days and peaked around 57032~MJD. 
The second one lasted for about 9 days and appeared 2 days after the first one. 
These flares have very different temporal evolution, 
thus, are not echoes of one another.

Again, the lack of detection of the time delay in the range between 20 and 30 days
shows that Flare 4 does not originate from the core region. 
The analysis method is sensitive for time delays $\lesssim 50$ days and there are no other detections.
The data show that the time delay must be greater  than $\sim50$~days and thus
the radiation  must  originate from a region located 
at projected distance from the core $\gtrsim  1.5$~kpc (see Figure~\ref{fig:jet_lens}).

%%%%%%%%%%%%%%%%%%%%%%%%%%%%%%%%%%%%%%
\section{Discussion}
\label{sec:discussion}
%%%%%%%%%%%%%%%%%%%%%%%%%%%%%%%%%%%%%%

\begin{figure}
%\vskip 1cm
\begin{center}
\includegraphics[width=8.9cm,angle=0]{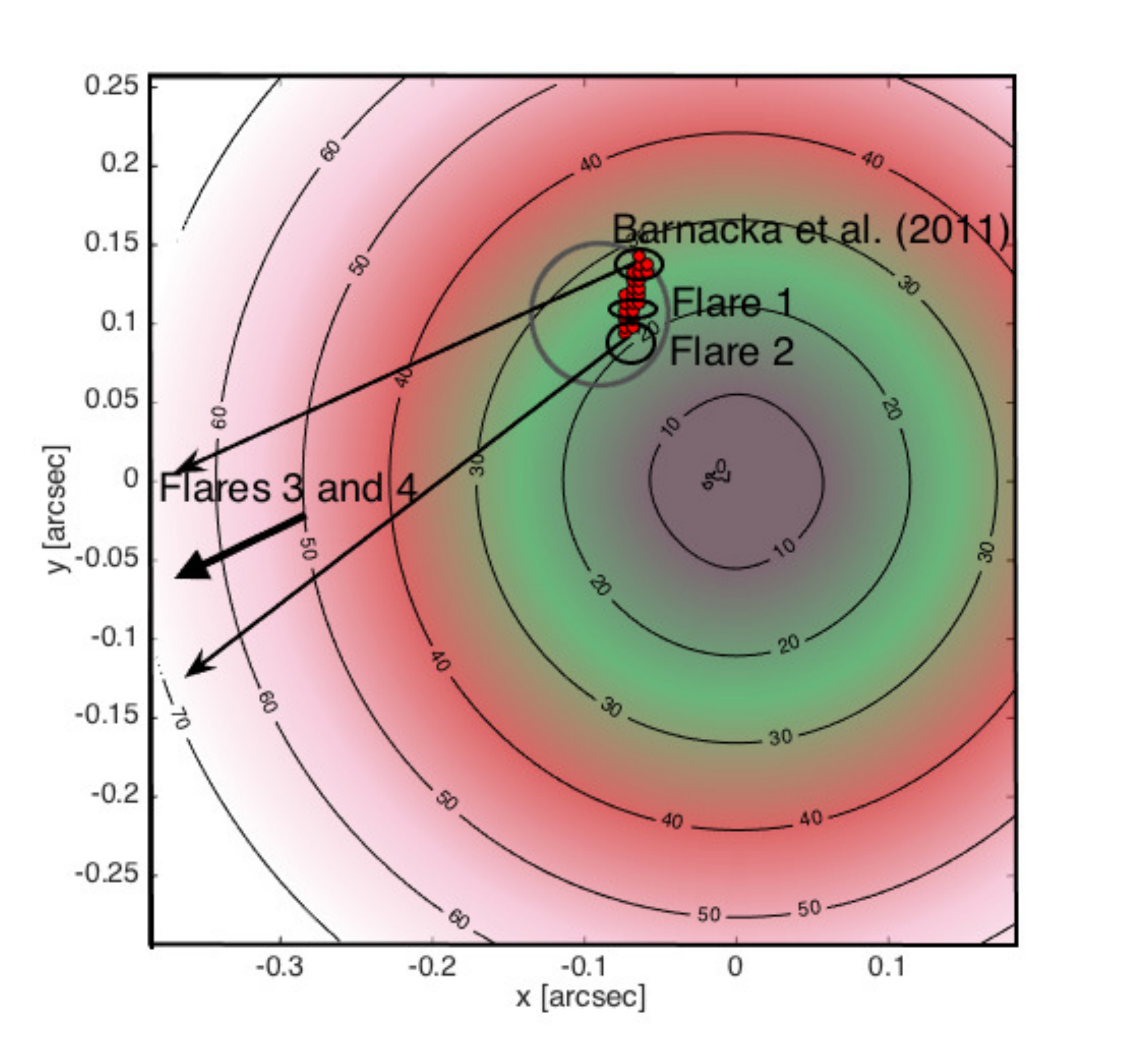}
\end{center}
\caption{\label{fig:jet_resolved} Resolved positions of gamma-ray flares. 
						 The color pallet and contours indicate the time delays in days.
						 The long arrows show the boundary of the jet alignment limited by well resolved radio observations.
						 Gray circle show the position of the core from \citet{Sridhar2013}.
						 Red circles  are further constraints  on the position of the core using the  lens model and 
						  the time delay and magnification ratio measurements by \citet{1998ApJ...508L..51L}.
						  Ellipses elucidate the spatial origin of 
						  the flares obtained through time delay measurement for Flare~1 and Flare~2; they are consistent with the core. 
						  The top ellipse shows the spatial origin of the time delay measured by \citet{2011A&A...528L...3B}
						  using the gamma-ray light curve in the quiescent state. 
						  The short arrow indicates constraints from Flares~3 and 4. The time delays $\gtrsim 50$~days  imply that  
						  the emitting region must be located at projected distance of $\gtrsim$ 1.5~kpc from the core.}
\end{figure}

Lensing resolves the gamma-ray emission of PKS~1830-211 during its flaring periods 
and limits the origin  to the core and to regions displaced by $\gtrsim$ 1.5 kpc along the jet. 
Flares~1 and~2 originate from a region of $\sim$100~pc around the core.
At the redshift of $z=2.507$, where PKS~1830-211 is located, a projected distance of 100~pc corresponds to $\sim0.02\,$arcsecond. 
Thus,  this lens improves the angular resolution at gamma-ray $\sim10000$ times
(Figure~\ref{fig:jet_resolved}).

Resolving the high energy universe using cosmic lenses relies on the ability to measure time delays and to model the mass distribution of the lens. 
The  localization to  100~pc in this gravitationally-lensed system corresponds to an uncertainty  in time delay measurement of 5 days.
The DPS method is an effective approach for 
measuring the time delay (Section~\ref{sec:DPS} and Appendix~\ref{app:DPS}). 
This method  can extract time delays from gamma-ray light curves  with an accuracy down to 0.5~days. 
In principle, this accuracy can provide a localization of the source to $\sim 10$ pc.

A limiting  factor in any lensing analysis is the precise  model of the lens and alignment of the jet.
We have used a very conservative position of the core and the jet alignment.
In principle, more detailed analysis 
of all of well-resolved radio images can yield better constraints, but this analysis is beyond the scope of this paper. 

For these complex sources, there are always puzzles.
The lack of detection of time delays following Flare~3 and Flare~4 could indicate
some other physical source for increased emission.
The first possibility is microlensing. 
Flare~3 and Flare~4 are unlikely to be microlensing events
because the typical time scale of a caustic crossing microlensing event  is of the order of months to years \citep{2001ASPC..239..351W}; 
Flares~3 and 4 have  a typical duration of days and a time structure characteristic of gamma-ray flares. 

The size of the emitting region (the source size effect) might impact the  magnification ratio for Flares~3 and 4.
A spatially larger emitting region results in a larger magnification ratio  \citep[see Figure~2 in][]{2014arXiv1403.5316B}.
However, the minimum variability time scale of $\sim$ 1 day observed in these  flares 
constrains the emitting region to $\lesssim$ 0.01~pc or $\lesssim0.001\%$ of the Einstein radius of the lens. 
In other words, the size of the emitting region is small enough to have a negligible effect on the magnification ratio.  

The final issue is $\gamma-\gamma$ absorption. 
Gamma-ray emission of lensed blazars  passes through the lens where low energy photons
may absorb the gamma rays of one of the images passing through the more luminous region of lensing galaxy.
The absorption may affect gamma-ray photons with energies larger than a few GeV.
{\it Fermi}/LAT detects a majority of photons in the energy range $>$ 100~MeV. 
In addition,  \citet{2014arXiv1404.4422B} show that the luminosity of a single galaxy is too low to cause significant absorption of the gamma-ray flux. 
If all four active periods originated from the same region, absorption would affect all of them in the same way. 
However, we detect time delays for half of the flaring periods,
suggesting that $\gamma-\gamma$ absorption is irrelevant. 

 We have checked the position of the Sun relative to the PKS 1830-211.  
For flares 2,3, and 4, the Sun was located outside the Region of Interest (ROI). 
For flare 1, the smallest separation between position of the Sun and PKS 1830-211 was $\sim 2.5$ deg in the period MJD~55558  - 55561, 
which is marginal fraction of the Flare 1. 
Thus, the Sun does not affect our analysis.  

A second gravitationally-lensed source, B2~0218+35, shows behavior similar to PKS~1830-211.
The bright flaring periods result in time delays consistent with origination from the core. 
The time delay measured from flaring period at gamma rays is $11.46\pm0.16\,$days  \citep{BucCheung}.
In the radio, \citet{1999MNRAS.304..349B} measured the time delay of $10.5 \pm0.2\,$days and \citet{2000ApJ...545..578C} obtained a time delay of $10.1\pm0.8\,$days.
However, the most recent gamma-ray flare of B2~0218+35 does not show delayed counterparts suggesting that in this source, 
flares also have multiple spatial origins.

%\citet{1990MNRAS.246..263S} shows that the radio observations at 2~cm of mirage images
%the NE and SW components appear decomposable into a dominant core and a jet containing a knot. 

%%%%%%%%%%%%%%%%%%%%%%%%%%%%%%%%%%%%%%
\section{Conclusions}
\label{sec:conclusions}
%%%%%%%%%%%%%%%%%%%%%%%%%%%%%%%%%%%%%%

Strong gravitational lensing is a  powerful tool for resolving the high energy universe. 
As a prototypical example of the power of lensing combined with long, uniformly sampled light curves in the gamma-ray regime, we  investigate the spatial origin of  flares from PKS~1830-211 observed with  {\it Fermi}/LAT. Despite the poor angular resolution of gamma-ray detectors,
gamma-ray flares can be the basis for spatial resolution of a source thanks to the unique observational strategy of {\it Fermi}/LAT. 

Analysis of four active periods in PKS~1830-211 shows that the gamma-ray radiation during two flaring periods originated from 
a region  spatially  coincident with the radio core. The effective spatial resolution we achieve is $\sim 100$ pc.
 
Two more recent flares apparently do not originate from the core region because the time delay must be $\gtrsim$~50 days.  
This delay and the lens properties derived from observations at lower energies indicate 
that these flares must originate at a  distance $\gtrsim 1.5\,$kpc from the massive black hole powering the blazar. 
Our analysis demonstrates that variable emission can originate from regions essentially
coincident with the core, or from regions substantially displaced along the jet.
These flares of short duration originating from regions within the jet challenge our understanding of particle acceleration along with the physical conditions along the jet. 

We lay out and apply three methods of time delay estimation from unresolved light curves:
(1) the  Autocorrelation Function, (2) the Double Power Spectrum, and (3) the Maximum Peak Method. 
We Monte Carlo simulations to investigate the strengths and weaknesses of these methods and 
the probability of detecting the gravitationally-induced time delays. We provide details of their application in the appendices.

Monte Carlo simulations demonstrate the power of signal processing (the DPS) in increasing the probability of time delay detection over the more standard autocorrelation methods. The MPM enables us to evaluate the consistency of the detected time delays with the magnification ratios expected from a model consistent with data at shorter wavelengths. For the long active periods consistent with origination in the core, the 
gamma-ray time delays,  $23\pm0.5\,$days and $19.7\pm1.2\,$days, are consistent with the radio,  $26^{+4}_{-5}\,$days. 

This analysis lays a foundation for future use of unresolved light curves of lensed sources
to enhance the effective angular resolution of the detectors and to elucidate the physics of radiation from distant sources. 
 {\it Fermi}/LAT has detected at least one additional lensed blazar which could be analyzed along similar lines. The techniques outlined here can be applied to long-term monitoring data at any wavelength. For example,  Euclid, LSST, or SKA, will monitor many distant blazars and we can expect that many  of them will be lensed thus enabling a probe of the nature and evolution of radiation from these sources. 
Long-term, uniformly sampled light curves are the critical input for these signal processing methods.

%The monitoring and methods design to measure these time delays should take into account possibility of multiple time delays.
 
%%%%%%%%%%%%%%%%%%%%%%%%%%%%%%%%%%%%%%
\acknowledgments
%%%%%%%%%%%%%%%%%%%%%%%%%%%%%%%%%%%%%%
We thank Benjamin Bromley,  Eric Charles, Scott Kenyon, Christopher Kochanek, Sebastien Muller, and Dan Schwartz
 for the valuable comments and discussions. 
We thank the referee for providing comments  that have led to a tighter discussion. 

A.B. is supported by the Department of Energy Office of Science, 
NASA \& the Smithsonian Astrophysical Observatory 
with financial support from the NCN grant DEC-2011/01/M/ST9/01891.
MJG is supported by the Smithsonian Institution.
%We thank the referee for valuable comments on the manuscript.

%%%%%%%%%%%%%%%%%%%%%%%%%%%%%%%%%%%%%%
\bibliography{PKS1830_gamma}
%%%%%%%%%%%%%%%%%%%%%%%%%%%%%%%%%%%%%%

%%%%%%%%%%%%%%%%%%%%%%%%%%%%%%%%%%%%%%
\appendix
\section{Signal Processing in The Double Power Spectrum Method}
\label{app:DPS}
%%%%%%%%%%%%%%%%%%%%%%%%%%%%%%%%%%%%%%

The importance of signal processing has been demonstrated in a wide range of applications 
including analysis of speech, imaging, or video and seismic events. Signal processing has
played a crucial  rule in the development of these fields and it is applied extensively in astronomy for spectroscopy, synthetic imaging, and radio astronomy, among others \citep[e.g. XCSAO,][]{1992ASPC...25..432K}. 

\begin{figure}[ht!]
%\vskip 1cm
\begin{center}
\includegraphics[width=17.5cm,angle=0]{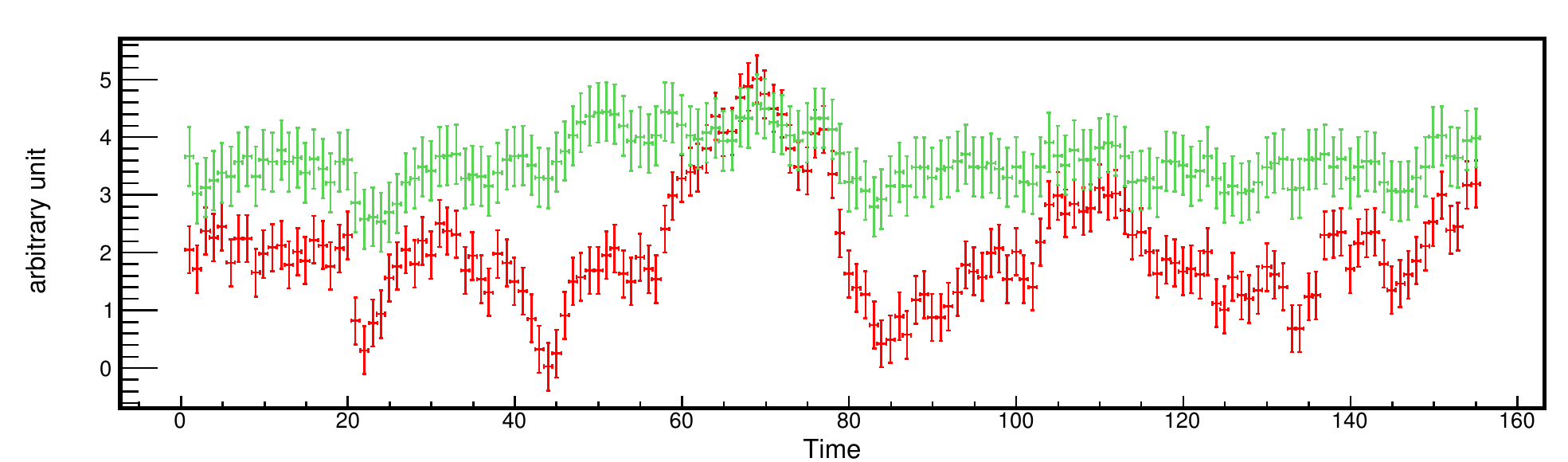} 
\end{center}
\caption{\label{fig:lc_MC} 
                          Artificial light curves of red noise. The red points represent pure noise. 
                          The green light curve is the sum of two components: 
                          the first component is the same as the red light curve, 
                          and the second component is the identical  light curve 
                          shifted by 20 days and with an a magnification ratio of 1.3. 
                          For visualization purposes, we have added error bars of 20\% of the average flux.  }
\end{figure}

We describe the signal processing method, the Double Power Spectrum, step-by-step.
The steps in this signal processing are based on widely used methods \citep{1975dsp..book.....O,1971A&A....13..169B}.
We begin by preparing the light curve, an input time series.
For demonstration purposes, we produce an artificial light curve of  red noise as a representation of the flaring state.
Figure~\ref{fig:lc_MC}  shows (red points) the artificial light curve of red noise of  duration 155 units.
The green light curve shows the same red noise, 
but including time delays and corresponding magnification ratios that simulate gravitational lensing of the source. 
We induced a time delay of 20 days and a magnification ratio of 1.3, following the procedure in Section~\ref{sec:Signal}. 

\begin{figure}[ht!]
%\vskip 1cm
\begin{center}
\includegraphics[width=5cm,angle=-90]{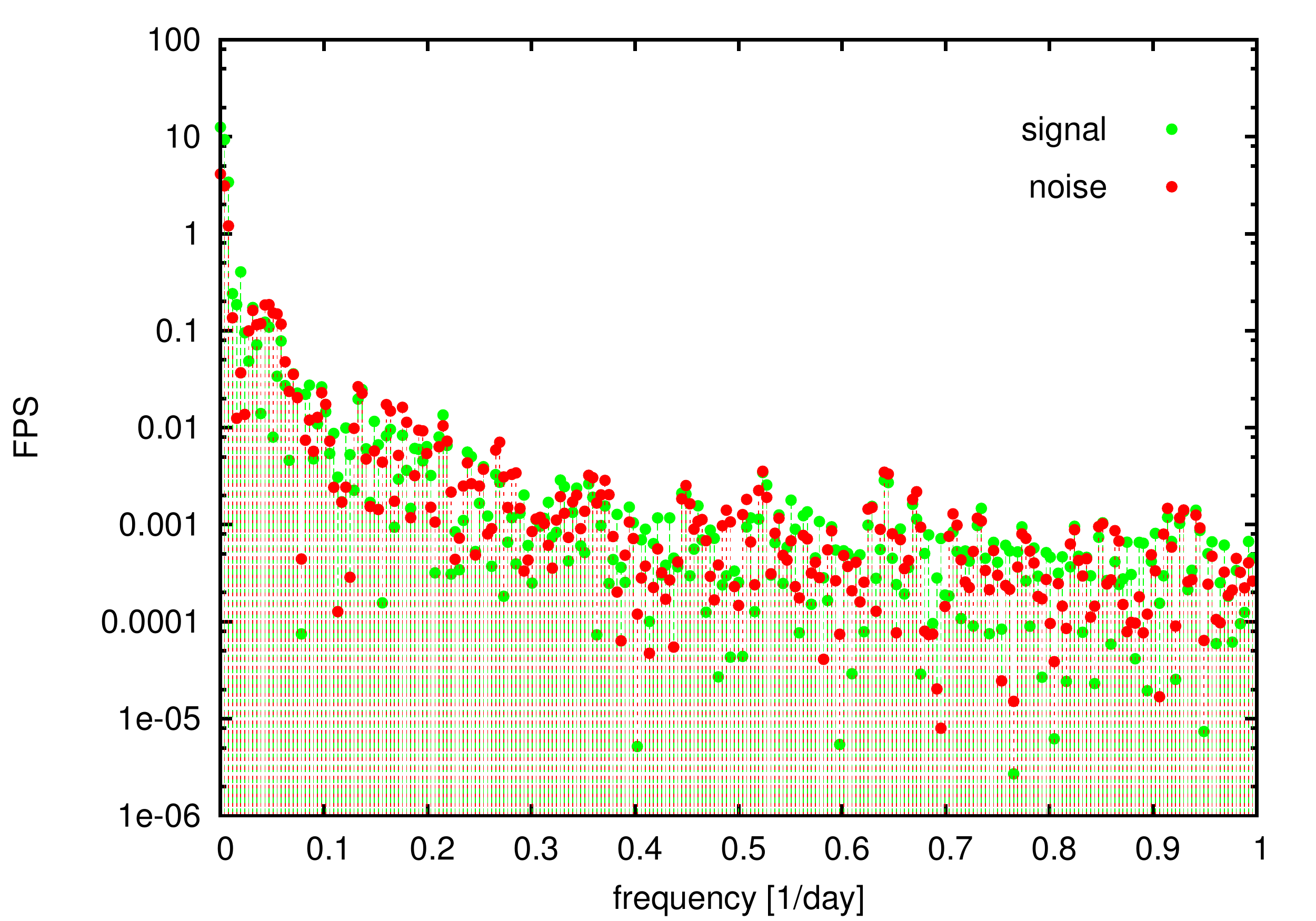} 
\includegraphics[width=5cm,angle=-90]{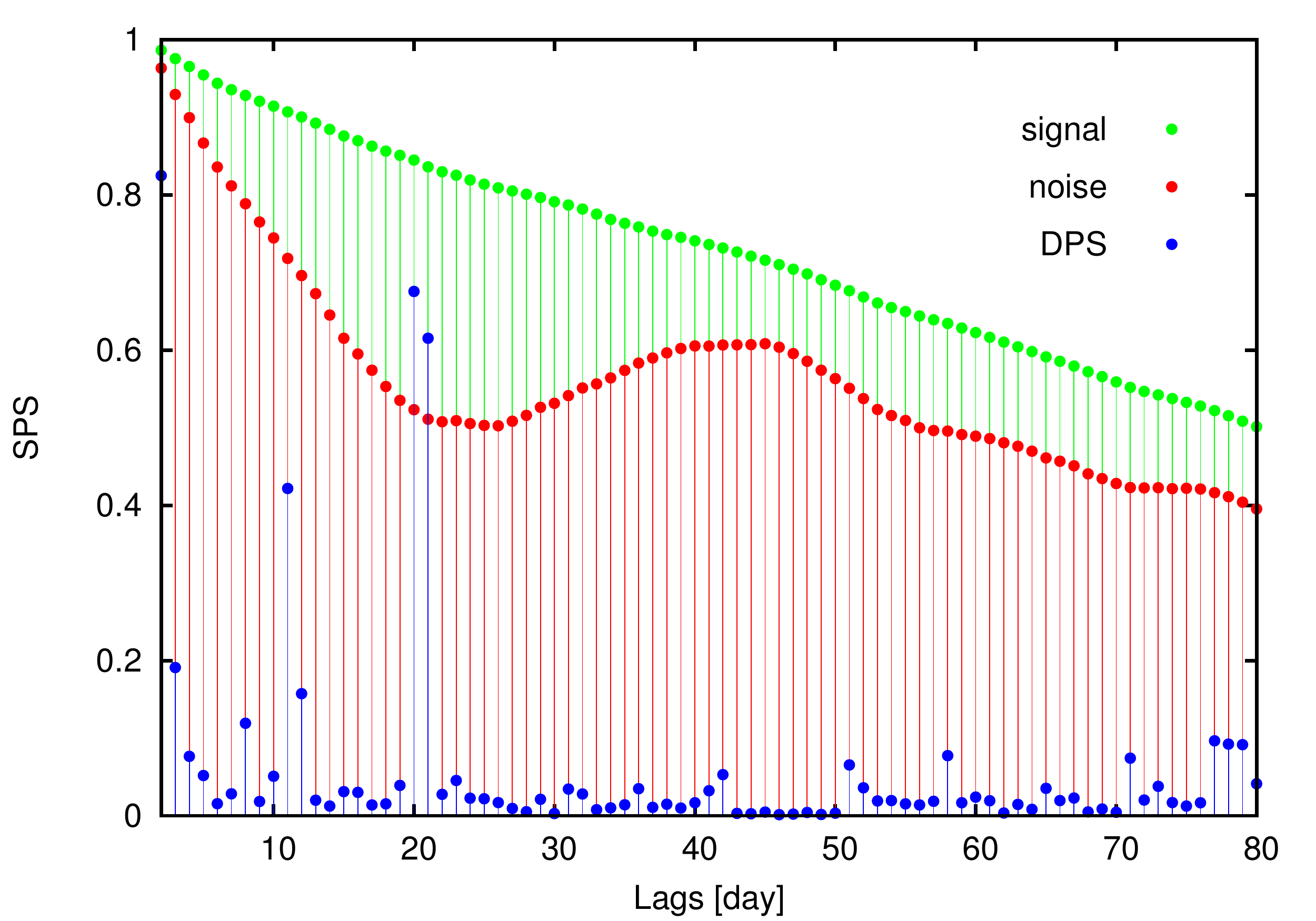} 
\end{center}
\caption{\label{fig:SP_no} 
                          No Signal Processing.
                          Right: The First Power Spectrum. 
                          Left: The Second Power Spectrum. 
                          The SPS is normalized by the maximum values in the spectra. 
                          The red points represents the FPS and SPS calculated for the red noise light curve (Figure~\ref{fig:lc_MC}).
                          The green points show the results for the light curve with an artificially induced time delay of 20 days.
                          The blue points in the SPS indicate the result of the full signal processing procedure applied to green light curve that simulates the impact of gravitational lensing. }
\end{figure}

The DPS method consists of three major stages. 
The first is a preparation of the input time series, the light curve.
The second stage is a calculation of the First Power Spectrum (FPS), 
and the last stage is a calculation  of the Second Power Spectrum (SPS). 

Figure~\ref{fig:lc_MC} shows
an example of the light curves, used here as the input. 
For demonstration purposes, we have used these light curves and calculated the first and second power spectra, 
without applying any signal processing. 
Figure~\ref{fig:SP_no} shows the result of this analysis with the FPS on the left,
and the SPS on the right. 
Note that the light curve (see Figure~\ref{fig:lc_MC}) serves as an input to the FPS, 
 the FPS is used as an input for the DPS. 

The green light curve shows the time series with an artificially induced time delay of 20 days. 
The DPS for this signal does not show a hint of the time delay detection when there is no signal processing applied. 
For comparison, in blue, we show the DPS of the same light curve, but after the full signal processing described below.

%%%%%%%%%%%%%%%%%%%%%%%%%%%%%%%%%%%%%%
\subsection{The First Power Spectrum}
%%%%%%%%%%%%%%%%%%%%%%%%%%%%%%%%%%%%%%

%%%%%%%%%%%%%%%%%%%%%%%%%%%%%%%%%%%%%%
\subsubsection{Step 1: Removing the Mean and Windowing}
%%%%%%%%%%%%%%%%%%%%%%%%%%%%%%%%%%%%%%

The input time series in the First Power Spectrum is the light curve shown in Figure~\ref{fig:lc_MC}. 
We start the signal processing by preparing the input.
In the first step, we subtract the mean from the time series. 
This step eliminates the large power in the first bin of the first power spectrum. 
%signal relevant to the determination of the time delay in the behavior of the mean.
In the next step, we apply a window function to the input.  
Windowing is induced to balance  the sharpness of the peak of a periodic signal 
with the spectral resolution. 
If a time delay is present in the time domain, 
 it will also manifest its presence  in the frequency domain (the FPS)
as a periodic pattern with a period inversely proportional to the time delay (see equation~(\ref{eq:fps})). 
Thus, we must preserve  the maximum resolution of the FPS; 
for this purpose, we use a rectangular window. 

%%%%%%%%%%%%%%%%%%%%%%%%%%%%%%%%%%%%%%
\subsubsection{Step 2: Zero Padding}
%%%%%%%%%%%%%%%%%%%%%%%%%%%%%%%%%%%%%%

To avoid the large power at low frequencies caused by discontinuity at the beginning and the end of the time series, 
we apply zero padding to the time series.

\begin{figure*}[ht!]
%\vskip 1cm
\begin{center}
\includegraphics[width=5cm,angle=-90]{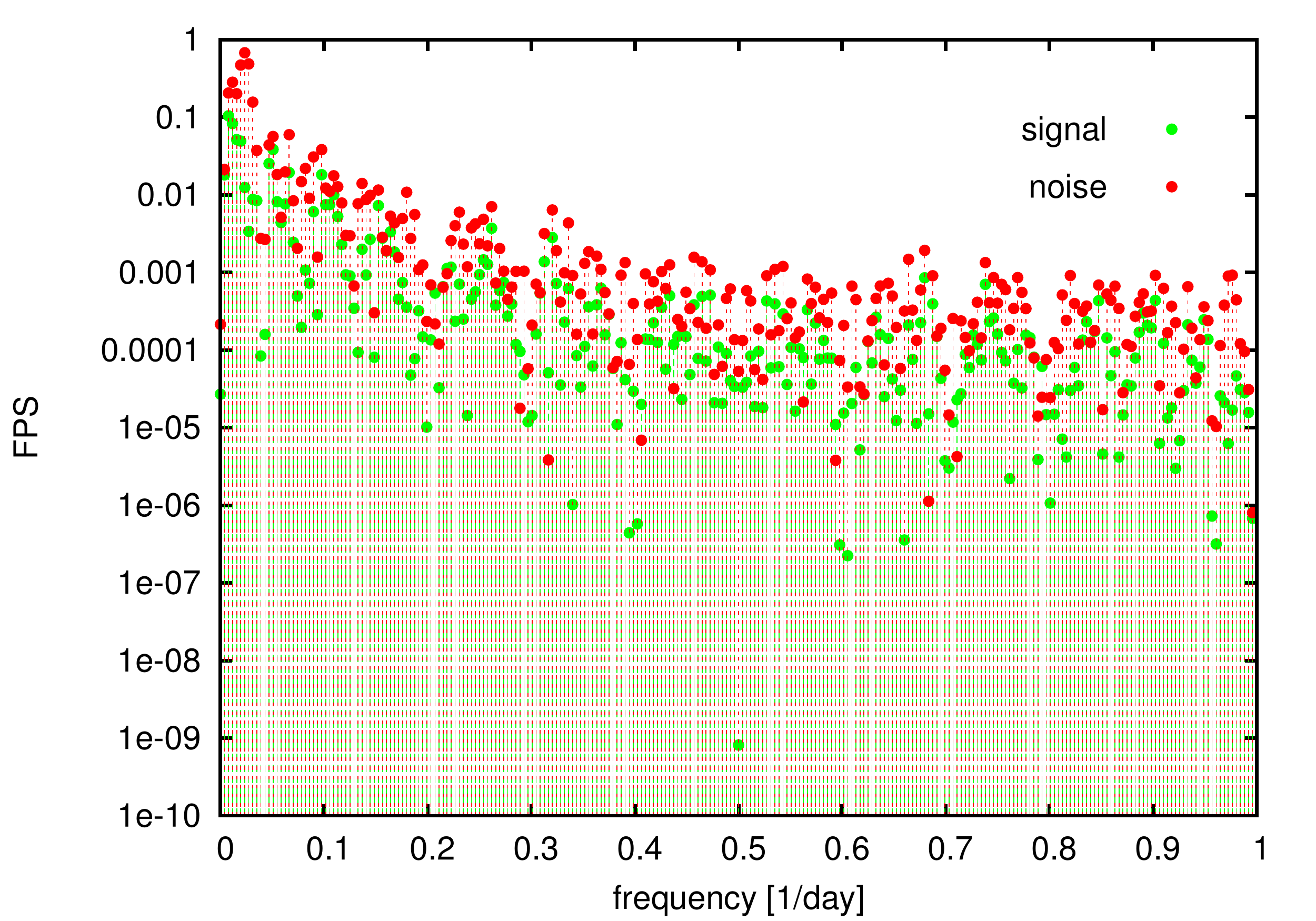} 
\includegraphics[width=5cm,angle=-90]{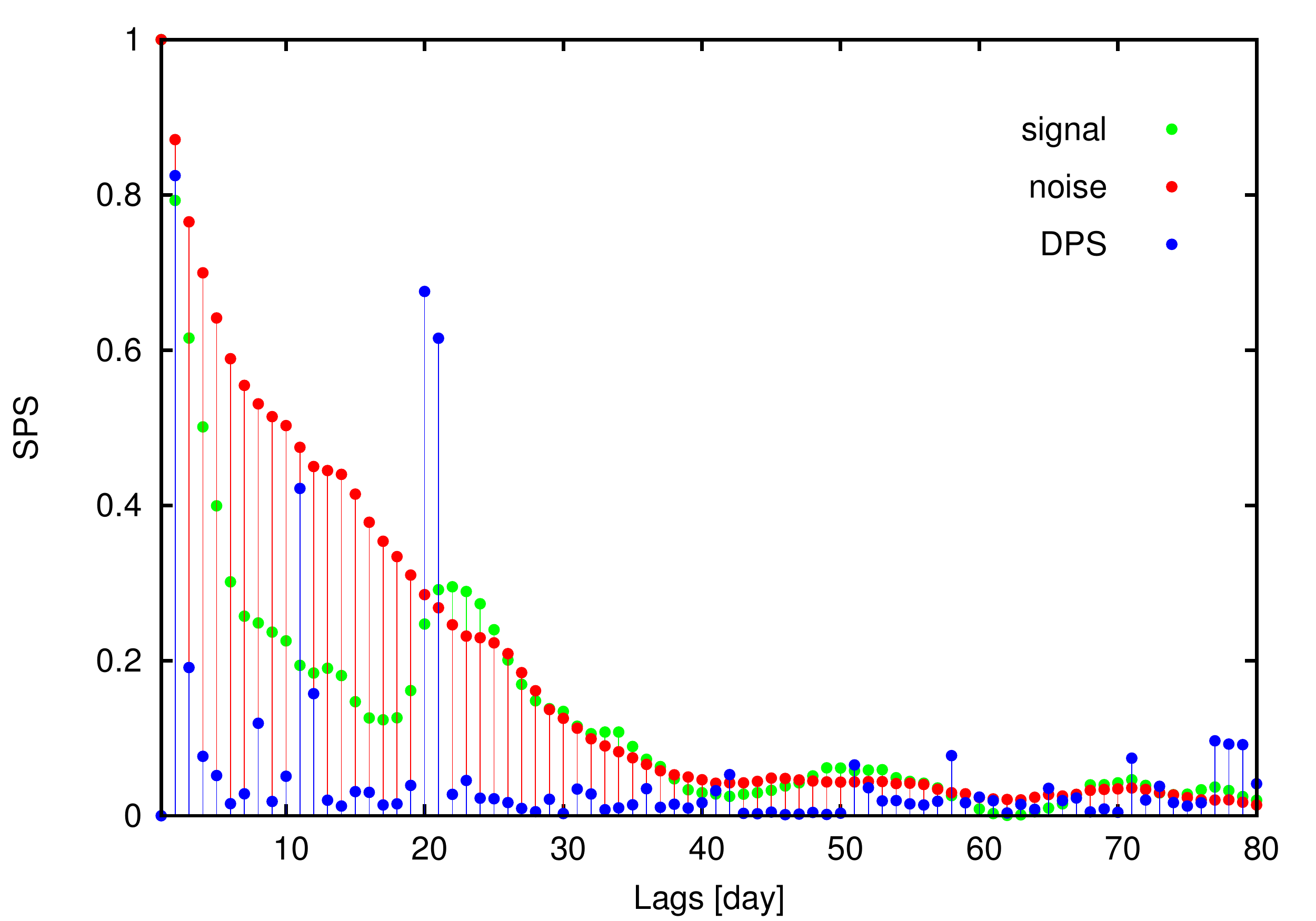} 
\end{center}
\caption{\label{fig:SP_FPSstep2} 
                          Right: The First Power Spectrum. 
                          Left: The Second Power Spectrum.
                          The input to the FPS are light curve in Figure~\ref{fig:lc_MC} after applying Steps 1,2 and 3 of  the FPS signal processing.
                          The input to the DPS is the FPS without any signal processing. 
                          The colors are the same as in Figure~\ref{fig:SP_no}.}
\end{figure*}

%%%%%%%%%%%%%%%%%%%%%%%%%%%%%%%%%%%%%%
\subsubsection{Step 3: Doubling the Points}
%%%%%%%%%%%%%%%%%%%%%%%%%%%%%%%%%%%%%%

Next, to avoid aliasing, we double the points.
Doubling the points does not introduce additional signal,
but, it shifts the Nyquist frequency and therefore allows the power of the spectrum to go to zero
when the frequency approaches the Nyquist frequency. 
We apply the Fourier transform and calculate the power spectrum of the time series. 

Figure~\ref{fig:lc_MC} shows 
the results of these steps of signal processing. 
The time delay in the green light curve appears as a broad peak around the true value of simulated time delay.

%%%%%%%%%%%%%%%%%%%%%%%%%%%%%%%%%%%%%%
\subsection{The Second Power Spectrum}
%%%%%%%%%%%%%%%%%%%%%%%%%%%%%%%%%%%%%%

The resulting FPS (see Figure~\ref{fig:SP_FPSstep2}, left)  serves as an input for the SPS.
We again process the input before applying the Fourier transform.

%%%%%%%%%%%%%%%%%%%%%%%%%%%%%%%%%%%%%%
\subsubsection{Step 1: Flattening and Mean Extraction}
\label{app:SPS1}
%%%%%%%%%%%%%%%%%%%%%%%%%%%%%%%%%%%%%%

\begin{figure*}[ht!]
%\vskip 1cm
\begin{center}
\includegraphics[width=5cm,angle=-90]{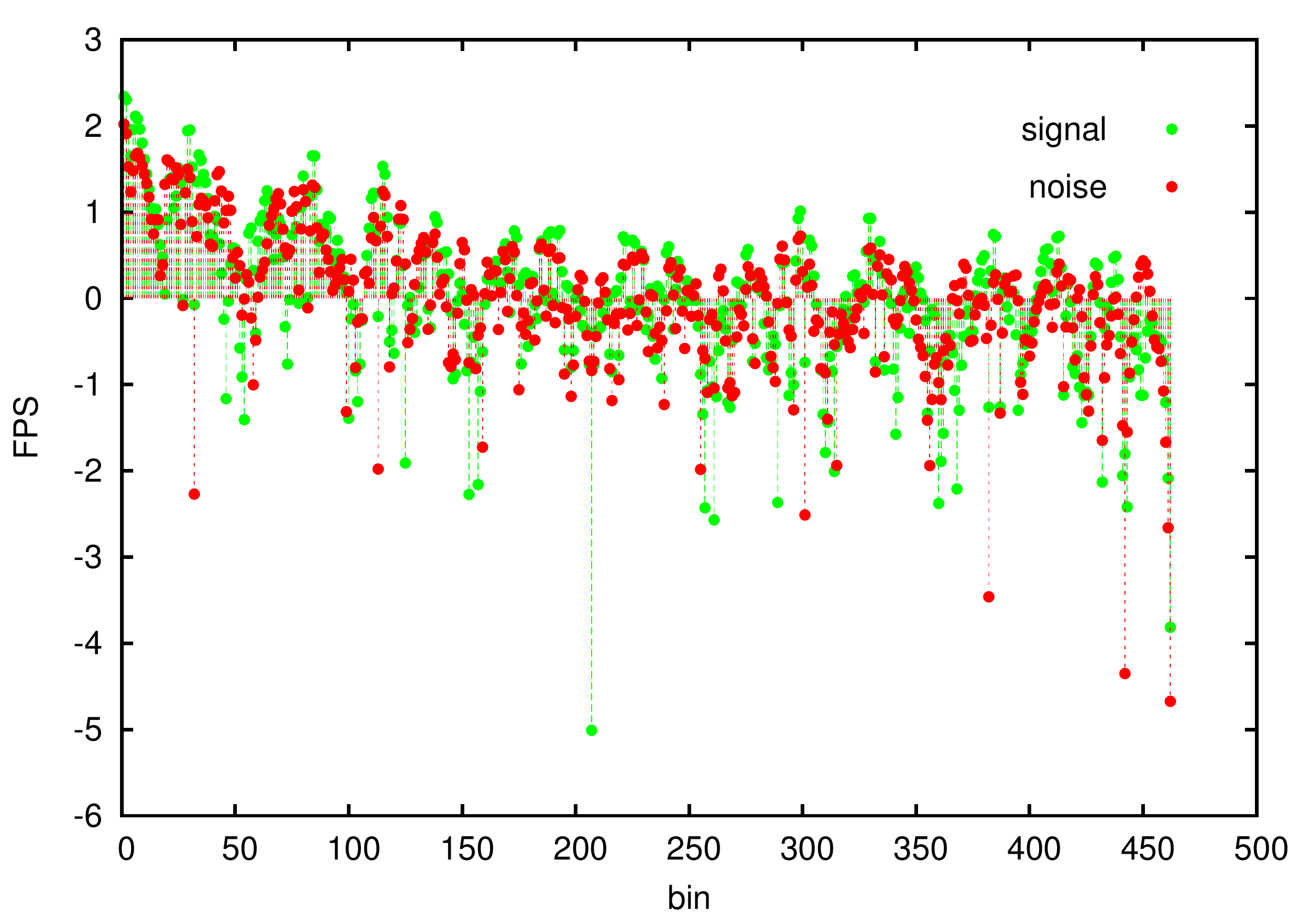} 
\includegraphics[width=5cm,angle=-90]{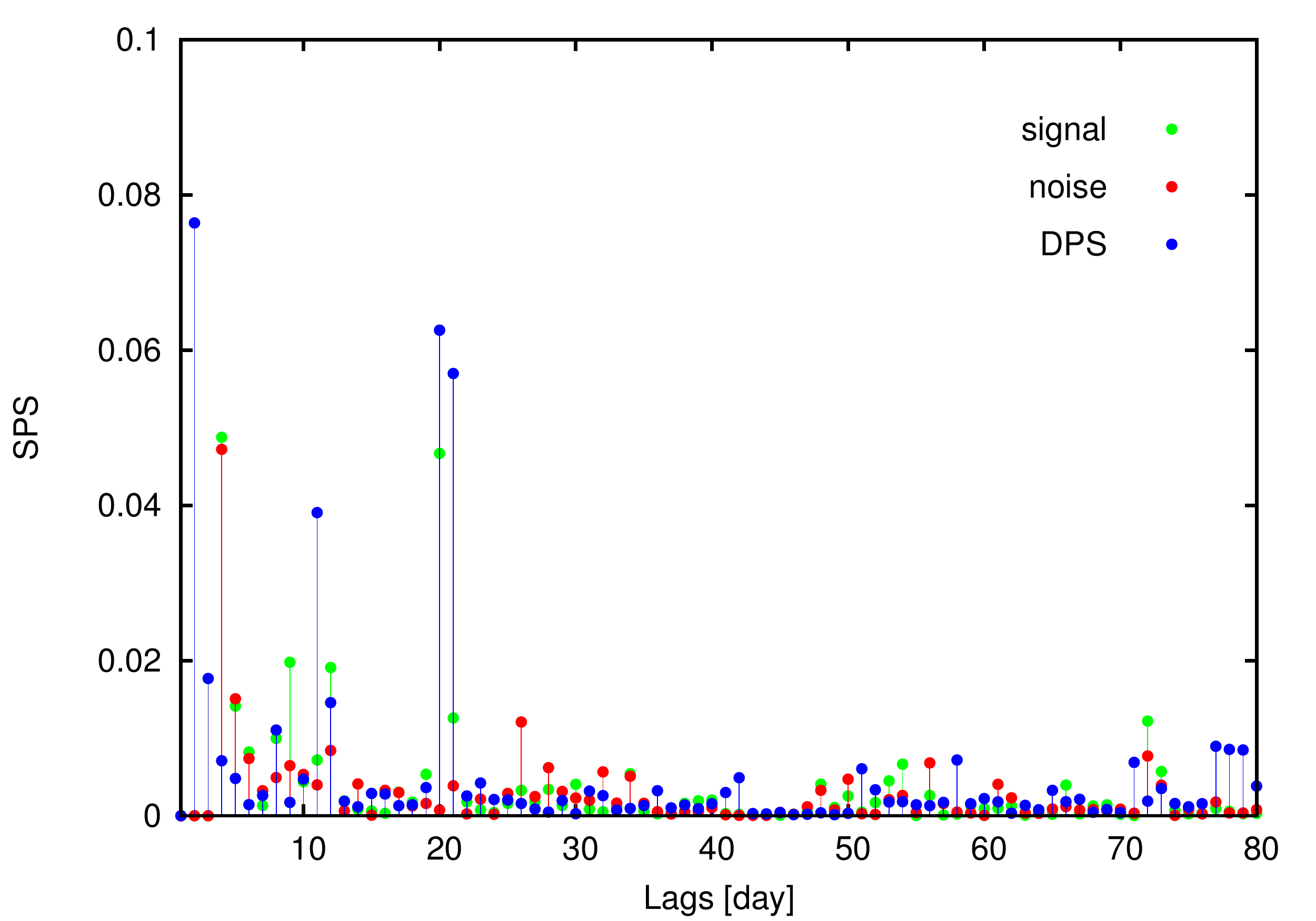} 
\end{center}
\caption{\label{fig:SP_SPSstep1} 
                          Right: The First Power Spectrum. 
                          Left: The Second Power Spectrum.
                          The spectra are calculated after applying signal processing to the light curve,
                          and then processing the FPS by flattening the spectrum and removing a mean
                           (Section~\ref{app:SPS1}).
                          }
\end{figure*}

The observed light curve of blazars can be characterized by power law noise.
This type of signal has large power at low frequencies
%which will manifest its presence with a large amplitude at high frequencies, 
and thus  the signal is not stationary (there is a trend in the data). 
To "flatten" the signal, we take the logarithm of the power spectrum \citep{Bogert1963}. 
Then we can remove the part of the spectrum at low frequencies with large amplitude resulting from power law noise. 
Next, we remove the average from the series. 
Figure~\ref{fig:SP_SPSstep1} (left) shows the input and
and the corresponding power spectrum is shown on the right.
These steps successfully reduce the high amplitudes at short (and spurious) time delays and 
sharpen the peak around the true time delay. 

%%%%%%%%%%%%%%%%%%%%%%%%%%%%%%%%%%%%%%
\subsubsection{Step 2: Windowing and Zero Padding}
%%%%%%%%%%%%%%%%%%%%%%%%%%%%%%%%%%%%%%

In the SPS,
the signal of interest is characterized as a peak around the true value of the time delay. 
Thus, successful processing should sharpen the peak. 
Here we use the Bingham window\footnote{http://www.vibrationdata.com/tutorials/Bingham\_compensation.pdf},  
a combination of a rectangular and a Hanning window. 

\begin{figure*}[ht!]
%\vskip 1cm
\begin{center}
\includegraphics[width=5cm,angle=-90]{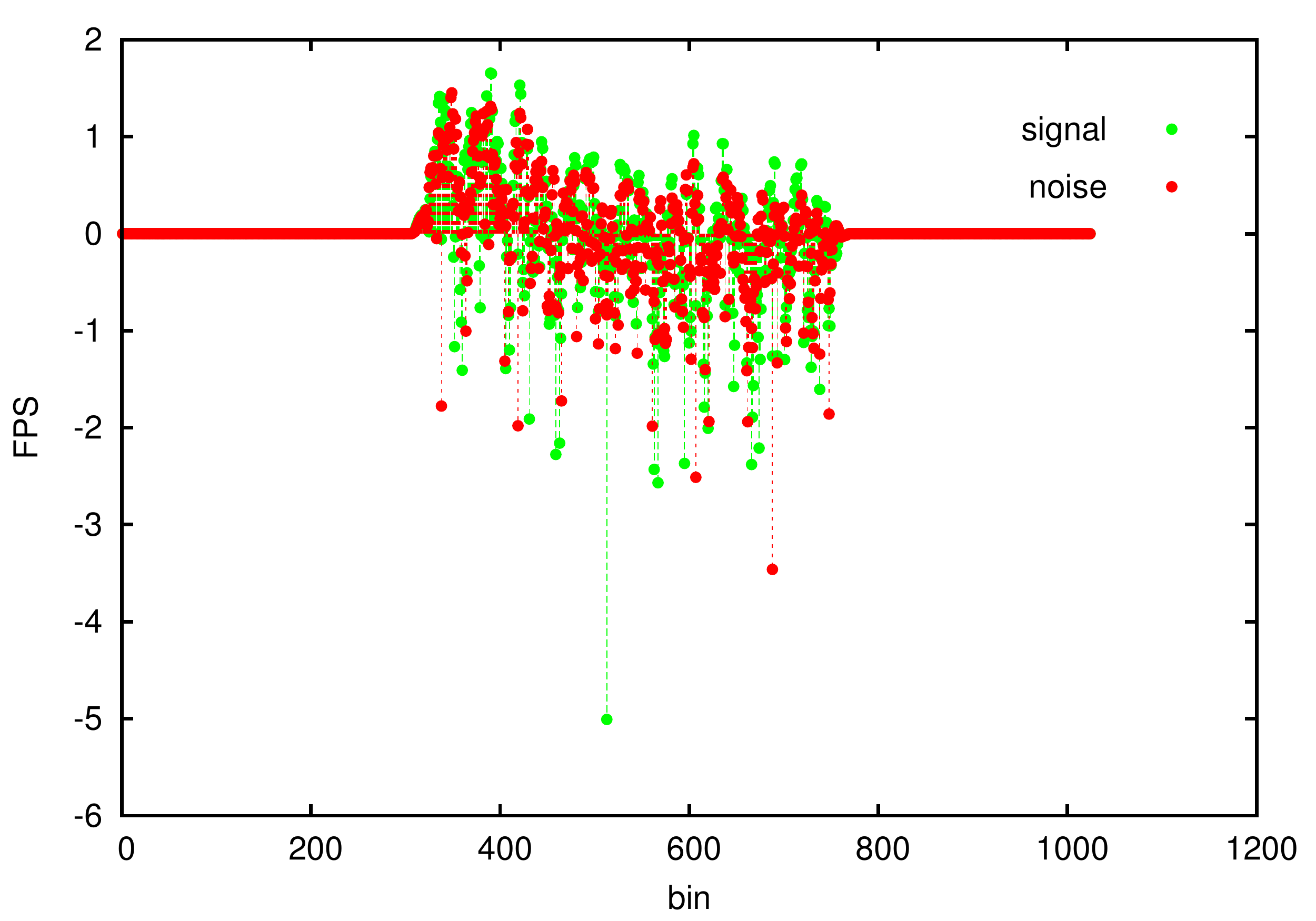} 
\includegraphics[width=5cm,angle=-90]{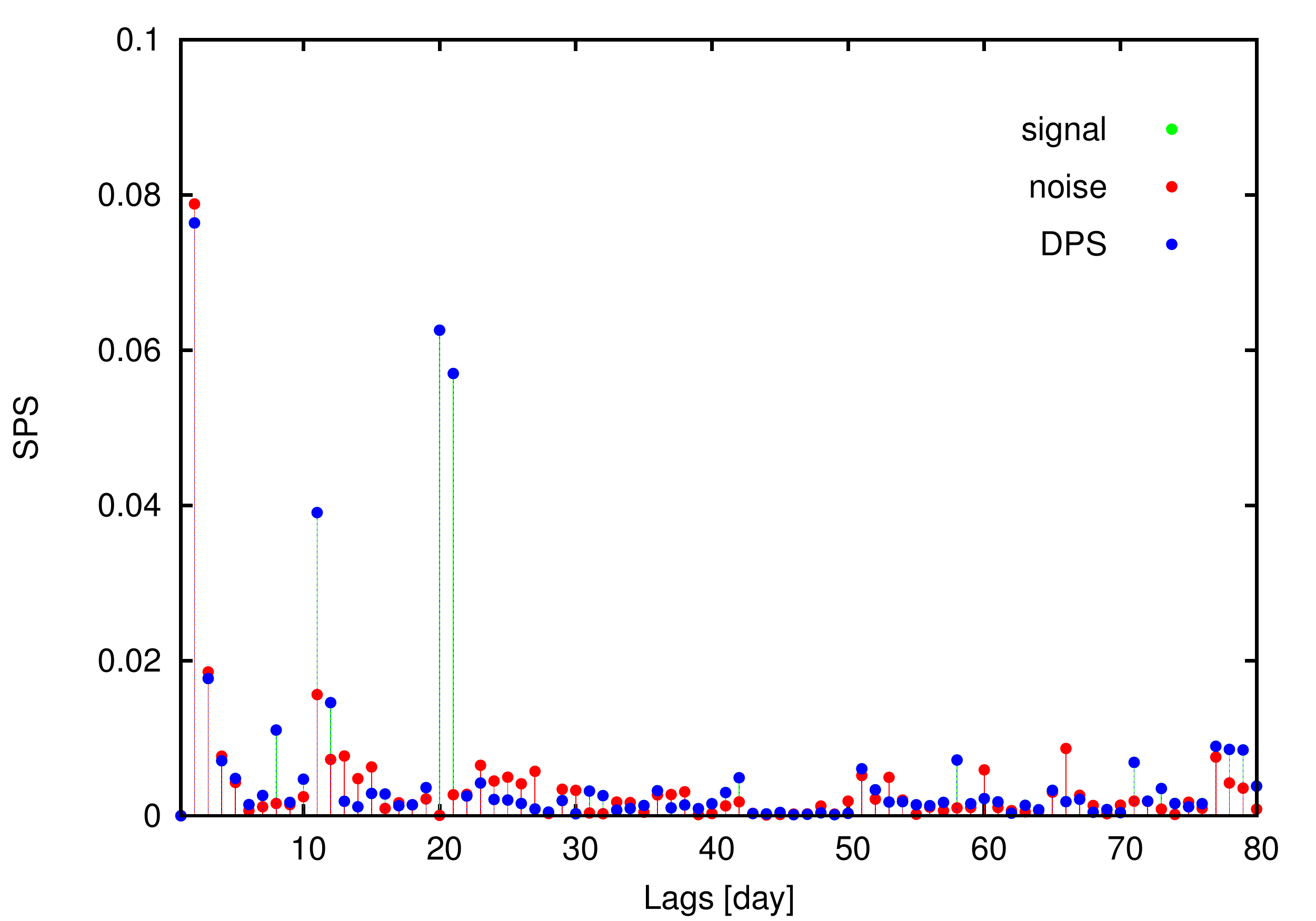} 
\end{center}
\caption{\label{fig:SP_SPSstep3} 
                          Right: First Power Spectrum. 
                          Left: Second Power Spectrum.
                          Results obtained after applying the full signal processing. }
\end{figure*}

Finally, we apply zero padding to the signal and calculate the final SPS (see Figure~\ref{fig:SP_SPSstep3}). 
Monte Carlo simulation demonstrate the effectiveness of the signal processing we have applied.
Figure~\ref{fig:SP_SPSstep3} displays  the results  for one possible realization of the light curve.
The DPS method  is efficient in detecting time delays 
independent of the character of the signal (whether the light curve is  white or red noise)
and 
the method is also very resistant  to producing  spurious detection of time delays even in very structured 
time series of red noise.

%%%%%%%%%%%%%%%%%%%%%%%%%%%%%%%%%%%%%%
\section{Maximum Peak Method - Simulations}
\label{app:MPM}
%%%%%%%%%%%%%%%%%%%%%%%%%%%%%%%%%%%%%%

The time series of gamma-ray light curves consist of short duration flares with variability time scales on the order of hours to days.
We can use this characteristic of the time series to  identify the most prominent flares in the light curves;
we calculate the ratio between the flux during the flare and the flux in the successive bins. 
The resulting  flux ratio is a proxy for the magnification ratio. 

The mirage image arriving first has a larger magnification than the echo images.
Thus, the most luminous  flares can be associated with the first mirage image.  
The echo flares should appear with a flux diminished by a factor corresponding to the magnification ratio. 
Thus we can look for time periods (time bins) when the flux ratios (normalized to brightest flares) are consistent
with the magnification ratios predicted for a particular projection of the jet and lens model. 

The active periods may consist of a series of flares (Flare~1 and Flare~2),
or they can consist of  a single outburst (Flare~3 and Flare~4). 
Analysis of light curves containing single flares is difficult because of the low photon statistics before and after the flare. 
Thus the light curve calculated for short time bins may consist primarily of upper limits. 
Such light curves are useless for the Autocorrelation Function, or even for the Double Power Spectrum methods. 
However, light curves  containing single isolated flares can still reveal the echo flares and their time delays and magnification ratios.
We demonstrate the Maximum Peak Method applied to a single simulated flare in Section~\ref{app:single}.
 
Long active periods with good photon statistics allow construction of  gamma-ray light curves where only a small fraction of the bins have upper limits only.
These light curves are perfectly suited for methods like the Autocorrelation Function and the Double Power Spectrum. 
In this case, the Maxim Peak Method is a powerful consistency check if the detected time delays are consistent with 
the parameters predicted by lens model. 
We demonstrate the performance of this method on simulated superimposed series of flares in Section~\ref{app:superimposed}.

The Maximum Peak Method is complementary to the Autocorrelation Function and the Double Power Spectrum
in the search for gravitationally-induced time delays in unresolved flaring light curves.

%%%%%%%%%%%%%%%%%%%%%%%%%%%%%%%%%%%%%%
\subsection{Single Flares}
\label{app:single}
%%%%%%%%%%%%%%%%%%%%%%%%%%%%%%%%%%%%%%

Flare~3 and Flare~4 are single isolated outbursts. 
We base our simulations on Flare~3. 

\begin{figure}[ht!]
\begin{center}
\includegraphics[width=18cm,angle=0]{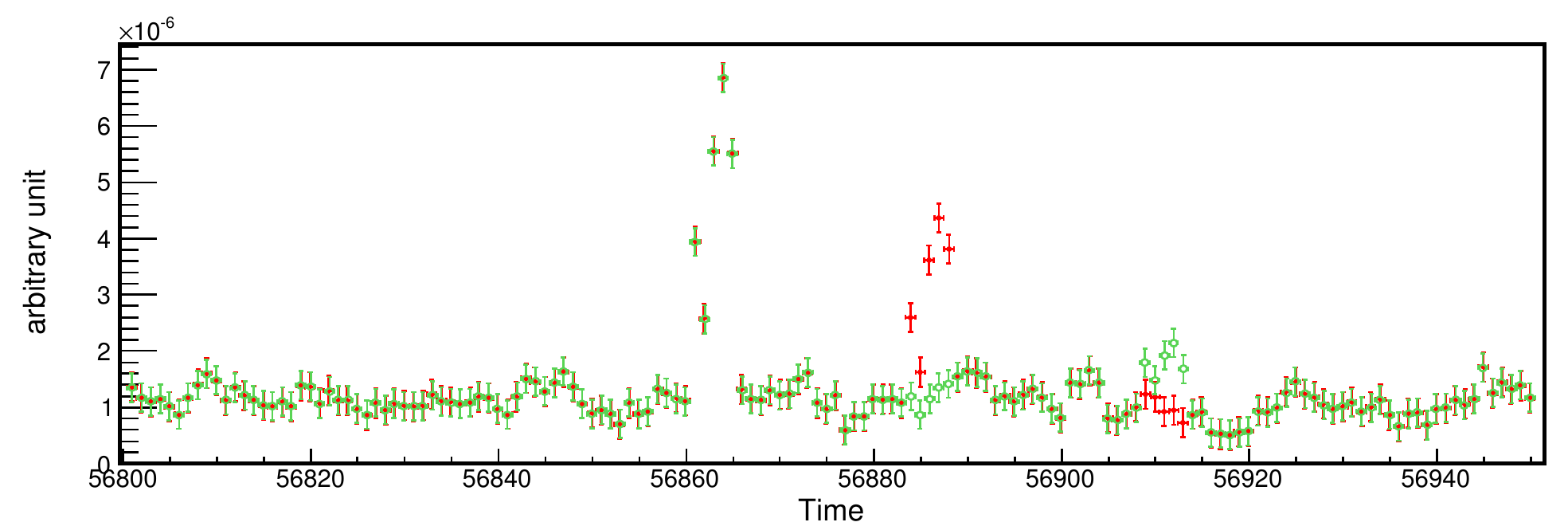} 
\end{center}
\caption{\label{fig:MC_flare} 
                         The artificial light curve generated as pink noise with a flare like structure with time delay of 48~days.
                         This light curve simulates Flare~3. 
                         We include an echo flare with a time delay of 48 days and magnification ratio of 4.5 (green points).
                         Red points represent the light curve with an echo flare  at a time delay of 23~days and magnification ratio of 1.8.  }
\end{figure}

Flares are observed when the emission  significantly exceeds the level typical of the quiescent state.
For single isolated flares the emission before and after the flare is consistent with the average flux.
Therefore, the average flux originates from the site of quiescent emission. 
The temporal behavior of quiescent emission is accurately represented by pink noise. 
We thus start our Monte Carlo simulations by producing light curve composed of pink noise. 
The flaring episode with a flux increase by a factor of 5 is inconsistent with pink noise. 
We add a flare to the light curve with a time structure and  flux per each bin similar to Flare~3.
Then we introduce an echo flare with a time delay of 48~days and magnification ratio 4.5 (Figure~\ref{fig:MC_flare}, green points).
The choice of these parameters  relays on the results of the Maximum Peak Method  for Flare~3. 
We also simulate an echo flare with time delay of 23~days and magnification ratio 1.8 (Figure~\ref{fig:MC_flare}, red points).
We simulate the echo flare at 23~days to demonstrate that if the flare would originate from the region consistent with the core 
then the echo flare would be detectable.
%Figure~\ref{fig:MC_flare} shows the simulated light curve.  

Figure~\ref{fig:SP_SPSstep3} shows the result of applying the Maximum Peak Method to the simulated light curve shown in  Figure~\ref{fig:MC_flare}.
The method shows that the ratio we obtain  between the flux of the flare peak and the flux in the bin corresponding to echo flare 
agree with model predictions. 
The method rejects the majority of time delay ranges where there is no consistent magnification ratio. 

\begin{figure}[ht!]
\begin{center}
\includegraphics[width=5cm,angle=-90]{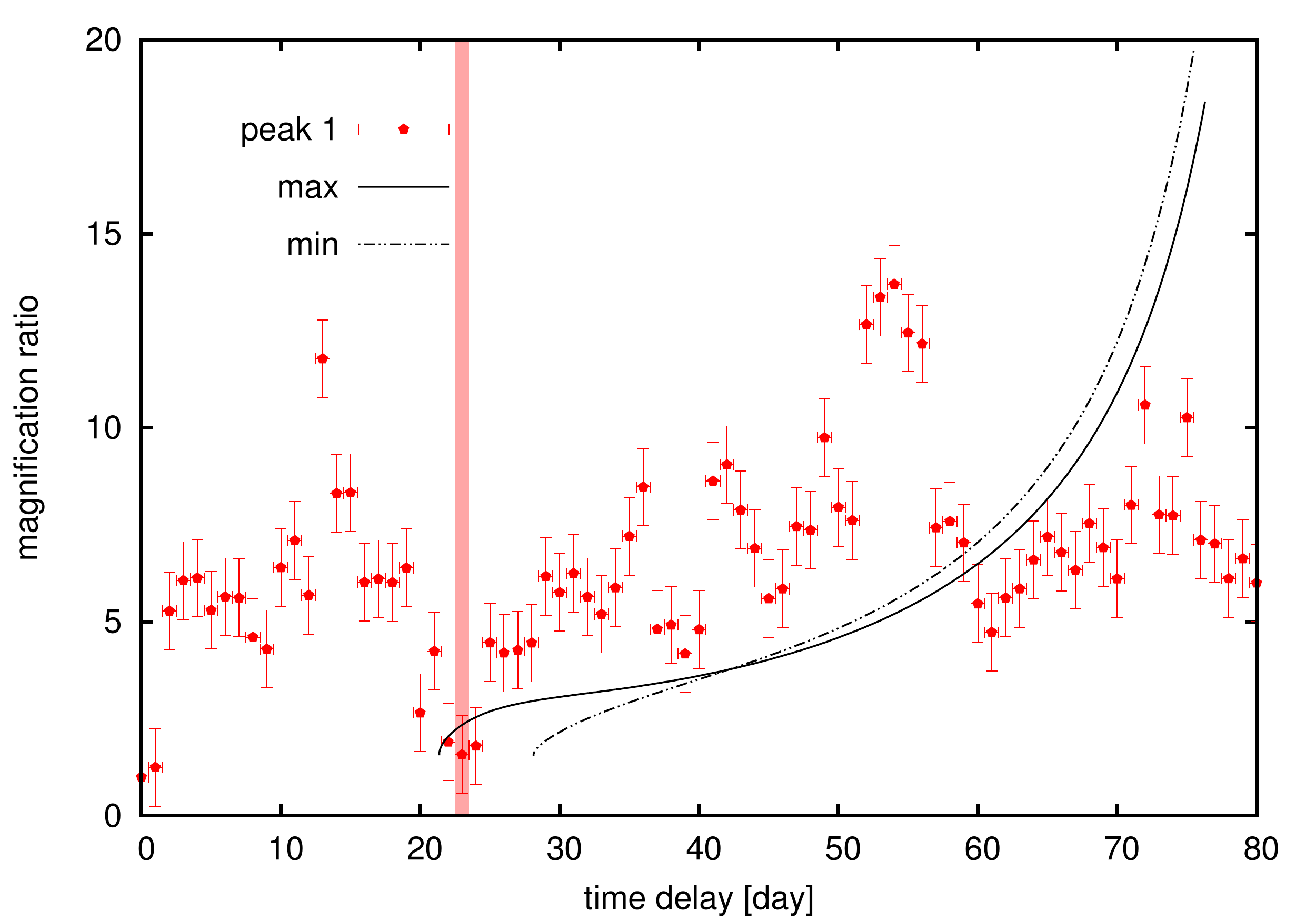} 
\includegraphics[width=5cm,angle=-90]{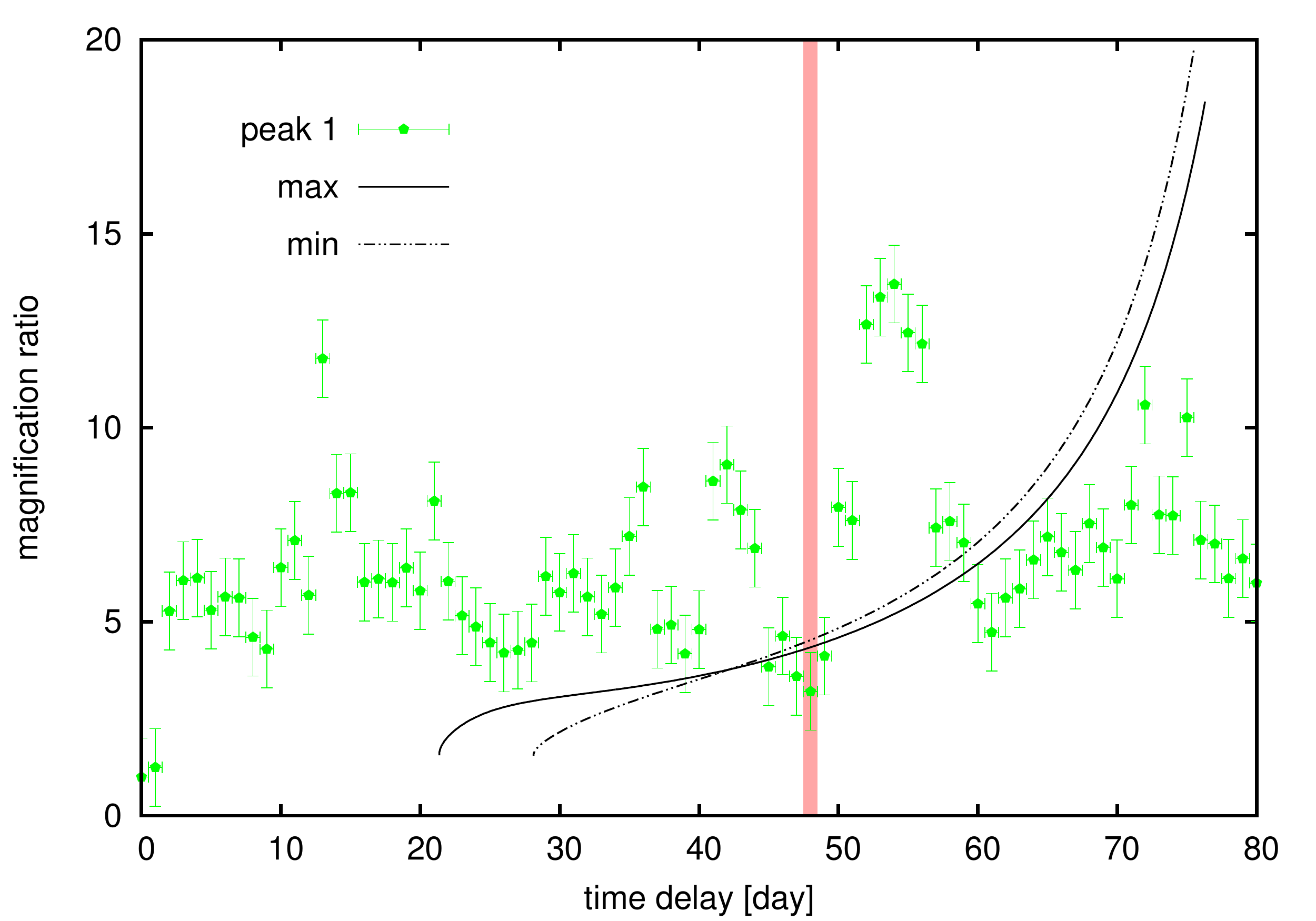} 
\end{center}
\caption{\label{fig:SP_SPSstep3} 
                          Maximum Peak Method applied to the simulated light curve Figure~\ref{fig:MC_flare}. 
                          The red area indicates the time bin corresponding to the simulated time delay.
                          Solid and dashed lines indicate the model predictions for magnification ratio as a function of the time delay 
                          for boundary alignments of the jet constrained by the radio observations (Figure~\ref{fig:core}). 
                          Left: Results for simulated light curve with induced echo flare with time delay of 
                          23~days and magnification ratio 1.7.
                           Right: Simulated light curve with induced time delay of 48~days and magnification ratio 4.5 }
\end{figure}

Note that the maximum distance of the emitting region along the jet which we can constraint 
depends on the ratio between the observed flux of the flare and the flux relative to quiescent state. 
Flare~3 and the simulated flare exceed the  average flux by a factor of $\sim5$. 
In this example, if the predicted magnification ratio is larger than $\sim$5, we do not expect to be able to detect the echo flare; 
the flux of the echo flare is then below the average of the quiescent state. 
Thus, we can only test expected  magnification ratios for Flare-3 in the range from 1 to 5.

For Flare-3 
the magnification ratio $\sim5$ corresponds to a region along the jet located at least 1.5~kpc from the core  (see Figure~\ref{fig:jet_lens}). 
Detection of a consistent magnification ratio f $\sim 5$ still does not provide  clear evidence of echo flare detection because the data are also consistent with a flare from a region at distances $\geq1.5\,$kpc from the core. In other words, the observed magnification ratio sets a very interesting limit on the distance between the core and the origin of the flare, but it does not pinpoint it location.

%%%%%%%%%%%%%%%%%%%%%%%%%%%%%%%%%%%%%%
\subsection{Superimposed Series of Flares}
\label{app:superimposed}
%%%%%%%%%%%%%%%%%%%%%%%%%%%%%%%%%%%%%%

Here we investigate the  performance of the Maximum Peak Method applied to a light curve which consists of a series of superimposed flares. 
Flares~1 and~2 are examples. 
As an input time series we use the simulated light curve  shown in Figure~\ref{fig:lc_MC}.

Figure~\ref{fig:SP_SPSstep3} shows the result of the Maximum Peak Method applied to the series of superimposed flares.
The method confirms that the time delay detected with Double Power Spectrum is consistent with the predictions and
the method excludes significant ranges of possible time delays.

\begin{figure}[ht!]
\begin{center}
\includegraphics[width=5cm,angle=-90]{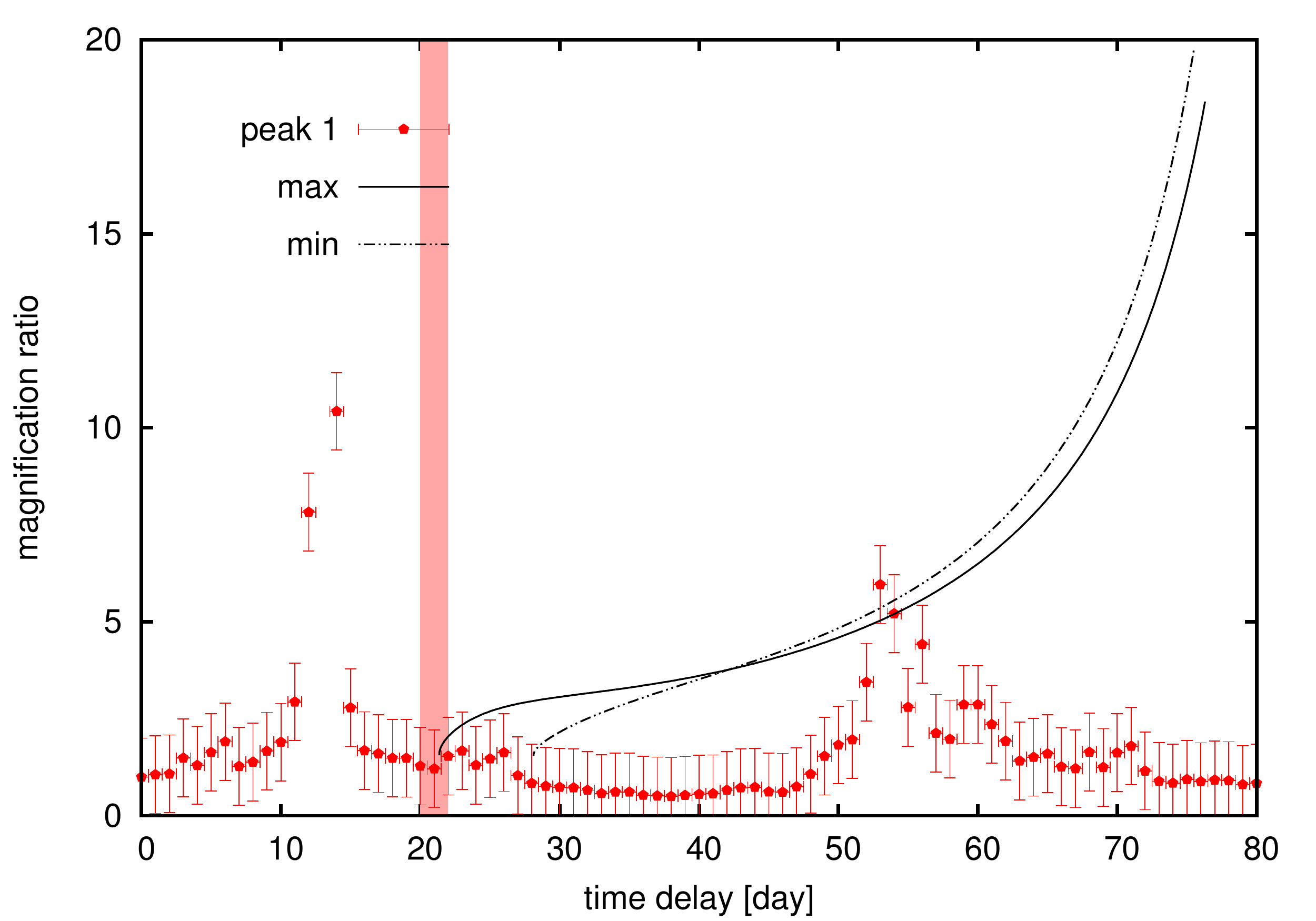} 
\end{center}
\caption{\label{fig:SP_SPSstep3} 
                          Maximum Peak Method applied to the simulated light curve shown in Figure~\ref{fig:lc_MC}. }
\end{figure}

\end{document}